\documentclass[onecolumn,superscriptaddress]{revtex4}

 \usepackage{amsfonts}
\usepackage{amsmath}
\usepackage{amssymb}
\usepackage{graphicx}
\usepackage{dcolumn}
\usepackage{dsfont}
\usepackage{times}
\usepackage{epsfig}
\usepackage[]{units}

\usepackage{lipsum}

\usepackage{xcolor}

\newcommand{\R}{{\mathbb R}}

\newcommand{\C}{{\mathbb C}}

\def\epsilon{\varepsilon}

\def\q{\mathbf{q}}
\def\F{\mathbf{F}}

\def\beq{\begin{equation}}
\def\eeq{\end{equation}}
\def\l{\ell}

\def\hatd{\hat{\delta}}

\newcommand{\subfigimg}[3][,]{%
  \setbox1=\hbox{\includegraphics[#1]{#3}}
  \leavevmode\rlap{\usebox1}
  \rlap{\hspace*{5pt}\raisebox{\dimexpr\ht1-.1\baselineskip}{#2}}
  \phantom{\usebox1}
}

\begin{document}

\title{Nonlinear Localized Modes in Two-Dimensional Hexagonally-Packed Magnetic Lattices}

\author{C. Chong*}
\affiliation{Department of Mathematics, Bowdoin College, Brunswick, ME 04011, USA}

\author{Yifan Wang}
\affiliation{Division of Engineering and Applied Science California Institute of Technology
Pasadena, CA 91125, USA}

\author{Donovan Mar\'echal}
\affiliation{Division of Engineering and Applied Science California Institute of Technology
Pasadena, CA 91125, USA}

\author{E. G. Charalampidis}
\affiliation{Mathematics Department, California Polytechnic State University, San Luis Obispo, CA 93407-0403, USA}

\author{Miguel Moler\'on}
\affiliation{Institute of Geophysics, Department of Earth Sciences, ETH Zurich, 8092 Zurich, Switzerland}

\author{Alejandro J. Mart\'inez}
\affiliation{Computational Biology Laboratory, Fundaci\'on Ciencia \& Vida, Santiago, 7780272, Chile}
\affiliation{Universidad San Sebastian, Santiago, 7510156, Chile}

\author{Mason A. Porter}
\affiliation{Department of Mathematics, University of California, Los Angeles, CA 90095, USA}

\author{P. G. Kevrekidis}
\affiliation{Department of Mathematics and Statistics, University of Massachusetts, Amherst, MA, 01003, USA}

\author{Chiara Daraio}
\affiliation{Division of Engineering and Applied Science California Institute of Technology
Pasadena, CA 91125, USA}

%%%%%

\begin{abstract}

We conduct an extensive study of nonlinear localized modes (NLMs), which are temporally periodic and spatially localized structures,
in a two-dimensional array of repelling magnets. 
%{\bf map: in terms of the definition of NLMs, aren't they allowed to be quasiperiodic or chaotic temporally? but the above 'definition' doesn't allow this. CC: see response to similar comment below}
In our experiments, we arrange a lattice in a hexagonal configuration with a light-mass defect, and %we drive the center of the 
we harmonically drive the center of the chain with 
%lattice with a harmonic drive with 
a tunable excitation frequency, amplitude, and angle. We use a 
%simple 
damped, driven variant of a vector Fermi--Pasta--Ulam--Tsingou lattice to model our experimental setup. Despite the idealized
%simple
nature of the model, we obtain good qualitative agreement between theory and experiments for a variety of dynamical behaviors.
We find that the spatial decay is direction-dependent and that
%We find that the spatial tails decay exponentially
%PGK: Chris: is this REALLY exponential or is it a modulated
%exponential as one might expect in 2d (given the Green's function)??
%Can you please check and if needed amend?
%in all directions with direction dependent decay coefficients. 
%with different rates in different directions. 
%{\bf map: above, you write "rates", but don't you really mean with respect to space? ('rates' makes me think of time)}
drive amplitudes along fundamental
%two different 
displacement axes
%the vertical or horizontal axis 
lead to
nonlinear resonant peaks in frequency continuations that are similar
to those that occur in one-dimensional damped, driven lattices. 
%{\bf map: "vertical" and "horizontal" don't have any context here (before, it was $y$ and $x$, which have even less context); do we really mean "orthogonal to the chain" and "parallel to the chain"? we need a description based on the physical object for this statement in the abstract to be meaningful [there is a note below that they don't want "horizontal" and "vertical", which I agree is bad; but $x$ and $y$ are also terrible, so we need to find some other way to phrase things}
%{\bf map: $x$ and $y$ axes are not going to work here as phrasings (that is extremely confusing and it only gets given a meaning several pages into the manuscript); reading various comments elsewhere in the .tex file, I understand the problem with the words "horizontal" and "vertical"; I therefore tried to make a change in the spirit of that one to remove the terms entirely, but we should iterate on phrasing here; I am flagging this one for further consideration and iteration}
However,
driving along other directions leads to the creation of asymmetric
NLMs that bifurcate from the main solution branch, which consists of
symmetric NLMs. When we vary the drive amplitude, we observe such
behavior both in our experiments and in our simulations.
We also demonstrate that solutions that appear to be time-quasi-periodic
%solutions
bifurcate from the branch of
symmetric time-periodic NLMs. 
%PGK: How sure are you of this Chris? Can you measure the
%quasi-periodicity somewhat definitively or is that you observe a Hopf
%and you believe it should be quasi-periodic (it'd just be good to
%know the degree of certainty on the statement which seems fairly
%definitive at the moment.

%{\bf map: I'll need to look below at Chris's answer to my comment about 'periodic' in the definition}

\end{abstract}


\maketitle

%%%%%%%%

%%%%%%%%

\section{Introduction}

 Discrete breathers are spatially localized, time-periodic solutions of nonlinear lattice differential equations.
% {\bf map: I don't think it has to be periodic in time, as concerns a definition; I don't see why it can't e.g. be time-quasiperiodic. CC: I think it would be better to leave the definition as time periodic. Papers I am familar with use this definition. MAP, if you have reference(s) where breather and NLM are non time-periodic we can add a sentence like"while some relax the condition of time periodicity in the definition of a breather \cite{} in this article we will always assume a breather and NLM are time periodic. }
 They have been studied in a host of scientific areas, including
optical waveguide arrays and photorefractive crystals~\cite{moti}, 
Josephson-junction ladders~\cite{alex,alex2}, layered antiferromagnetic crystals~\cite{lars3,lars4},
halide-bridged transition-metal complexes~\cite{swanson}, dynamical models of the 
DNA double strand \cite{Peybi}, molecular lattices \cite{2Dreview2015}, 
Bose--Einstein condensates in optical lattices~\cite{Morsch}, and many others.  

Most of the immense volume of work --- now spanning more than three decades --- on discrete breathers
%in over three decades now 
has been in one-dimensional (1D) lattices \cite{Flach2007,pgk:2011,Dmitriev_2016}.   
%{\bf map: what about research from the past 13 years? is there also a
%more recent review? (otherwise, I think we need to rephrase the
%sentence above a bit). CC: I added two more. Maybe Panos has other
%suggestions...} PGK: There is no new major review that I am aware
%of. But I did dig up something nonetheless, from 2016 S. Dmitriev.
Most relevant to the present article is research on discrete breathers in Fermi--Pasta--Ulam--Tsingou (FPUT)
lattices, which have nonlinear inter-site coupling \cite{FPU55,FPUreview}. Moreover, FPUT-like lattices with power-law potentials have been used to model a variety of mechanical systems, such as granular crystals \cite{Nester2001,granularBook,yuli_book,gc_review}
and (more recently) magnetic lattices \cite{moleron,Mehrem2017,Marc2017}.

There have also been some studies of breathers in two-dimensional (2D) lattices, although there are many fewer such studies than of 1D lattices. Example physical settings in 2D include crystal lattices \cite{Marin2000, Dmitriev_2016}, electric circuits \cite{English2013}, and dusty plasmas \cite{DustyPlasma,Koukouloyannis2010}.
Breathers in 2D lattices have been analyzed with both asymptotic methods \cite{Wattis_2Dreview} and numerical methods 
in both homogeneous \cite{Flach2D_1997,Marin98} and heterogeneous media \cite{Qiang2009,Koukouloyannis2009}. See \cite{2Dreview2007,2Dreview2015} for overviews of results about 2D breathers.

The 2D setting of the present work is a mechanical system in which
each magnet has
%both
%a vertical and horizontal displacement,
%PGK: Chris, there is an important issue here too. By
%vertical/horizontal you mean x-y I believe. However, one things for
%planar systems z as vertical so this is misleading. If you concur, I
%would make sure to eliminate the use of that term (also in abstract etc.
two displacement fields, which
distinguishes it from many studies of scalar 2D lattices, such as
those that describe electrical
circuits~\cite{English2013}.
%{\bf map: I don't understand what distinction is being made above; it is 2D, so of course we have two axes of displacement; I am sure something specific is meant, but I don't know what is it. CC: in some lattices, the equations don't describe displacement and so the equations are scalar}
Specifically, we examine a lattice of repelling magnets that are arranged in a hexagonal configuration. The choice of a hexagonal arrangement
is motivated by the experimental setup, as hexagonal configurations are more robust structurally than other arrangements (such as square ones).
%{\bf map: "stable" in what sense of the word; this needs some unpacking (or at least a pointer to a discussion later in the paper where we say what we mean by this). CC: I will
%leave this to one of the experimentalists to answer.}
At the center of the lattice is a light-mass defect, which introduces a localized defect mode into the spectrum of the linearization of the system. To excite the system experimentally, we drive the center of the lattice by a force that results
from the current that flows along a wire that we suspend above the lattice. We model damping using a dashpot term. Putting everything together, the proposed model for the experimental setup is a damped, driven variant of a vector FPUT lattice.

Although a breather is defined as a spatially localized and time-periodic structure, it is useful to label different types of breathers. 
%{\bf map: I don't see why a breather can't be e.g. quasiperiodic. CC: see response above}
 Linear systems with an impurity or a defect (e.g., with a particle of lighter mass than the other particles) have 
 isolated points in their spectra that lie above the spectral edge. These modes are called ``defect localized modes''~\cite{Mara}.
 In the presence of nonlinearity, breathers can bifurcate from these 
 %localized defect 
 modes and can exist
 for frequencies other than the linear defect frequency. Breathers that manifest in this way are called ``nonlinear localized modes'' (NLMs) \cite{Theocharis2009} and are the subject of the present work.
By contrast, breathers that do not manifest via a defect or an impurity are called
 ``intrinsic localized modes'' (ILMs). 
% One example of how 
One way for ILMs, which we do not investigate in the present paper, to manifest themselves is via a modulation instability of plane waves \cite{Flach2007}.  
%By contrast, breathers that manifest via a modulation instability of plane waves are called
% ``intrinsic localized modes'' (ILMs) \cite{Flach2007} (and are not the subject of this paper).  
% {\bf map: is the above always true? to me, the notion of 'intrinsic' and MI don't have to be tired together; this is being written as a 'definition', and that doesn't seem right; shouldn't these, instead, be \emph{examples} of ILMs? CC: see my re-wording.} 
In addition to breathers, other kinds of orbits --- such as quasiperiodic and chaotic ones --- can also occur in nonlinear lattices.
%be identified in nonlinear lattices, . 
For example, such orbits have
been identified in strongly nonlinear damped, driven granular chains \cite{Nature11,hooge12}, suggesting that such solutions may also be present
in damped, driven magnetic lattices.
%, too, in line with what we will explore in what follows. 
In the present work, we examine such NLM states, their stability, and the modes that arise 
%through bifurcations 
as a result of instabilities.

Our paper proceeds as follows. We present our experimental setup in Sec.~\ref{sec:exp_setup}, and
we detail the corresponding model equations, 
%basic 
linear theory, and numerical methods in Sec.~\ref{sec:model}.
We give the main numerical and experimental results in Sec.~\ref{sec:results}, where we explore NLM profiles, spatial decay, parameter continuations, and nearly time-quasiperiodic orbits. 
%{\bf map: above is the first time we mention quasiperiodic orbits; I think we need to bring it up briefly earlier in the introduction if we're going to have it here. CC see new text at end of previous paragraph.}
We conclude and discuss future
challenges in Sec.~\ref{sec:theend}.

%%%%%

\section{Experimental Setup} \label{sec:exp_setup}

   \begin{figure}
    \centering
   \begin{tabular}{@{}p{0.4\linewidth}@{}p{0.27\linewidth}@{}p{0.3\linewidth}@{}  }
  \rlap{\hspace*{5pt}\raisebox{\dimexpr\ht1-.1\baselineskip}{\bf (a)}}
\includegraphics[height= 5cm]{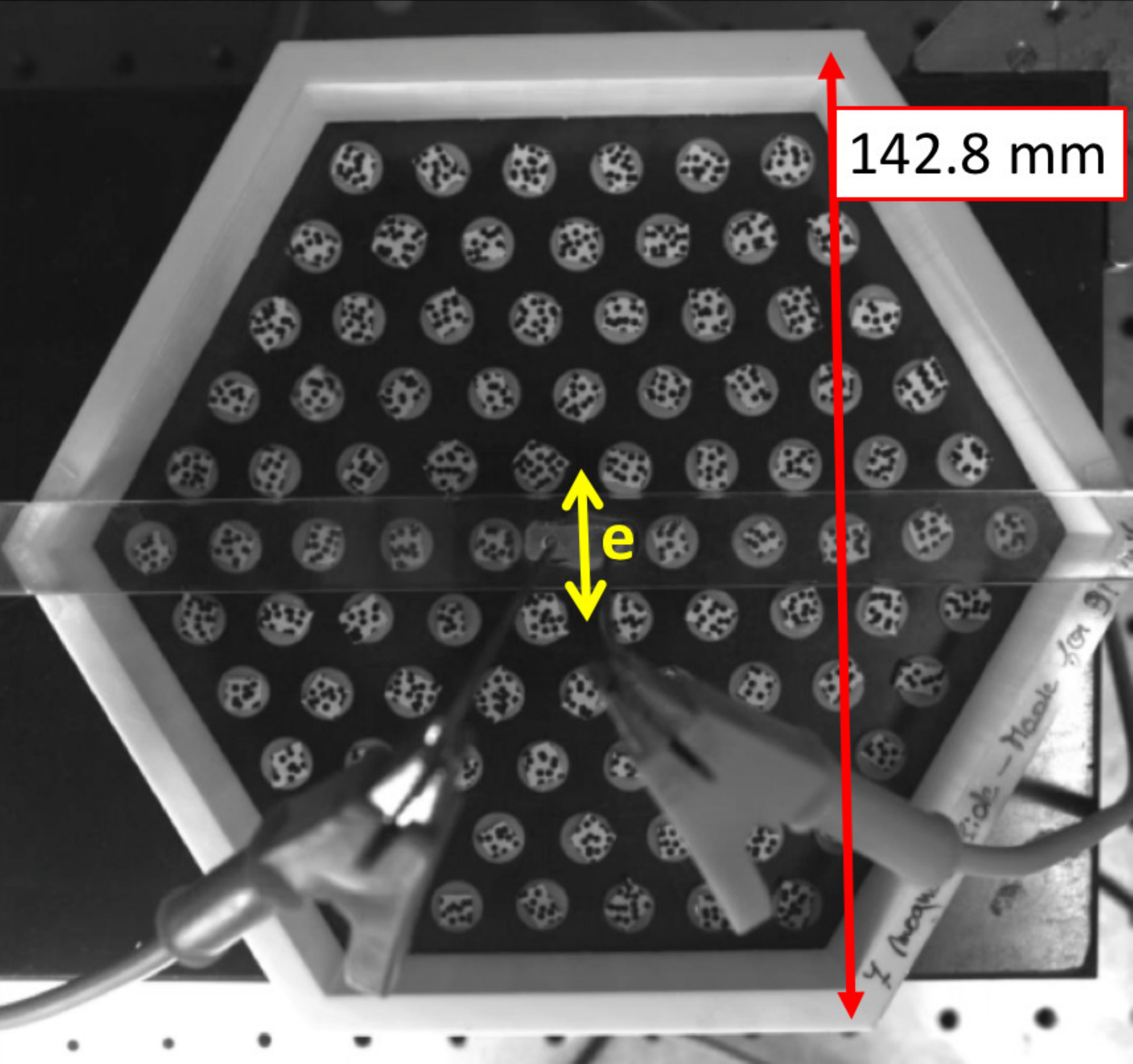} &
  \rlap{\hspace*{5pt}\raisebox{\dimexpr\ht1-.1\baselineskip}{\bf (b)}}
\includegraphics[height= 5cm]{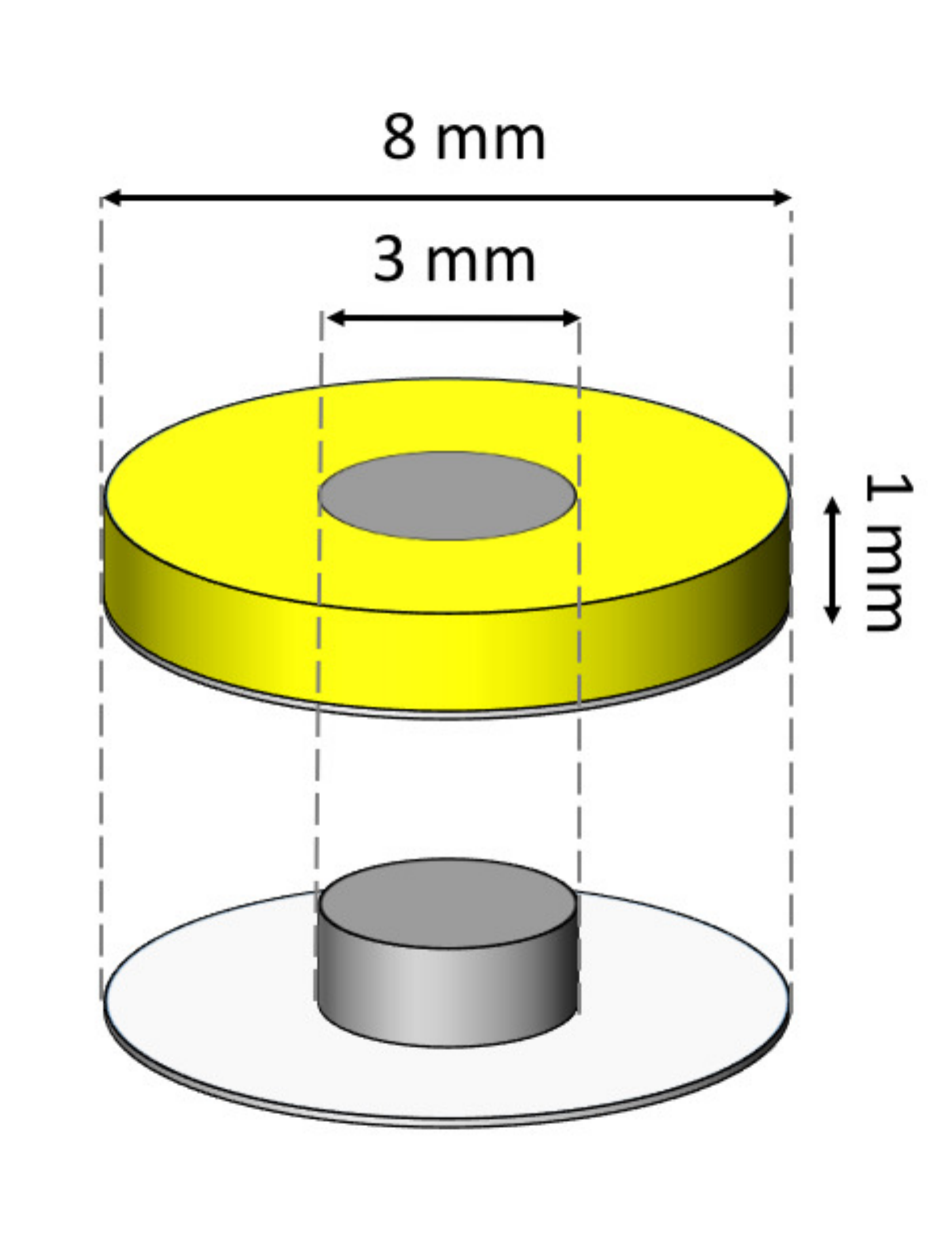}&
  \rlap{\hspace*{5pt}\raisebox{\dimexpr\ht1-.1\baselineskip}{\bf (c)}}
\includegraphics[height= 5cm ]{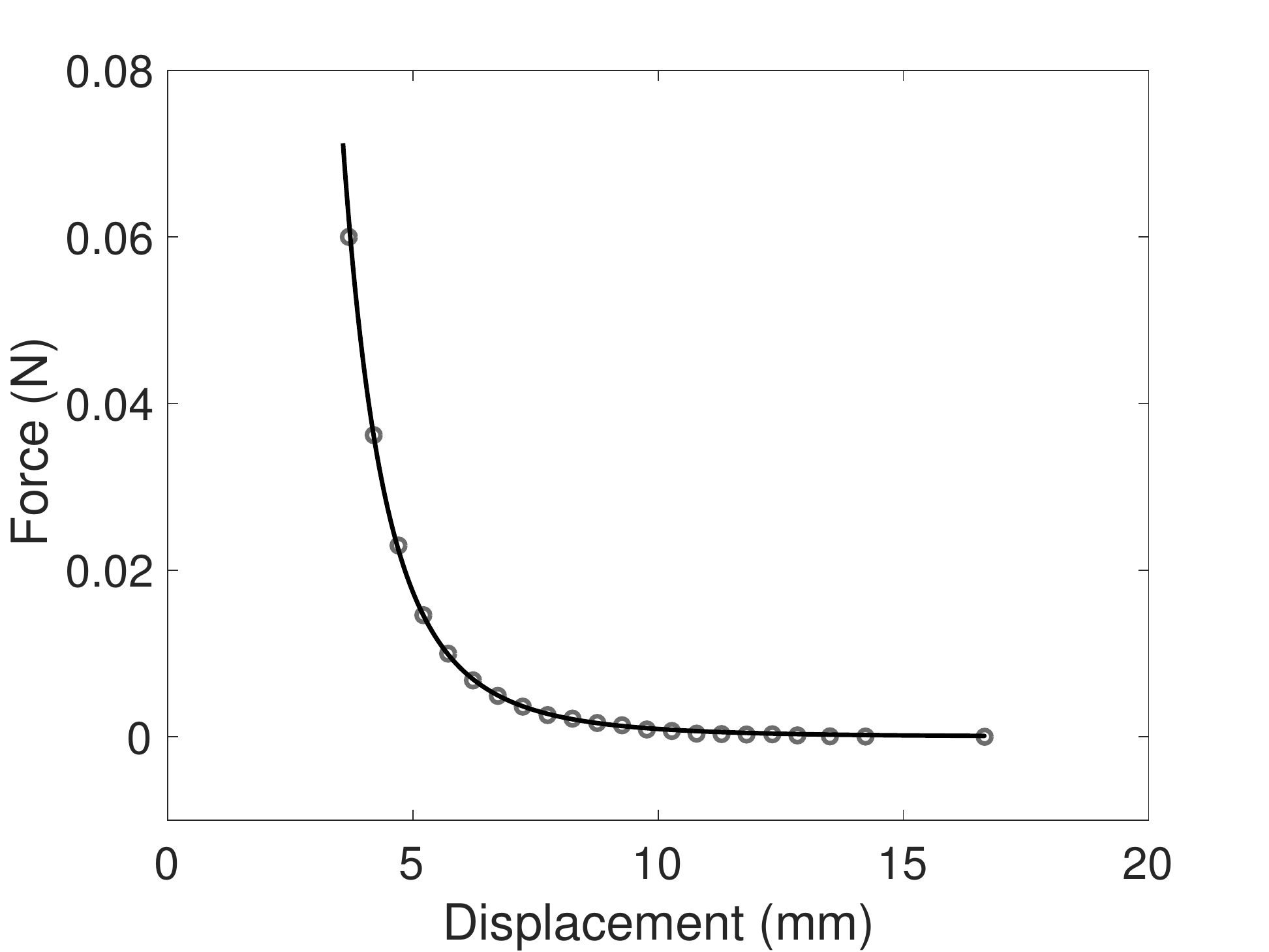}
%
%      \subfigimg[width=\linewidth]{\bf (a)}{setup_picture}  &
%    \subfigimg[width=\linewidth]{\bf (b)}{video_tracking} &
%    \subfigimg[width=\linewidth]{\bf (c)}{units_illustration} & 
%    \subfigimg[width= \linewidth]{\bf (d)}{magnetic_repulsion}
%    
%    
%   \newcommand{\subfigimg}[3][,]{%
%  \setbox1=\hbox{\includegraphics[#1]{#3}}
%  \leavevmode\rlap{\usebox1}
%  \rlap{\hspace*{5pt}\raisebox{\dimexpr\ht1-.1\baselineskip}{#2}}
%  \phantom{\usebox1}
%}
  \end{tabular}
 \caption{ \textbf{(a)} Picture (from a video frame that was used for particle motion tracking) of the experimental setup. 
 %We glue 36 magnets under the white plastic boundary. 
 The yellow arrow 
 %(with an accompanying ``e'') 
 indicates the direction of the external excitation. \textbf{(b)} Sketches of the (top) normal and (bottom) defect particles. \textbf{(c)} Magnetic dipole--dipole interactions in experiments (open grey circles) and fitted model using Eq.~\eqref{eq:magnet} (solid black curve).
% {\bf map: note to self: come back here and see if I want to cite the equation number for the model}
%\egc{Would it be possible to use different markers to highlight the distinction between experiments and numerical simulations in panel (d)? Perhaps, solid black line for numerical simulations and open circles for experiments? DONO: is that okay?}
 }
 \label{fig:expsetup}
\end{figure}

We place a 2D lattice on an air-bearing table to make the magnetic particles (which constitute the nodes of the lattice) levitate.
 %\textit{(I would not use the term friction here)}}. 
 The lattice consists of 127 magnetic particles that are hexagonally
packed. We glue 36 of these particles to the boundaries, and 91 of them are free to move [see Fig.~\ref{fig:expsetup}(a)]. Each particle is a 3D-printed disc with a hole in the center, where we attach a neodymium magnetic cylinder. We glue a thin piece of cover glass at the bottom to make the surface smoother and thereby improve the levitation of the particles. We build the defect particle, which is located in the center of the lattice, by directly attaching the magnet on the glass without the 3D-printed structure. This particle has a lighter mass and serves as a defect [see Fig.~\ref{fig:expsetup}(b)]. The mean mass of a normal disc particle is 138.2 mg $\pm$ 3.1 mg (where we measure the standard deviation from a sample of $20$ particles). The defect particle has a mass of 81.6 mg, which corresponds to 58.68\% of the normal particle mass.

We excite the defect particle using an external magnetic field that we generate using a conductive wire that we place over the particle at a height of $3$ mm. We generate the AC current that flows through the wire from a Lock-In amplifier (SR860 500kHz DSP Lock-in Amplifier), and we amplify it with an audio amplifier (Topping TP22, class D). 
%{\bf map: can we include references in the bibliography to websites for these pieces of equipment? DONO: Chiara and Yifan both say the model number is enough}
The equation that describes the force that the wire exerts on a magnet at distance $r$ from it can be calculated as: 
\begin{equation} \label{eq:wire}
	F_{\rm wire}(r) =  \frac{I \mu_0 \mathcal{M} }{2 \pi}  \frac{h^2 - r^2}{(h^2 + r^2)^2}\,,
\end{equation}
%{\bf DONO: Yifan asking reference of the equation}
where $h$ is the height of the wire from the plane of floating discs,
$I$ is the wire current, $\mu_0= 4 \pi \cdot 10^{-7}$ N A$^{-2}$ is the magnetic permeability, and $\mathcal{M} = 7.8\cdot 10^{-3} $ Am$^2$ is the magnetic 
moment of the floating disc. See the appendix for the derivation of
Eq.~\eqref{eq:wire}.
%PGK: Chris would it make sense to draw the quantities h and r e.g. in
%panel 1(c) ??
%PGK: ALSO, I think that we should say "something" about the potential
%effect of the wire to other magnets (e.g., comparing order of
%magnitude of the effect or something similar??). You may recall that
%I brought up this point before.
%PGK: BTW, as an aside now that I have read about the central magnet,
%I can't help but wonder whether the magnetic force of other beads
%with the central one may or may not be affected by the lack of the
%(sort of screening) 3D printed structure. But anyway I presume we
%don't wish to touch this now, so I 'll let it be...
We use harmonic excitations in our experiment, so the current through the wire is $I(t) = a I_0 \sin( 2 \pi f t)$, where $f$ is the drive frequency (in Hz), $a$ is the drive-voltage amplitude (in Volts), and $I_0 = 0.1$ A\,V$^{-1}$ is the current per unit voltage that we measure in the wire.

The magnets repel each other. In the ideal situation of a perfect dipole--dipole interaction, the magnetic force between two repelling magnets is
\begin{equation} \label{eq:magnet}
	F_{\mathrm{magnet}}(r) = A r^{p}\,,
\end{equation}
where $r$ is the distance (in meters) between the two center points of the magnets, $p = -4$, and 
$A =   3 \mu_{ 0 } \mathcal{M} ^ { 2 } /  2 \pi$. 
%where $\mathcal{M} = 7.8\cdot 10^{-3} $ Am$^2$ is the
%magnetic moment of the magnets and $\mu _ { 0 } = 4 \pi \cdot 10^{-7}$ N A$^{-2}$ is the magnetic permeability. 
Although Eq.~\eqref{eq:magnet} is reasonable for large separation distances, we obtain better agreement by empirically determining $A$ and $p$. 
%{\bf map: which of those two things is used for the model curve in Fig. 1? this needs to be indicated explicitly in the caption}
Because the force between two magnetic dipoles is too small to measure directly, we create a pair of plastic plates, with 25 magnets attached to each plate. We position the plates to align each pair of cylindrical magnets from opposite plates through their radial directions. 
%{\bf map: I don't understand "aligned between plates"; this needs to be expanded - additional sentence proposed by Yifan}
We measure the repulsive force as a function of the displacement between these two plates in a materials tester (Instron ElectroPuls E3000). 
%{\bf map: it would be good to add a REF to the website for this product}
The distance between the magnets on each plate is large enough
(specifically, it is 2.5 cm) so that we can neglect interactions between magnets that are not aligned. 
%{\bf DONO: we aim to only have one dipole to one dipole interaction. So that one dipole of one plate is not affected by multiple dipole of the other plate. Having only one to one interaction for each pair of magnet, we can easily divide the total force by the number of pairs.}
%Indeed, t
%Note that t
The distance between a magnet on the first plate and the non-aligned magnets of the other plate is larger than 25 mm. As one can see in Fig.~\ref{fig:expsetup}(c) the interaction force already tends to $0$ for distances smaller that are significantly smaller than 25 mm.
%{\bf map: is there any particular reason we're using 'cm' in the text but 'mm' in the label on the plot; wouldn't it be better to make the same choice in both places?}
%to minimize interactions within the plate. 
%{\bf map: (1) "minimize" seems like the wrong phrasing; do we mean instead to ensure that they are small?; (2) "within" ? shouldn't it be "with"? I am confused here}
Consequently, the measured force is approximately equal to the sum of the repulsive force of the 25 isolated magnet pairs.
%{\bf DONO: Previously: Consequently, the measured force is approximately equal to the repulsive force of the 25 magnet pairs. I would say this sentence is not accurate because the force measured is equal, precisely because measured, to the force for 25 pairs (located in the same space). It is however the approximation of 25 times the interaction for one isolated pair, neglecting the interaction between magnets that are not in front of each other.}
% 
%{\bf map: above previously said it was equal, but surely this is only approximate? also, how good is this approximation?}
We 
%then 
fit the data using Eq.~\eqref{eq:magnet}, which yields $p \approx -4.2$ and $A \approx 3.8 \cdot 10^{-12} N/m\textsuperscript{$p$} $ [see Fig.~\ref{fig:expsetup}(c)].
%\begin{equation} \label{eq:magnet_fitted}
%%F_{\rm magnet \ exp.}(r) = 15.08 r^{-4.2}
%F_{\rm magnet \ exp.}(r) = 3.7879 \cdot 10^{-12} r^{-4.2}
%\end{equation}
%\textcolor{blue}{\textit{(the value of A in equation (3) is different from the value given in table I)}: CC: the value was with r in mm, but converted to meters you obtain the value in the table } 

%{\bf map: why do we have different numbers of significant digits for $p$ and $A$ above? that is strange}

We monitor the motion of the central particle using a laser vibrometer (Polytec CLV-2534), and we record the remainder of the lattice using a digital camera (Point Grey GS3-U3-41C6C-C) with a frame rate of 90 fps. We 
%then 
analyze the images using digital-image-correlation (DIC) software (VIC-2D) to determine each particle's velocity. 
%{\bf map: it would be good above to include a ref in the bibliography to the websites for the full specs of these products}
We inspect half of the lattice, as the cables that are connected to the driving wire block most of the system's other half [see Fig.~\ref{fig:expsetup}(a)]. 
Due to 
%Because of 
imperfections at the bottom of the glass disks (e.g., dust, scratches, and so on) and the fact that mass is not distributed evenly on a disk, a few particles start to rotate when they are levitated by the air that flows out of the air-bearing table. The image correlation software then loses track of them. We ignore these rotating discs in our subsequent analysis.
%{\bf map: but we still look at the other units? I am trying to figure out if it is a subset of units that rotate and we ignore those, or instead if we ignore all of them once rotation starts; please clarify - DONO: We ignore these rotating discs in subsequent analysis. The question was:  but we still look at the other units? First, the disks usually rotate from the beginning (remember we wait that the system is stationary before recording). When they rotate slowly, they are tracked. If they rotate too fast, we loose track of them (actually we loose track of them from instant 0). So yes, we still look at the other units. Sometimes we loose track of a particle during recording for some reason. If the time of recording is long enough, the RMS of the velocity for that particle is calculated with the data we have. Otherwise (time too short), the particle is ignored.}
To estimate the value of the damping coefficient, we excite the center
particle and record the resulting temporal amplitude decay once we
switch off the excitation. We fit the decay to an exponential
function, which we then use to calculate the damping coefficient
$\gamma \approx 10.52 \cdot10^{-3 }$N s/m.
%PGK: we were sure that doing this to the rest of the units would NOT
%change \gamma??  CC: This is possible...
%In particular, we match the decay rate with the real part of the eigenvalue
%that is associated to the linear defect mode. 
%{\bf map: I don't understand the use of the term "In particular" above; how does that fit in with the fitting? is this a separate thing that we're doing here or part of the same thing? please clarify}
%In Sec.~\ref{sec:disp}, we detail how we calculate the eigenvalues.
The lattice particles are always in
%in constant 
motion with at least small speeds, even in the absence of excitation.
%{\bf map: why must it be constant? surely there is some heterogeneity?}
%{\bf map: I had trouble with the sentence above as phrased previously; did I interpret the meaning correctly?}
 This is due to interactions with the air flow from the table and to imperfections (e.g., nonaxisymmetric mass distributions) of the particles. We use this motion to estimate the noise in the system.
 %{\bf map: the word used before above was "represents", but I don't understand it in the above context; does my version give the correct meaning? Before: This motion constitutes
 %represents 
 %the noise in our system.}
  To evaluate the amount of noise, we record the lattice motion without excitation as a comparison. 
  %{\bf map: above: I replaced "it" by "amount of noise"; is this precisely correct? ('it' is ambiguous, and I am not 100\% sure that I am right)}
  We summarize the values of the parameters in Table \ref{t:measured_params}.

%{\bf map: we're citing the subpanels of Fig. 1 in a different order than what is in the figure; currently, we do (a,c,d,b); shouldn't the ordering in the float match that in the text?}

\begin{table}[pht]
\caption{Summary of the parameter values in our experimental setup.} %{\bf map: do we want to distinguish in this table between fit and measured quantities? I think we should}}
\centering
\begin{tabular}{ | l | l | l || l | l | l |}
\hline
\textrm{Description}   &  \textrm{Symbol}   &  \textrm{Value (measured)}       &  \textrm{Description}   & \textrm{Symbol}  &  \textrm{Value (fitted)}     \\  \hline \hline 
         Mass of bulk magnet   & $M_b$  &  $138.2$ mg    & Magnetic coefficient& $A$     &  $3.8\cdot 10^{-12}  $ N/m\textsuperscript{$p$}   \\ \hline
         Defect mass & $M_\delta$ & $81.6$ mg     & Nonlinearity & $p$     &  $-4.2$ \\ \hline
         Equilibrium distance & $\delta$     &  $ 13.7$ mm  & Damping coefficient    &  $\gamma$      & $10.52 \cdot10^{-3 }$N s/m     \\ \hline
	 Wire height & $h$ & $3$ mm &    Magnetic moment & $\mathcal{M} $ & $7.8\cdot10^{-3}$ A m$^2$\\ \hline
       \hline
\end{tabular}
\label{t:measured_params}
\end{table}

%\centering
%\begin{tabular}{ | l | l | l || l | l | l |}
%\hline
%\textrm{Description}   &  \textrm{Symbol}   &  \textrm{Value (measured)}       &  \textrm{Description}   & \textrm{Symbol}  &  \textrm{Value (fitted)}     \\  \hline \hline 
%         Magnetic coefficient& $A$     &  $3.8\cdot 10^{-12}  $ N/m\textsuperscript{$p$}    & Wire height & $h$ & $3$ mm   \\ \hline
%         Nonlinearity & $p$     &  $-4.2$     & Magnetic moment & $\mathcal{M} $ & $7.8\cdot10^{-3}$ A m$^2$\\ \hline
%         Defect mass & $M_\delta$ & $81.6$ mg & Equilibrium distance & $\delta$     &  $ 13.7$ mm      \\ \hline
%	 Mass of bulk magnet   & $M_b$  &  $138.2$ mg  &    Damping coefficient    &  $\gamma$      & $10.52 \cdot10^{-3 }$N s/m\\ \hline
%       \hline
%\end{tabular}

%%%%%%

\section{Theoretical Setup} \label{sec:model}

 \begin{figure}
 \centerline{
  \includegraphics[width=.8 \linewidth]{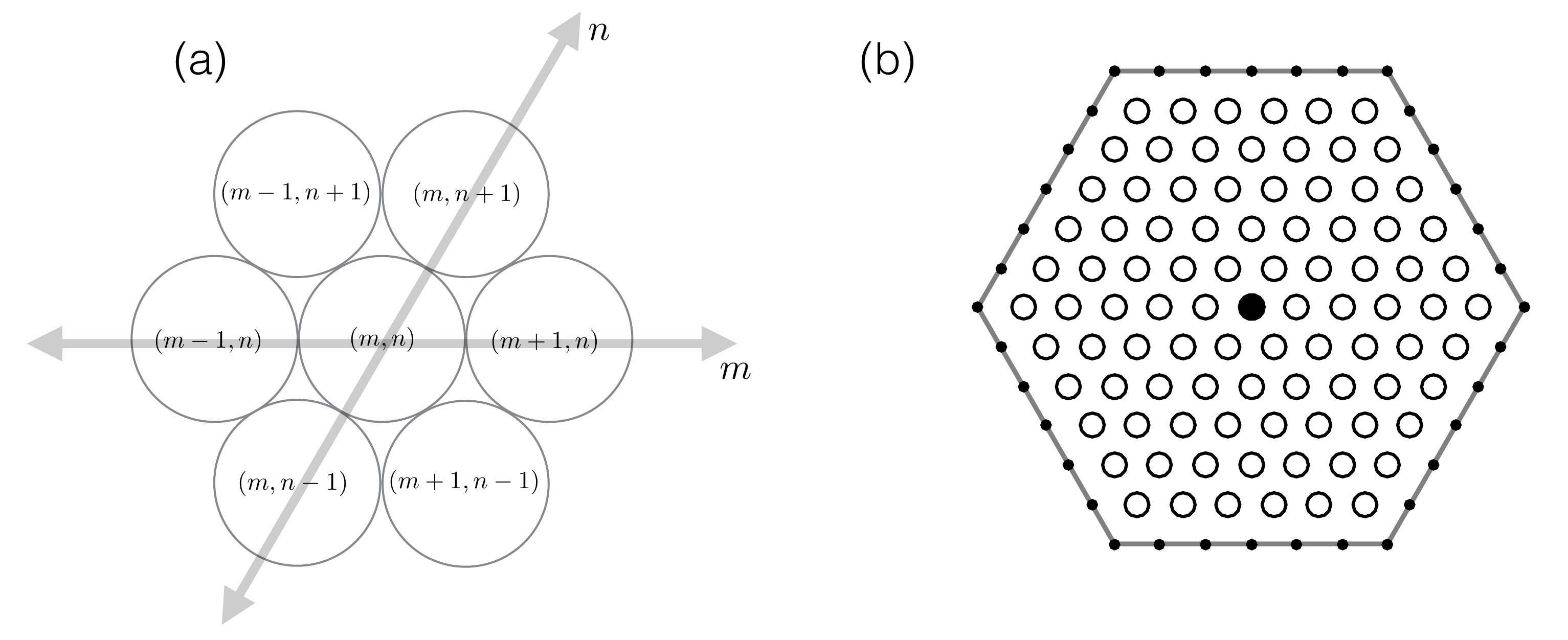}
   }
%    \centering
%  \begin{tabular}{@{}p{0.45\linewidth}@{\quad}p{0.45\linewidth}@{}}
%    \subfigimg[width=\linewidth]{\bf (a)}{the_lattice} &
%     \subfigimg[width=\linewidth]{\bf (b)}{biglattice.eps}
%  \end{tabular}
  \caption{ \textbf{(a)} Orientation for our convention of indexing particles in a hexagonal lattice. 
  %The $m$ axis represents the horizontal direction, and
  The $m$ axis and $n$ axis meet at an
angle of $\theta = \pi/3$. \textbf{(b)} A hexagonal lattice with $w=6$
%, which is the number of
magnets along each edge of the lattice. The empty circles and solid center circle
represent the locations of the magnets in equilibrium. When in equilibrium, the center-to-center distance between any two adjacent magnets
is $\delta$. The outer hexagonal boundary is the solid gray hexagon that encloses the lattice. On the boundary,
the solid points represent fixed (i.e., immovable) magnets. There are $w+1$ such magnets along each edge
of the 
%outer hexagonal 
boundary. The solid circle represents the defect particle, which has index
$(m,n) = (0,0)$.
%{\bf map: left panel (cosmetic): (1) the gray arrows are a bit dark for the text that goes on top of them; I find that I am straining my eyes a bit; maybe if the gray arrows are just slightly lighter but also hatched? (2) more painful: maybe I am imagining things, but it looks like the angular arrow isn't going through the diameters of the circles? [this one seems painful enough that we may want to wait until after referee reports when we are going through another set of rounds anyway]}
%{\bf map: for this figure, is it possible to change the labels in the graphics to put "$m$" and "$n$" in math mode so that our notation is consistent? CC:done
} 
 \label{fig:lattice}
\end{figure}

%%%%%%

\subsection{Model Equations}

Our goal is to study NLMs in a 2D hexagonal lattice. In selecting equations to model the system that we described in Sec.~\ref{sec:exp_setup}, we seek the simplest possible model that incorporates the ingredients (nonlinearity, discreteness, and dimensionality) that are essential for NLMs and also yield reasonable agreement with experimental data. 
%{\bf map: should the "e.g." actually be "specifically,"? if not, what factors are also true but not listed in the parenthetical comment above?}
It is in this spirit that we develop our model equations. After doing so (at the end of this section), we briefly discuss model simplifications. 
%and we discuss model validation through direct comparisons of theory and experiment in the discussions
%sections {\bf map: I think this statement is not accurate, as there are also ones in this section (but in different subsections}
%that follow.

%{\bf map: last sentence above: please add explicit pointers (with numerical identifiers) to appropriate sections and subsections to help readers. CC: model simplifications are only discussed at the end of this section}

We consider a hexagonally packed lattice of magnets. We use the lattice basis vectors $e_1=(1,0)$ and $e_2=(1/2, \sqrt{3}/2)$.
Let $\mathbf{q}_{m,n}(t)= (x_{m,n}(t),y_{m,n}(t)) \in\R^2$ denote the displacement from the static equilibrium of the magnet at 
position $\textbf{p} = \delta (m e_1 + n e_2)$ in the plane [see Fig.~\ref{fig:lattice}(a)],
where $\delta$ is the center-to-center distance between two particles at equilibrium.
%is the initial separation distance (which we take to be the equilibrium distance). {\bf map: I am confused by the word 'initial' here; it seems to me that $d$ needs to be defined in the same way in all places}
The lattice indices $m$ and $n$ take the values $m,n \in \{-w , -(w-1), \ldots,0,\ldots, w-1, w \}$,
where $w$ is the number of magnets along an 
%given 
edge of the hexagon.
The lattice boundary is given by the hexagon with magnets at positions $(w \delta \cos( j \pi/3),
w \delta \sin( j \pi/3))$, where $j=0,1,\ldots, w-1$  [see Fig.~\ref{fig:lattice}(b)]. 
%\egc{Chris, please check.}
% and $w$ is the number of magnets along a given edge of the hexagon [see Fig.~\ref{fig:lattice}(b)].
%The lattice boundary is given by the hexagon that inscribes a circle of radius $w \delta$
%with vertices at the angles $j \pi/3$ for $j=0,1,\ldots5$ 
For our fixed
%the fixed zero {\bf map: the word 'zero' is confusing without some further description to go with it}
boundary conditions along the edge of the hexagonal boundary, 
$\mathbf{q}_{m,n}(t)=0$ if $|m+n| > w$. 

One can express the distance between the magnet with index $(m,n)$ and
one of its nearest neighbors in terms of the displacements $x_{m,n}$ and
 $y_{m,n}$
%PGK: notice that I changed it here
 of the magnets from their respective equilibrium positions. Once we determine the distance,
we compute the resulting force using Eq.~\eqref{eq:magnet}. Summing
the forces from each of the six nearest neighbors and applying
 Newton's second law leads to the following equations of motion:
\begin{equation} \label{eq:model}
\begin{array}{ll} \displaystyle
	M_{m,n} \ddot{\mathbf{q}}_{m,n} = 
&- \mathbf{F}_0(\mathbf{q}_{m+1,n} - \mathbf{q}_{m,n}) 
 - \mathbf{F}_1(\mathbf{q}_{m,n+1} - \mathbf{q}_{m,n}) 
+\mathbf{F}_{-1}(\mathbf{q}_{m,n} - \mathbf{q}_{m-1,n+1}) 
 \\
& +\mathbf{F}_0(\mathbf{q}_{m,n} -\mathbf{q}_{m-1,n})
+\mathbf{F}_1( \mathbf{q}_{m,n} - \mathbf{q}_{m,n-1})
- \mathbf{F}_{-1}(\mathbf{q}_{m+1,n-1} -\mathbf{q}_{m,n})
- \gamma \dot{\mathbf{q}}_{m,n} + \mathbf{F}^{\mathrm{ext}}_{m,n}(t) \,.
	\end{array}
\end{equation}
The vector functions $\mathbf{F}_j(\mathbf{q}) = \mathbf{F}_{j}(x,y) \in \R^2$ have a magnitude of
\begin{equation*}
	|\mathbf{F}_{j}(x,y)| =   A \left[  \sqrt{ (\delta \cos(\theta_j) + x )^2   + (  \delta \sin(\theta_j) + y )^2}  \,    \right]^{p}\,, \quad \theta_j = \frac{j \pi}{3}\,, \quad j \in \{-1,0,1\} \,.
\end{equation*}	
The mass of the magnet with index $(m,n)$ is $M_{m,n}$. The dashpot $\gamma \dot{\mathbf{q}}_{m,n}$ is a phenomenological
term that we add to account for damping. Using such a term has yielded reasonable agreement with experiments in other, similar lattices \cite{moleron,cantilevers,magBreathers}. The quantity $\mathbf{F}^{\mathrm{ext}}_{m,n}$ is the external force that we apply to the magnet at $(m,n)$.
%{\bf map: here it would be good references to prior uses of a linear dashpot for damping in such systems and when it is reasonable/unreasonable to do so}
In the present article, we 
%only 
consider excitations via a wire that is directly above the center of the lattice. The magnitude
of the excitation is given by Eq.~\eqref{eq:wire}. Therefore,
\begin{equation} \label{eq:excite}
	\mathbf{F}^{\mathrm{ext}}_{0,0}(t) =  \ \  a \sin(2 \pi f t) \, \frac{I_0 \mu_0 \mathcal{M}}{2\pi}  \begin{pmatrix} \cos(\phi)   \frac{h^2 - x_{0,0}^2}{(h^2 + x_{0,0}^2)^2}  \\ \sin(\phi)  \frac{h^2 - y_{0,0}^2}{(h^2 + y_{0,0}^2)^2}  \end{pmatrix} \,,
\end{equation}
%\ajm{It puzzles me the election of $\phi$ to be such that when $\phi = 0$ the external force point in the direction $e_1$, whilst when $\phi = \pi/2$ it points in the direction $e_2$. Shouldn't be better to write $\cos(3\phi/2)$ and $\sin(3\phi/2)$ in equation (4), so that $\phi$ has a geometric interpretation as the angle between the wire and the x-axis?. Maybe I am missing something here. CC: $\phi$ is simply the angle indicating the direction of the force, which was more natural to me when coding it. I don't think it is wise to change it now, as it would require relabeling many things.}
where $\phi$ is the angle of the excitation and $\mathbf{F}^{\mathrm{ext}}_{m,n} = 0$ when $m\neq0$ and $n\neq 0$. 
%{\bf map: "$m\neq0\neq n$": are $m$ and $n$ allowed to be equal or
%not? please replace this with something clearer. CC : see change}
%PGK: again sth should be said here about the wire NOT affecting the
%rest of the magnets, if possible.
In our experiments and in most of our numerical computations,
the excitation angle is $\phi = \pi/2$, so we excite only the $y$-component of the center magnet.
  We will also explore some other excitation angles.
 As we discuss in the appendix, the lattice forces dominate the dynamics. The wire 
has only a small effect on magnets 
that are beyond the center of the lattice. For example, at equilibrium, the force that is exerted on
the center magnet by the wire is two orders of magnitude larger than the force that the wire exerts on the center magnet's nearest neighbors. Compare Eq.~\eqref{eq:wire} with $r=0$ and $r=\delta$.

In our model, we ignore effects beyond nearest-neighbor coupling of the magnetic interactions. It is known that such long-range effects can alter the structure of localized modes. For example, it was shown in
\cite{Flach_long} that the spatial decay of breathers can transition from exponential spatial decay to algebraic decay in lattices with
algebraically decaying interaction forces (as is the case in our model) for lattices with sufficiently many sites. More recently, Moler\'on et al. \cite{magBreathers} studied NLMs in a 1D magnetic lattice using a model with long-range interactions. 
%{\bf map: the previous phrasing (specifically, the use of "or" in parentheses) was confusing, and it seemed to directly contradict the terminology that we previously established in this paper}
%Their experimental results suggest that the transition in the tail of a breather occurs for lattice
%indices that are larger than those that we consider. 
Although the differences between long-range and nearest-neighbor lattices that were considered in \cite{magBreathers} are detectable, they are 
%nonetheless 
still small. For example, at equilibrium, the force that is exerted on
the center magnet by its nearest neighbors is one order of magnitude larger than that exerted by its next-nearest neighbor. (Compare Eq.~\eqref{eq:magnet} with $r=\delta$ and $r=2\delta$.)
%{\bf map: but isn't the setup somewhat different? we need to be careful in these comments about accounting for the differences in these systems with respect to what claims we make}
%Therefore, we expect the effect of long-range interactions to be small --- likely barely noticeable or even not noticeable --- for a lattice with an edge length of $w = 6$. 
Therefore, 
%in order 
to keep our model as simple as possible, we
ignore such small long-range effects.
% in the present work.
%PGK: we somehow need to be a bit more definitive here, esp. since we
%claimed that these effects are small already above.
% we expect the effect of long-range interactions to be small --- likely barely noticeable or even not noticeable --- for a lattice with an edge length of $w = 6$. 
%The study of localized modes in larger hexagonal lattices with long-range effects will be left for future work. {\bf map: this statement is very unhelpful}
%% Formula (4) from the NJP paper would probably be different for the 2D situation and would be interesting to investigate in its own right

In our analysis of experimental data we ignore magnets that are rotating, so our model does not account for rotation.
%and thus in our model we do not account for rotation. 
This leaves air resistance as the primary source of damping. 
%\ajm{Simply curiosity
%  here, what about the dissipation through energy transference between
%  translational and rotational degrees of freedom? I wonder if one can
%  actually backup the assumption (intuition?) that air resistance is
%  the dominant source of damping. Considering that in the experiments
%  rotating discs are ignored, as the software loses track of these
%  disks, I think it gets very tricky to analyze, for instance, the
%  orders of magnitude between the different effects involved in the
%  dynamics.} {\bf map: following up on Alejandro's comment, can we
%  provide a brief statement and some kind of reference in support,
%  please?CC: I tried to reword things to address this issue, but the
%  experimental team may wish to add more.}
%PGK: In principle one can think of ways in which things might be
%measurable e.g. to measure kinetic and potential energy over time
% and the amount of energy lost over dissipation in order to infer the
% amount of energy transfer into rotational dof's. But this is
% probably tough work...
Given the size of the magnets and velocities that we consider,
we employ a linear dashpot \cite{SteidelBook}.  
%{\bf map: I think the above sentence needs some more specificity to make the validity of this assumption more concrete; (this is one I have questioned in past papers, as you know). CC: I added additional references above.}
We also assume that the magnets stay in a plane. We validate the many assumptions that we have made in formulating the model in Eq.~\eqref{eq:model} via a direct comparison with experimental results in Sec.~\ref{sec:results}. 

For the remainder of the manuscript, we fix all parameters of the model (and we summarize them in Table \ref{t:measured_params}), except for the excitation amplitude $a$, frequency $f$, and angle $\phi$. We will specify these in our various examples.
%, which will be specified.
%{\bf map: which of the three things above do we specify? (need to be explicit); and where do we specify them? (please also be explicit)}
In all cases, we examine a lattice with a single defect particle in the center and a hexagonal boundary with a length of $w=6$ magnets (see Fig.~\ref{fig:lattice}). 
Importantly, we do not fit the parameter values to the reported experimental results. Instead, we determine them beforehand using the procedures
that we detailed in Sec.~\ref{sec:exp_setup}.

%%%%%

\subsection{Linear Analysis} \label{sec:disp}

\begin{figure}
    \centering
  \begin{tabular}{@{}p{0.3\linewidth}@{}p{0.3\linewidth}@{}p{0.3\linewidth}@{}}
    \subfigimg[width=\linewidth]{\bf (a)}{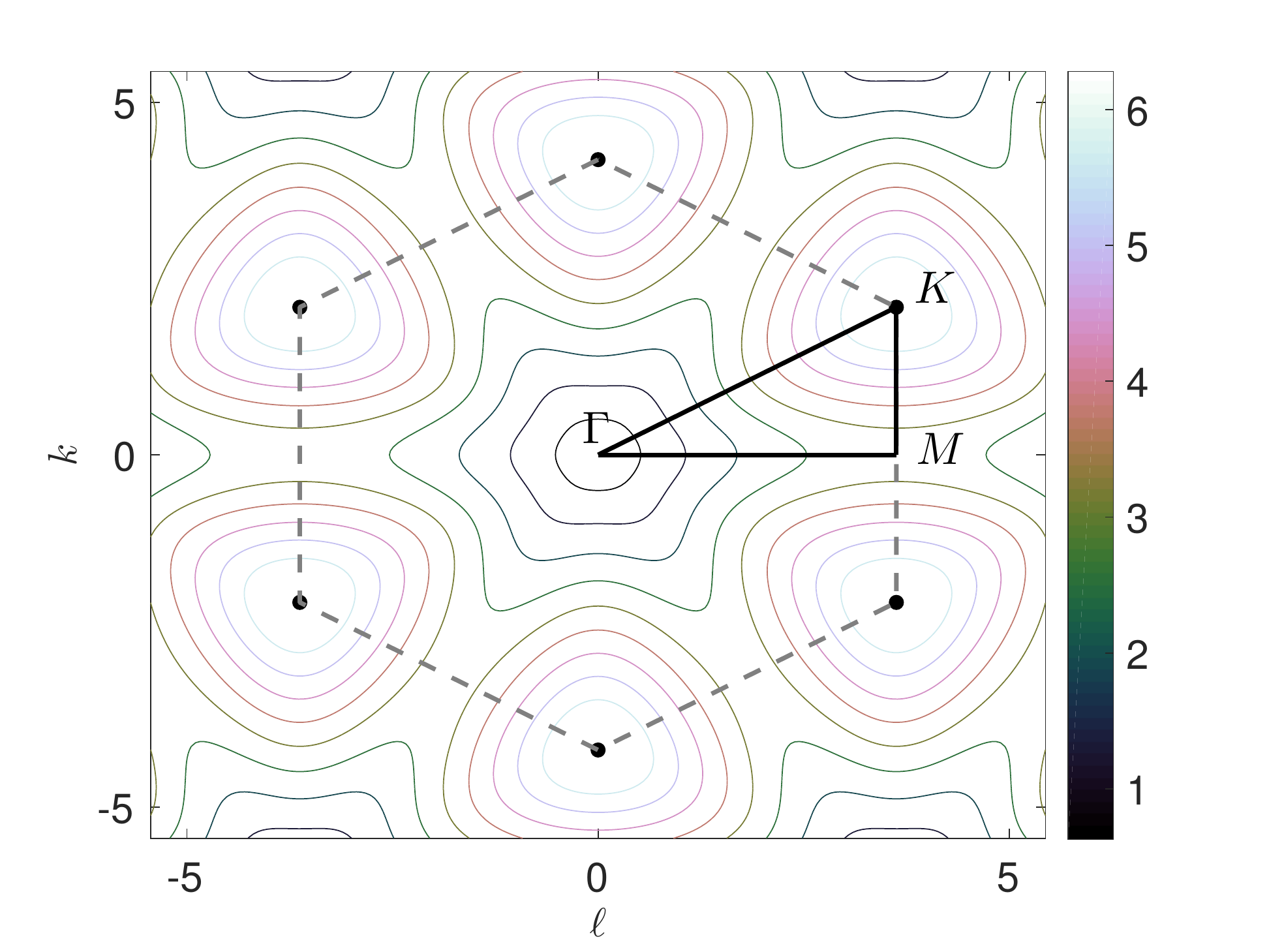} &
     \subfigimg[width=\linewidth]{\bf (b)}{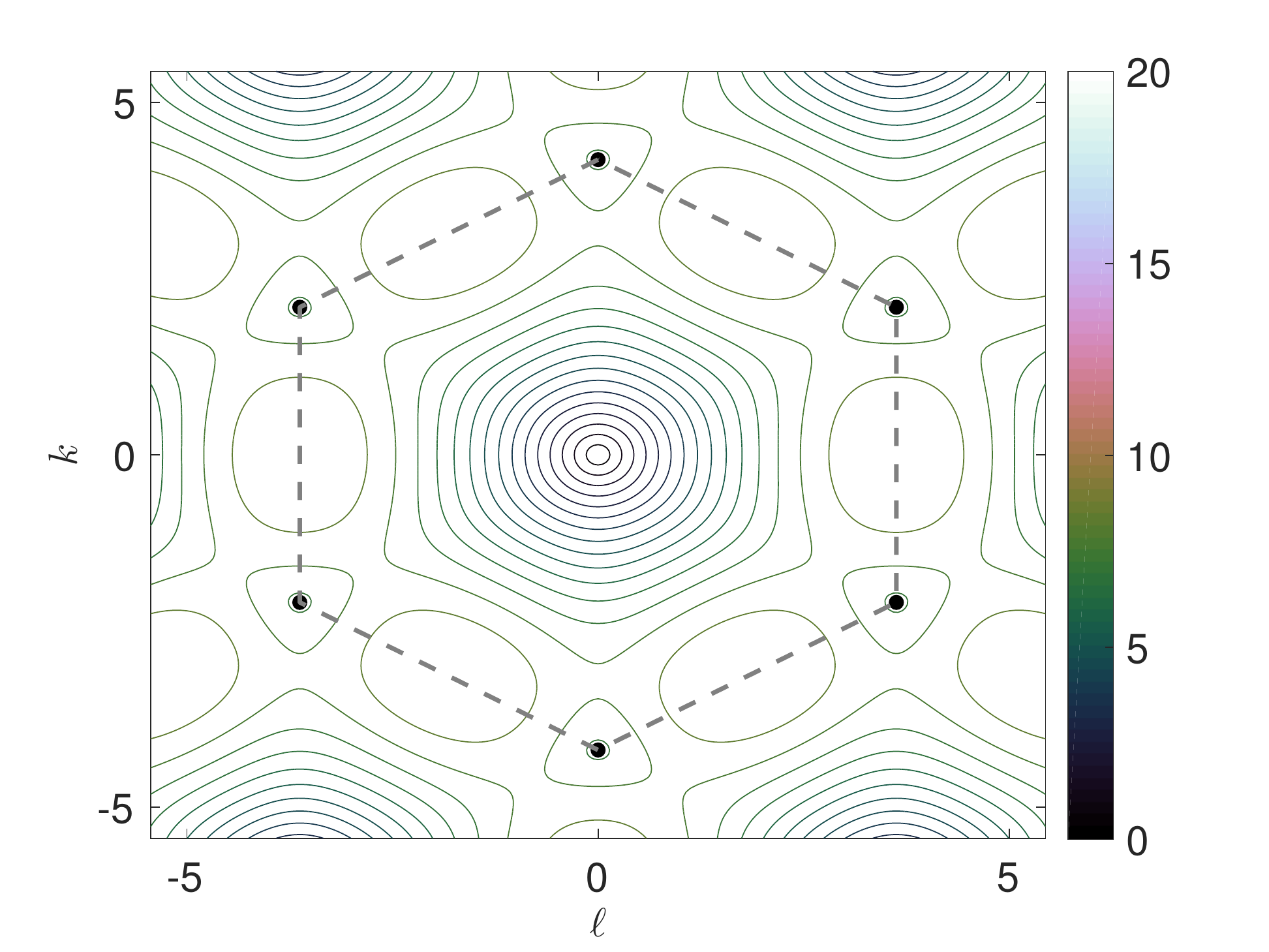} &
   \subfigimg[width=\linewidth]{\bf (c)}{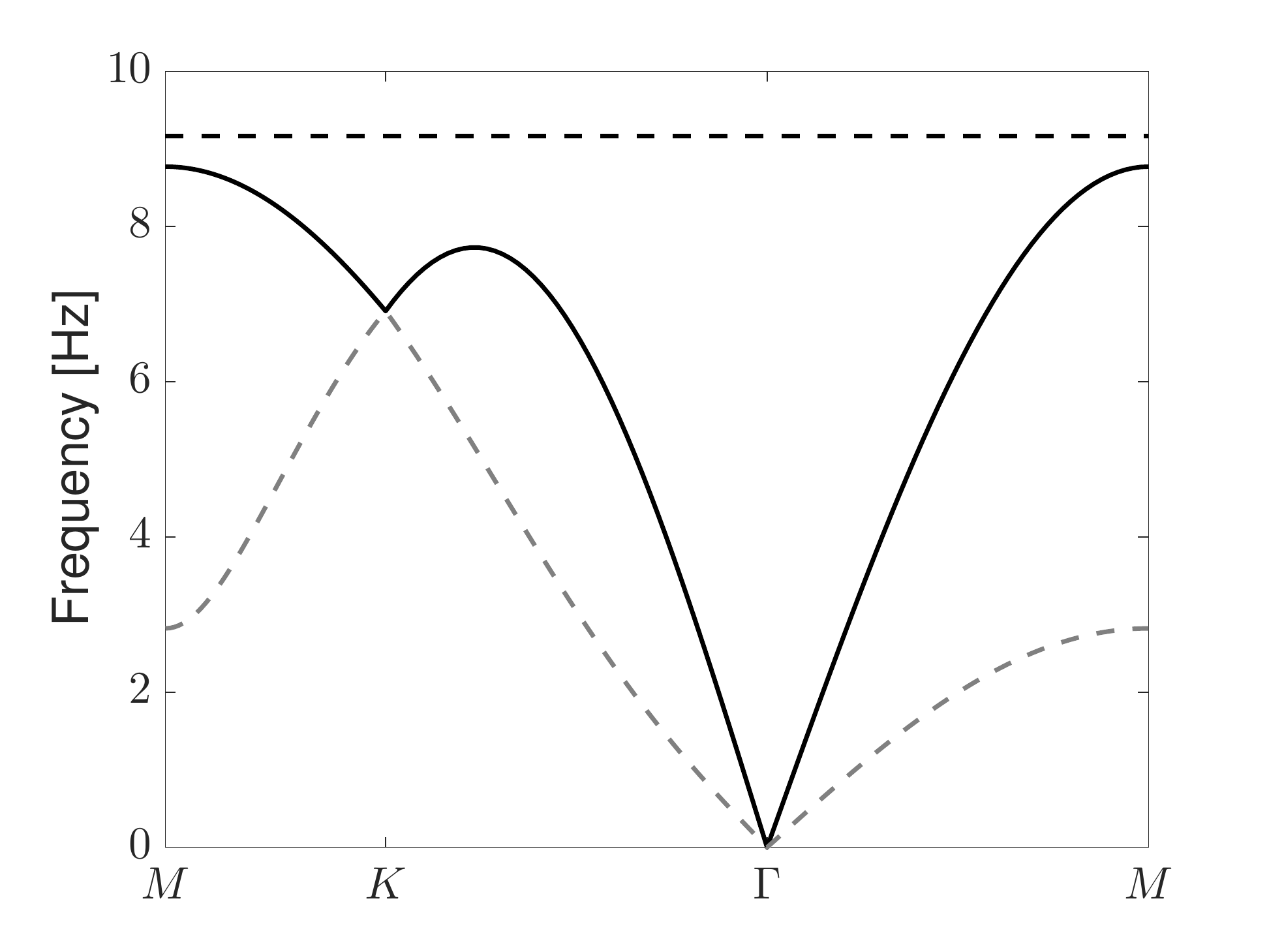}
  \end{tabular} 
 \caption{(Color Online)  \textbf{(a)} Contour plot of the bottom dispersion surface. We show the irreducible Brillouin zone as the triangle
 with magnets at the points that we mark by $\mathbf{M}$, $\mathbf{K}$, and $\mathbf{\Gamma}$. 
 %{\bf map: (1) can we use math mode in the figure for these symbols so that the notation matches? (2) can we move the $\Gamma$ label so that it is closer to the point? (3) the "k" needs to be in math mode; 
 %(4) "$\ell$ appears to be different notation from what we used in the text; we need to harmonize the notation throughout the text}
 \textbf{(b)} Contour plot of top dispersion surface. 
  \textbf{(c)} Band structure along the edge of the irreducible Brillouin zone [also see the triangle in panel (a)] for the bottom (dashed gray curve)
  and top (solid black curve) dispersion surfaces. The horizontal dashed curve corresponds to the defect-mode frequency $f \approx 9.17$ Hz
  in a finite-dimensional system with a mass ratio of $M_\delta/M_b \approx 0.5868$, where $M_\delta$ denotes the mass of the defect magnet and $M_b$ denotes the mass of the other (``bulk'') magnets.
  %Notice the Dirac cone at the vertex $K$ and the pair of Dirac cones at the vertex $\Gamma$.
 % {\bf map: the subpanel labels are too close to the graphics themselves}
 % {\bf map: the section pointer here goes to something blank, so this needs to be fixed}
% {\bf map: (1) in panels (a) and (b), it is desirable for the $k$ and $\ell$ labels to be larger, if possible; (2) in panel (a), the $\gamma$ is hard to read above the contours; additionally, the $\mathbf{M}$, $\mathbf{K}$, and $\mathbf{\Gamma}$ are different symbols (not bold style) in panels (a) and (c), so we should harmonize the notation here anyway (and making $\mathbf{\Gamma}$ bold will also hopefully help with readability)}
  }
 \label{fig:spectrum}
\end{figure}

%\subsection{The Infinite Dimensional Hamiltonian Case with No Defects}

We start with the basic linear theory of localized modes for our hexagonal magnetic lattice.
We are particularly interested in modes with frequencies that lie above the cutoff frequency
of the pass band. We first derive an analytical expression for the cutoff frequency, which is
straightforward for an infinite-dimensional Hamiltonian system (i.e., with all integers $m$ and $n$, along with $a = 0$ and $\gamma = 0$).
%{\bf map: the \in\mathbb{Z} wasn't right, as it's not merely that it's an element of the integers, but instead that we have all of them}
We then numerically estimate the frequency of a linear mode that is associated with the defect for in the finite-dimensional Hamiltonian system.
Finally, we compute linear localized modes in the associated finite-dimensional damped, driven system.

Assuming small strains, such that % {\bf Note: $u$ is replaced by $\mathbf{q}$ below.}
\begin{equation} \label{eq:smstrain}
 	\frac{| \mathbf{q}_{m \pm 1,n} - \mathbf{q}_{m,n} |}{\delta} \ll 1\,, \quad  \frac{| \mathbf{q}_{m,n \pm 1} - \mathbf{q}_{m,n} |}{\delta} \ll 1\,, \quad   \frac{| \mathbf{q}_{m \pm1,n \mp 1} - \mathbf{q}_{m,n} |}{\delta} \ll 1 \,,
\end{equation}
one can Taylor expand,
\begin{equation*}
	\mathbf{F}_j(\mathbf{q}) \approx \mathbf{F}_j(\mathbf{q}_0) + D \mathbf{F}_j(\mathbf{q}_0)   \mathbf{q} \,,
\end{equation*}	
where $D\F_j$ is the Jacobian matrix of $\F_j$. Using this notation, we write the linearized equations of motion:
\begin{align} \label{eq:linear}
	M_{m,n} \ddot{\q}_{m,n}  &=  -D \mathbf{F}_0 ( \q_{m+1,n} + \q_{m-1,n} )
- D \mathbf{F}_1 ( \q_{m,n+1} + \q_{m,n-1} ) \notag \\
 &\quad - D \mathbf{F}_{-1} ( \q_{m-1,n+1} + \q_{m+1,n-1} )
+ 2( D \mathbf{F}_0 + D \mathbf{F}_1 + D \mathbf{F}_{-1})  \q_{m,n}\,,
\end{align}
where the entries of the Jacobian matrices are denoted by,
\begin{equation*}
	D\mathbf{F}_j = \begin{pmatrix} a_j & b_j \\ c_j & d_j\end{pmatrix}\,, \qquad  j \in \{-1,0,1\}\,,
\end{equation*}
with
\begin{align*}
	&a_{-1}=p \hatd\,,                    &b_{-1}&=0\,,                                           & c_{-1}&=0\,,       & d_{-1}&=\hatd\,,\\
	&a_{0}= \frac{3+p}{4} \hatd \,, &b_{0}&=\frac{\sqrt{3}(p-1)}{4} \hatd \, ,& c_{0}&=b_{0}   \, ,& d_{0}&=\frac{1+3p}{4} \hatd\,,\\
	&a_{1}=a_0\,,                          &b_{1}&=-b_0\,,                                     & c_{1}&=-c_0  \, ,& d_{1}&=d_0 \,,
\end{align*}
where  $\hatd  \equiv  A \delta^{p-1}$. For a monoatomic system (in which all magnets are identical, such that $M_{m,n} = M_b$), the linear system has plane-wave solutions
\begin{equation*}
	\mathbf{q}_{m,n} = \mathbf{q}_0 \exp\left(i( k m + \frac{n}{2}( k +  \sqrt{3} \l  ) )\right) e^{i \omega t}\,, \quad \mathbf{q}_0\in\C^2\,, \quad k,\ell,\omega \in \R \,,
\end{equation*}
where the wavenumbers $k,\ell$ and angular frequency $\omega = \omega(k,\ell)$ satisfy the dispersion relationship
\begin{equation}\label{eq:disp}
	\left[\omega(k,\ell)\right]^2 =  \frac{\omega_a + \omega_d \pm \sqrt{ \rule{0pt}{.4cm} (\omega_a + \omega_d)^2 - 4(\omega_a\omega_d -\omega_b \omega_c )  }   }{2}\,, 
\end{equation}
where
\begin{equation*}
	\omega_\alpha(k,\ell) = (-2\alpha_{-1}\cos(k) - 2\alpha_{0}\cos(k/2+ \sqrt{3}/2 \l) - 2\alpha_1\cos(k/2 - \sqrt{3}/2 \l) + 2(\alpha_{-1}+\alpha_{0}+\alpha_1))/M_b
\end{equation*}
and $\alpha \in \{a,b,c,d\}$. 
%{\bf map: we used the notation $d$ to mean distance before, so we need to change notation to fix this issue. CC: distance symbol is now $\delta$}
In Fig.~\ref{fig:spectrum}(a,b), we show contour plots of the two dispersion surfaces from Eq.~\eqref{eq:disp}.
% in Fig.~\ref{fig:spectrum}(a,b).
%using the parameter values shown in Table \ref{t:measured_params}. 
%We show the dispersion curves along
%In particular, we present 
The dispersion curves along
the edge of the irreducible Brillouin zone are shown in Fig.~\ref{fig:spectrum}(c). The cutoff value of the pass
band has the wavenumber pair $(k,\ell) = (0, 2\pi/3)$, which is where the dispersion curve attains its maximum value. For the parameter values in Table \ref{t:measured_params}, the cutoff frequency is $f_c = \omega(0,2\pi/3)_+/(2\pi) \approx 8.77$ Hz, where $\omega_+$ corresponds to the top dispersion surface.

%{\bf map: the table above needs to have its reference be with a 'ref' command, instead of hard-wired}

%From inspection of Fig.~\ref{fig:spectrum}, one can see regions where the top and bottom dispersion surfaces form a downward and upward pointing cone respectively. The point where these
%two cones meet is the Dirac point. They have the values
%$ (\pm \, 4\pi/3, 0) $ and $ (\pm \, 2\pi/3, \pm \, 2 \pi / \sqrt{3} ) $. While the dynamics emerging
%from these points are interesting in their own right, see e.g Refs.~\ref{}, they are not a focal point of the present article.

%\subsection{The Finite Dimensional Case with Mass Defects} \label{sec:finite_spectrum}

\begin{figure}
    \centering
  \begin{tabular}{@{}p{0.25\linewidth}@{}p{0.25\linewidth}@{}p{0.25\linewidth}@{}p{0.25\linewidth}@{}  }
      \subfigimg[width=\linewidth]{\bf (a)}{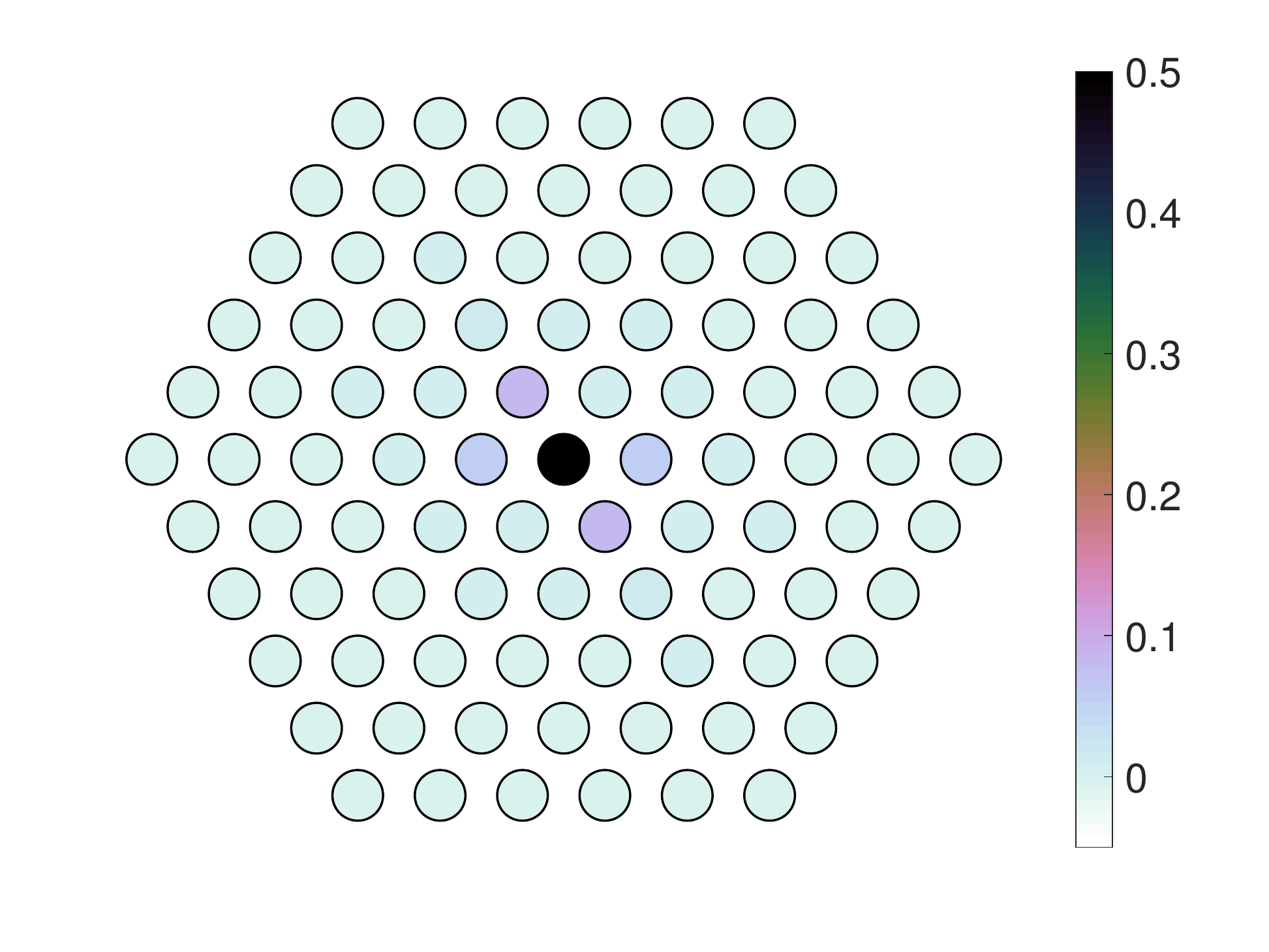}  &
      %\subfigimg[width=\linewidth]{\bf (a)}{unwrap_spectrum} 
    \subfigimg[width=\linewidth]{\bf (b)}{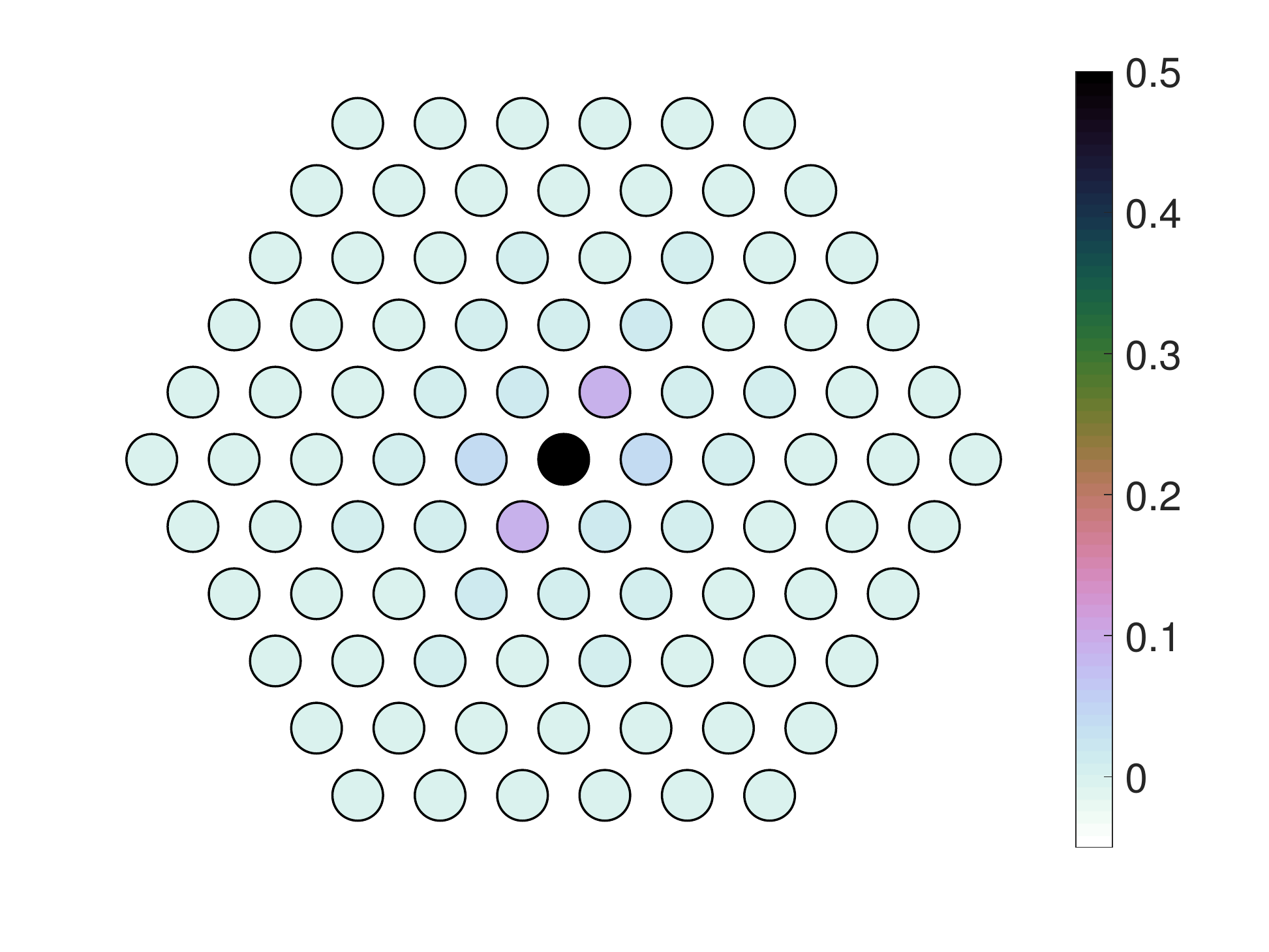} &
    \subfigimg[width=\linewidth]{\bf (c)}{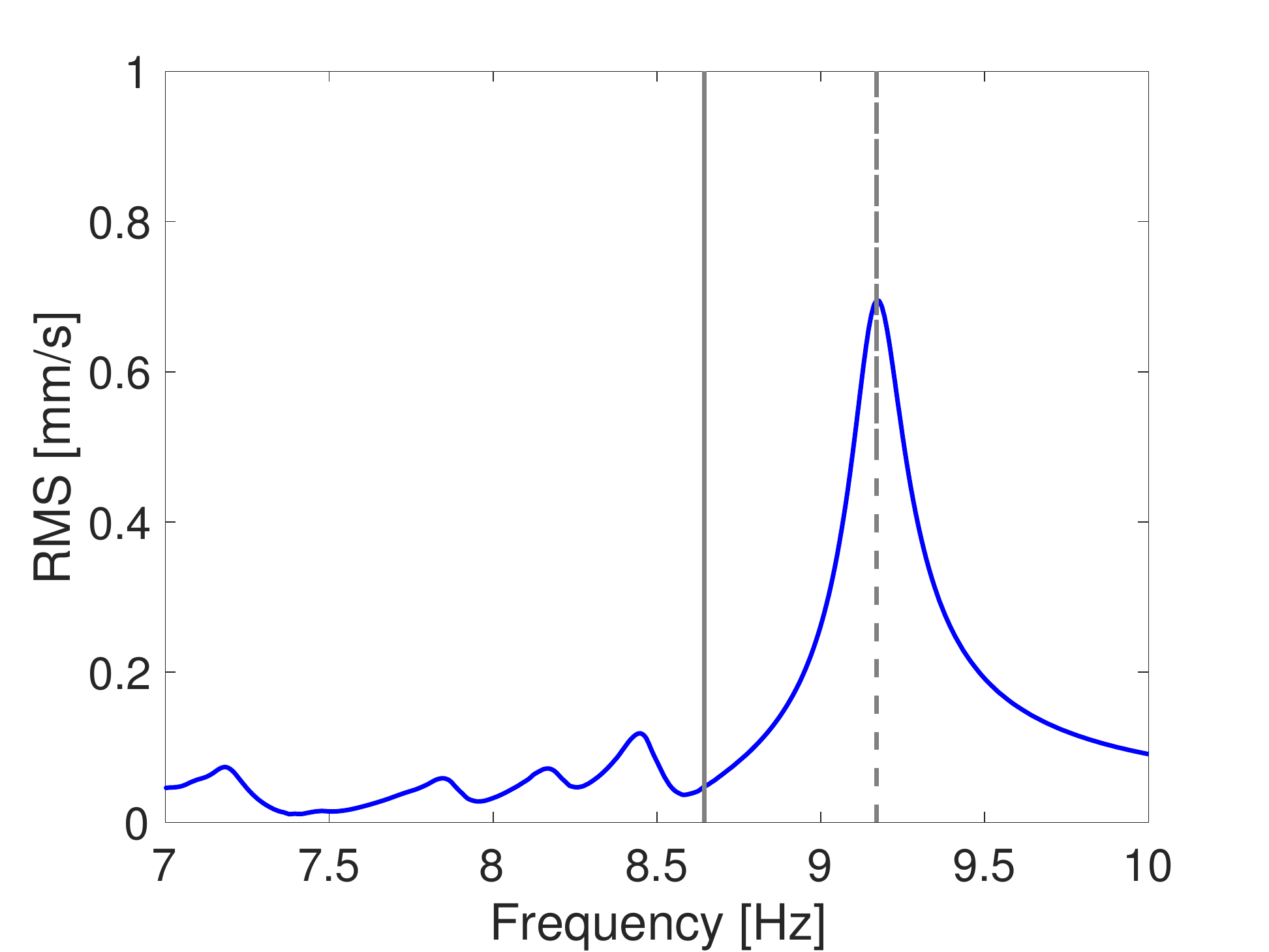} & %linear_sweep
    \subfigimg[width=\linewidth]{\bf (d)}{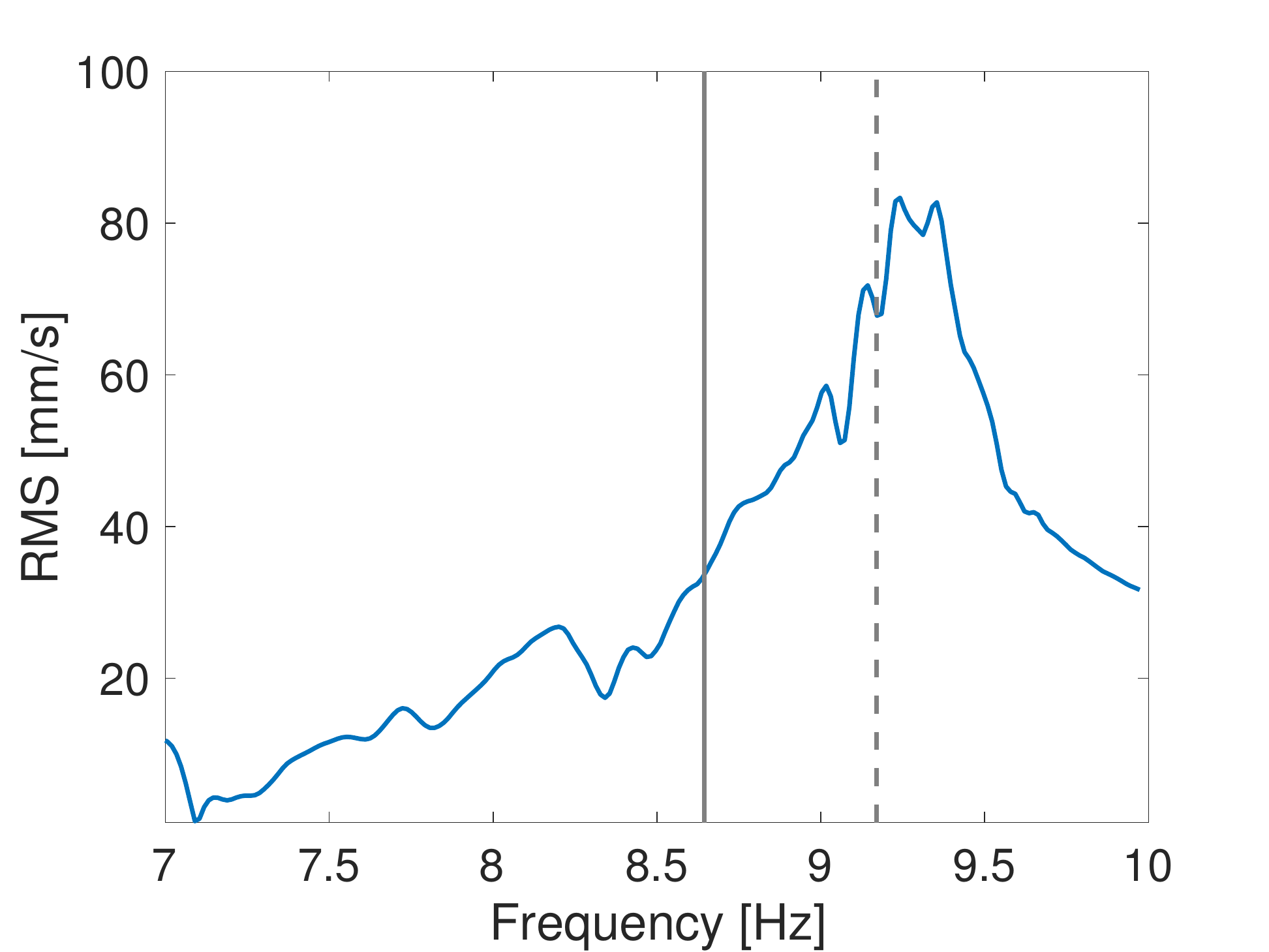}
  \end{tabular}
 \caption{ \textbf{(a,b)} Shape of the two modes with 
 %corresponding 
 defect frequency $f_d \approx 9.17$ Hz of the linear Hamiltonian system \eqref{eq:linear} with $w=6$ magnets along each edge of the boundary. 
 The color intensity at each point $(m,n)$ corresponds to 
 %the magnitude 
 $\sqrt{\dot{x}_{m,n}^2 +\dot{y}_{m,n}^2}$. 
 %{\bf map: notation "$x$" and "$y$"? where were these defined? CC: see page 3}
  \textbf{(c)} RMS of $\dot{y}_{0,0}$ of the linear damped, driven 
  %steady-state 
  solution~\eqref{eq:steady} as a function of the drive frequency $f$ with excitation amplitude $a=0.01$ mV 
  %(for which the force that is exerted by the wire is $F_{\mathrm{wire}} \approx 1.2414 \cdot10^{-9 }$\,N)
   and $\phi = \pi/2$. \textbf{(d)} RMS of 
 the velocity of the center particle of the experimental frequency sweep with $a=4$ mV 
 %(for which $F_{\mathrm{wire}} \approx 4.9656 \cdot10^{-7 }$\,N) 
 and $\phi = \pi/2$.
 %{\bf map: should the above be "speed", rather than velocity? CC: velocity is meant, but the quantity is squared, per the definition I fixed}
 %{\bf map: subpanel labels (c) and (d) are too close to the picture}
 }
 \label{fig:finite_spec}
\end{figure}

The presence of the lighter defect introduces a linear mode into the system that is localized in space and oscillates with a frequency above the cutoff frequency of the linear monoatomic system.  
%We now identify such linear localized modes of the magnetic lattice, which then can be used as seeds for the computation of nonlinear localized modes.  
With the light-mass defect at the center of the lattice, we write
\begin{equation}
	M_{m,n}= \left\{ \begin{array} { l l } { M_\delta  \,,} & { n = 0\, \text{  and  }\, m = 0 } \\ { M_b \,, } & { \text { otherwise}\,, } \end{array} \right. 
\end{equation}
where $(0,0)$ is the index of the mass defect with mass $M_\delta$, the quantity $M_b$ is the mass of a magnet in the ``bulk'' (i.e., the non-defect
mass), and $M_\delta < M_b$. We numerically compute the linear modes of the system with a mass defect, and we are thereby able to consider finite lattices. 
We use a hexagonal boundary with 
edge length $w=6$ magnets [see Fig.~\ref{fig:lattice}(b)].
One can embed this lattice into a square matrix of size $N\times N$, where $N = 2w-1$ is the number magnets along the $n=0$ line
of the lattice. Let $X(t)$ be the $N\times N$ matrix whose $(m,n)$th entry is $x_{m,n}(t)$, and let $Y(t)$ be the $N\times N$ matrix whose $(m,n)$th entry is $y_{m,n}(t)$.  We enforce the 
%(fixed)
fixed hexagonal boundaries by setting the displacements of magnets with indices $(m,n)$ such that $|m+n| > w$ to $0$.
We define the $N\times N$ matrix operators $L_\alpha$ through
 \begin{equation}
	L_\alpha Y  = \alpha_1 D Y + \alpha_2 Y D + \alpha_3( E^TYE^T + EYE - 2 Y )\,,
 \end{equation}
where $\alpha \in \{a,b,c,d\}$; the $N\times N$ tridiagonal matrix $D$ has $1$ entries on the super-diagonals and sub-diagonals, $2$ entries along the diagonal, and $0$ entries everywhere else; $E$ is an $N\times N$ matrix with $1$ entries along the super-diagonal and $0$ entries everywhere else; and $E^T$ is the transpose of $E$.
%{\bf map: need to fix the notational issue with "$d$" here}
%{\bf map: also, is there any additional structure for $E$, or can the entries except for those on the super-diagonal be anything? I suspect that some constraints are missing...}
With these definitions, Eq.~\eqref{eq:linear} becomes
 \begin{align}\label{this}
	M \circ \ddot{X}(t) = L_a X(t) + L_b Y(t)  \,, \notag \\
	M \circ \ddot{Y}(t) = L_c X(t) + L_d Y(t) \,,
 \end{align}
where $M$ is an $N\times N$ matrix in which all entries except the $(0,0)$ entry (which is equal to $M_\delta$) are equal to $M_b$. The operation $\circ$ denotes pointwise multiplication (i.e., the Hadamard product).
The system \eqref{this} has solutions of the form $X(t) = \tilde{X}e^{i \omega t}$ and $Y(t) = \tilde{Y}e^{i \omega t}$, where
$\tilde{X}$ and $\tilde{Y}$ are $N\times N$ time-independent matrices and 
 \begin{equation} \label{ev}
	-\omega^2 \begin{pmatrix}  M \circ \tilde{X}  \\  M \circ \tilde{Y}  \end{pmatrix} =  \begin{pmatrix}  L_a & L_b \\L_c & L_d   \end{pmatrix}
\begin{pmatrix}   \tilde{X}  \\  \tilde{Y}  \end{pmatrix} \,.
 \end{equation}
One can cast Eq.~\eqref{ev} as a standard eigenvalue problem by letting $\lambda = -\omega^2$ and unwrapping the $\tilde{X}$ and $\tilde{Y}$ matrices into equivalent 
%1D 
row vectors
and reshaping the block matrix (with entries given by $L_\alpha$) into a corresponding  $2N^2 \times 2N^2$ matrix. One can then numerically solve the resulting eigenvalue problem to obtain $2N^2$ eigenvalues and their corresponding modes. Using the values in Table \ref{t:measured_params} and $w=6$ (which yields %equivalently 
$N = 11$), we see that two eigenvalues (each with a frequency of $\sqrt{-\lambda}/(2\pi) = \omega/(2\pi) = f_d \approx 9.17$) lie above the cutoff frequency $f_c \approx 8.77$ Hz. We plot the value of $f_d$ as the horizontal dashed line
in Fig.~\ref{fig:spectrum}(c), and we show the two corresponding modes in Fig.~\ref{fig:finite_spec}(a,b). Both of these modes are spatially localized,
as desired.

%\subsubsection{Damed-driven}

Now suppose that there is driving and damping.
%In the presence of driving and damping, 
%the equations,
Near the background state, equations~\eqref{this} yield the following approximate system:
 \begin{align} 
	M \circ \ddot{X}(t) &= L_a X(t) + L_b Y(t)   - \gamma \dot{X}(t) + a \mathcal{A} \cos(\phi)\sin(2 \pi f t)\,,  \label{eq:lineardd1}  \\
	M \circ \ddot{Y}(t) &= L_c X(t) + L_d Y(t)   - \gamma \dot{Y}(t) + a \mathcal{A} \sin(\phi)\sin(2 \pi f t)\,, \label{eq:lineardd2}
 \end{align}
 where $\mathcal{A}$ is an $N\times N$ matrix that has all $0$ entries except for the single nonzero entry
 \begin{equation*}
	\mathcal{A}_{0,0} =   \frac{I_0 \mu_0 \mathcal{M}}{2\pi h^2 }  \,.
\end{equation*}	 
We obtain $\mathcal{A}$ by expanding the external drive function
$\mathbf{F}^{\mathrm{ext}}$
%in the vicinity of 
near the vanishing displacements and maintaining the leading, non-vanishing term.
We can then find solutions of the system (\ref{eq:lineardd1},\ref{eq:lineardd2}) 
%can then be found in the  form 
% PGK: These are NOT steady states!!!
in the form
 \begin{equation}\label{eq:steady}
	X(t) = \tilde{X}_1\cos( 2 \pi f t) + \tilde{X}_2\sin( 2 \pi f t)\,, \quad Y(t) = \tilde{Y}_1\cos( 2 \pi f t) + \tilde{Y}_2 \sin( 2 \pi f t) \,,
\end{equation}
 where we obtain the $N \times N$ matrices $\tilde{X}_1$, $\tilde{X}_2$, $\tilde{Y}_1$, and $\tilde{Y}_2$ by substituting
 Eq.~\eqref{eq:steady} into Eqs.~\eqref{eq:lineardd1} and \eqref{eq:lineardd2} and then solving the resulting system of linear equations.
%PGK: Chris there is sth here that troubles me. The linearized
%equation is *inhomogeneous*. It is NOT a linearization problem. I
%don't see how this can be. I do not see the derivative of the
%F^{ext} wrt. x_{0,0} in the first eqn. nor wrt to y_{0,0} in the
%second (BTW, I do wonder a bit whether only the x enters the r^2 of
%the x-component and y into the r^2 of the y-component. But even
%putting that aside. I don't see the proper Jacobian formed here. Nor
%is the result really a linear problem around the NLS, but rather an
%inhomogeneous one. It is for that reason that you don't really get an
%eigenvector eqn. as far as I see.

 % I Guess the only way that I can reconcile this is that this is NOT
 % a linearization. It is simply the inhomogeneous mode solution of
 % the original system near the 0 state. As such BTW, this is *approximate*.
 %It is NOT a true
 % linearization. If you agree please amend the text. If you don't
 % agree, let us discuss.

We use root mean square (RMS) quantities as our principal diagnostic for evaluating our results, such as in our bifurcation diagrams.
Most commonly, we compute the RMS of the velocity of the $y$-component of the center particle (i.e., $\dot{y}_{0,0}$).
%{\bf map: what is "$y$"? it looks like the notation has been switched from what it was before; notation needs to be harmonized. CC: see the definition of q on page 3. y is the vertical displacement}
In this case,
\begin{equation*}
	\mathrm{RMS} = \sqrt{ \frac{\int_0^T \dot{y}_{0,0}^2(t) \, dt  }{T} }\,,
\end{equation*}	
where $T = 1/f$ is the period of the excitation frequency.
%{\bf map: should there be an absolute value around $\dot{y}_{0,0}(t)$?. CC: square was missing}
We show a plot of the RMS of the linear 
%steady 
state~\eqref{eq:steady} in Fig.~\ref{fig:finite_spec}(c) as a function of the excitation frequency for a fixed amplitude of $a=0.01$ mV
%(which produces a harmonic force of amplitude $F_{\mathrm{wire}} \approx 1.2414 \cdot10^{-9 }$\,N on the central particle) 
and an excitation angle of $\phi = \pi/2$. The lone resonant peak above the cutoff point is close to the estimated defect frequency $f_d \approx 9.17$ Hz.

In Fig.~\ref{fig:finite_spec}(d), we show a frequency sweep in our experiment for $a=4$ mV 
%(for which the force that is exerted by the wire is $F_{\mathrm{wire}} \approx 4.9656 \cdot10^{-7 }$\,N) 
and $\phi = \pi/2$. 
%{\bf map: $F_{\mathrm{wire}}$ was defined as depending on $r$; what is the value of $r$ above? we defined this as a function, but now we're suddenly writing it as if it were a constant... ; this notation needs to be clarified}
 We show the theoretical values of the cutoff frequency $f_c \approx 8.77$ Hz and defect frequency $f_d \approx 9.17$ Hz that we found in Sec.~\ref{sec:disp} as vertical solid and dashed lines, respectively. We observe that the experimental resonant peak is close to the theoretical value. 
 %, demonstrating
 %that our linear analysis is consistent with
 %{\bf map: model prediction being what? simulation of the full nonlinear equations? (need to be more specific here)}
 To obtain a cleaner resonant peak, we use an excitation amplitude that is large enough to overcome the noise of the system. One such amplitude is $a = 4$ mV. 
 %(for which $F_{\mathrm{wire}} \approx 4.9656 \cdot10^{-7}$\,N).
 %{\bf map: "larger than" what? the other half of the comparison above is missing; please clarify}
 As we will see in Sec.~\ref{sec:results}, an excitation amplitude of $a = 4$ mV is already in the nonlinear regime of the system.

%Now that the linear theory has been established, the key question is how does the nonlinearity of the system affect  the localized modes of the system.

%%%%%%

\subsection{Numerical Methods for the Computation of Nonlinear Localized Modes and Their Stability} \label{sec:nummethods}

For the remainder of our paper, we focus on how the presence of nonlinearity affects the ``defect-induced" linear localized modes
of the system [see, e.g., Fig.~\ref{fig:finite_spec}(c)]. We refer to these solutions, which are localized in space and periodic in time, as
nonlinear localized modes (NLMs). Unlike in the linear equations,
we are unable to obtain
analytical solutions for NLMs using the ansatz \eqref{eq:steady}.
%above ansatz. 
%{\bf map: this is the equation we meant, right? we need to be citing a specific displayed equation here}
%{\bf map: also, in what sense are they 'semi' analytical? it would be good to add a sentence in the appropriate spot when we are departing from analytical things and the 'semi' part becomes in play. Changed wording}
Therefore, we turn to numerical computations. We compute time-periodic orbits of Eq.~\eqref{eq:model} with 
period $T = 1/f$ to high precision by finding roots of the map 
$G = \mathbf{x}(T) - \mathbf{x}(0)$, where $\mathbf{x}(T)$ is the solution of Eq.~\eqref{eq:model} at time $T$ with initial condition $\mathbf{x}(0)$.  
%{\bf map: need to check: was "$F$" used to denote something else previously? CC: we used bolded F before}
%{\bf map: we have $F$ with a subscript for forces; can we change this, please? what about $G$? [even with only bold $F$ used before, I still think it would be rather suboptimal}
Additionally, $\mathbf{x} \in \R^{4N^2}$ is the vector that results from reshaping the matrix with elements
%each matrix representing 
$x_{m,n}$, $y_{m,n}$, $\dot{x}_{m,n}$, and $\dot{y}_{m,n}$ into row vectors and concatenating them into a single vector. 
%{\bf map: can we be a bit clearer above; can we show a block of the matrix and the resulting vector block (after concatenation) or something? CC: see updated text
%to see if this makes the situation clearer. Not sure if this detail warrants a dedicated figure...} 
We obtain roots of the map $G$ using a Jacobian-free Newton--Krylov method \cite{nsoli}
with an initial guess of our linear 
%steady 
state~\eqref{eq:steady}. We compute bifurcation diagrams using a pseudo-arclength continuation algorithm \cite{Doedel} with the excitation frequency $f$ or amplitude $a$ as our continuation parameter. Once we obtain the branches of a bifurcation diagram, we determine the linear stability of each solution $\mathbf{x}$ by solving the variational equations $V' = DG\cdot V$
with initial condition $V(0) = I$, where $I$ denotes the identity matrix and $DG$ is the $4N^2 \times 4N^2$ Jacobian matrix 
of the right-hand side of Eq.~\eqref{eq:model} evaluated at the solution $\mathbf{x}$ \cite{HirschSmaleDevaney}. 
We calculate the Floquet multipliers, which are hereafter denoted by $\sigma$, for a solution by computing the eigenvalues 
of the matrix  $V(T)$.  %\egc{Chris, I added $\sigma$ in order to be consonant with Fig.~5(c)}
%{\bf map: note to self to check when we get to them in a relevant point of the paper: do the $\sigma$ actually have subscript (etc) in the notation that we employ?}
If all of the Floquet multipliers of a solution have an absolute value that is less than or equal to $1$,
 we say that the solution is 
%called
``linearly stable''.
%{\bf map: but aside from the marginal eigenvalues, surely equality to 1 also has the chance to go wrong? How do we know that such a situation can't happen? CC: MAP, I am not sure what you mean by "has the chance to go wrong"...}
%{\bf map: clarifying my prior comment: '1' is a marginal case in the linearization, so can't a nonlinear term put us in an unstable situation? I am puzzled by a linearly marginal case being written down as necessarily stable [wouldn't we get one "1" multiplier by construction, but if there are any additional ones with a magnitude of 1, then we are not guaranteed stability, depending on the nature of the nonlinear perturbation?]}
Otherwise, we say that the solution is ``unstable''.  Clearly, the Floquet multipliers only
give information about the spectral stability of the solutions, and marginal instabilities
associated with unit Floquet multipliers and nonlinear instabilities are possible.
Therefore, stability 
is verified through numerical simulations.
In our bifurcation diagrams, solid blue segments correspond to stable parameter regions and dashed red segments
correspond to unstable regions. 
%Note that w
We compute the Floquet multipliers after we obtain a solution with the Newton--Krylov method to avoid repeatedly solving the large variational system. 
%{\bf  map: I don't understand the previous sentence; it appears to be misconstructed; what is it trying to say? CC: word "avoid" was missing}
This computation would be necessary if we were using a standard Newton method, because the Jacobian of the map $G$ is $V(T)-I$.

%%%%%

 \section{Main Results}\label{sec:results}
%The first step in validating the model equations Eq.~\eqref{eq:model} is to perform a frequency sweep in the experiment to identify
%defect frequency.

%{\bf map: is there a reasonable more descriptive title that we can use for this section?}

%We now present the main results of our computations.

%    \begin{tabular}{@{\quad}p{0.33\linewidth}@{}p{0.33\linewidth}@{}p{0.33\linewidth}@{}  }
%  %\hspace{.2cm}
%\subfigimg[width=\linewidth]{\bf (a)}{breather_3D_a4} &
%\subfigimg[width=\linewidth]{\bf (b)}{breather_cont_a4} &
%\subfigimg[width=\linewidth]{\bf (c)}{FMa4} 
 % \end{tabular}
 \begin{figure}[t]
    \begin{tabular}{@{\quad}p{0.25\linewidth}@{}p{0.25\linewidth}@{}p{0.25\linewidth}@{}p{0.25\linewidth}@{}  }
  %\hspace{.2cm}
\subfigimg[width=\linewidth]{\bf (a)}{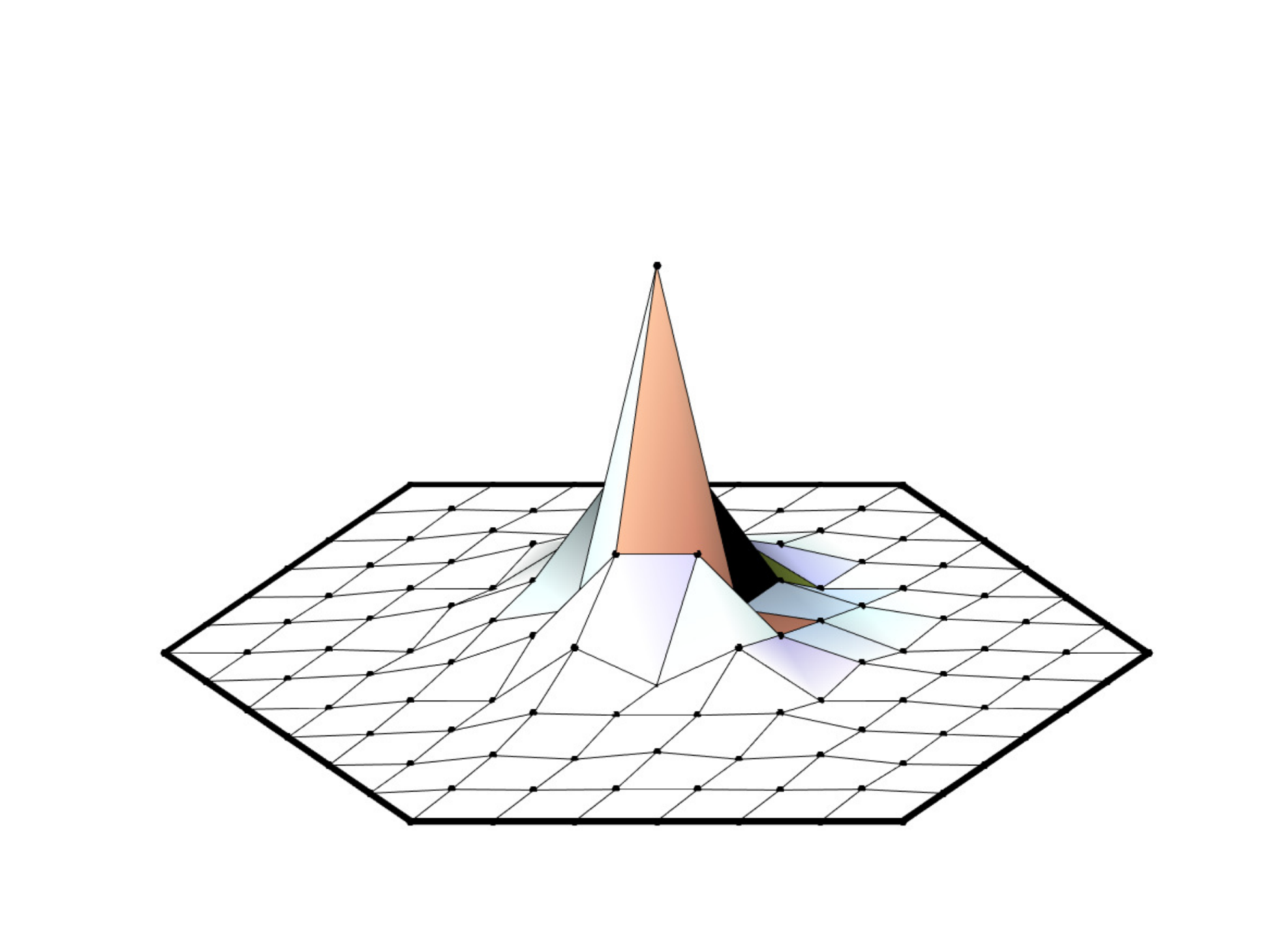} &
\subfigimg[width=\linewidth]{\bf (b)}{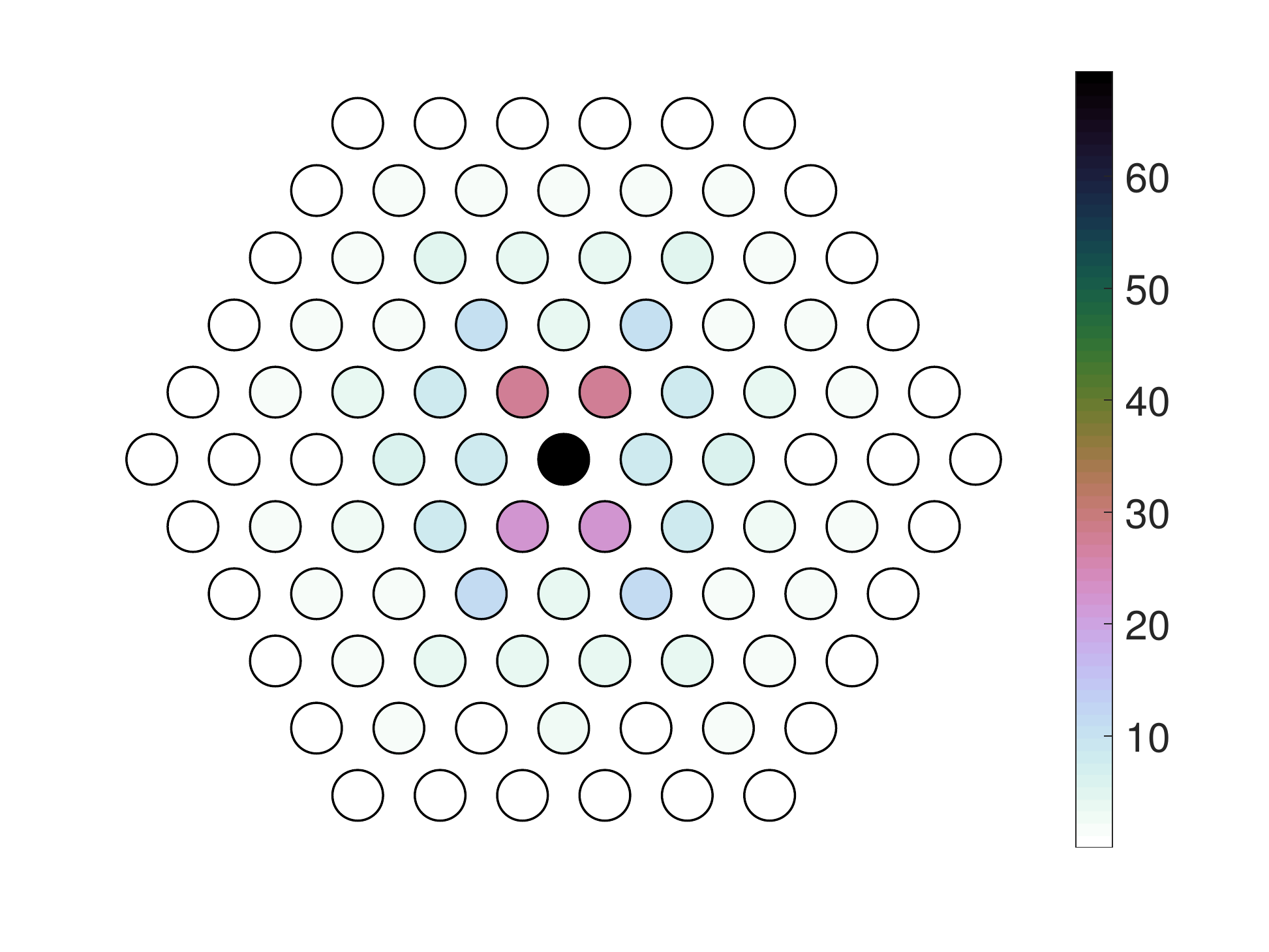} &
\subfigimg[width=\linewidth]{\bf (c)}{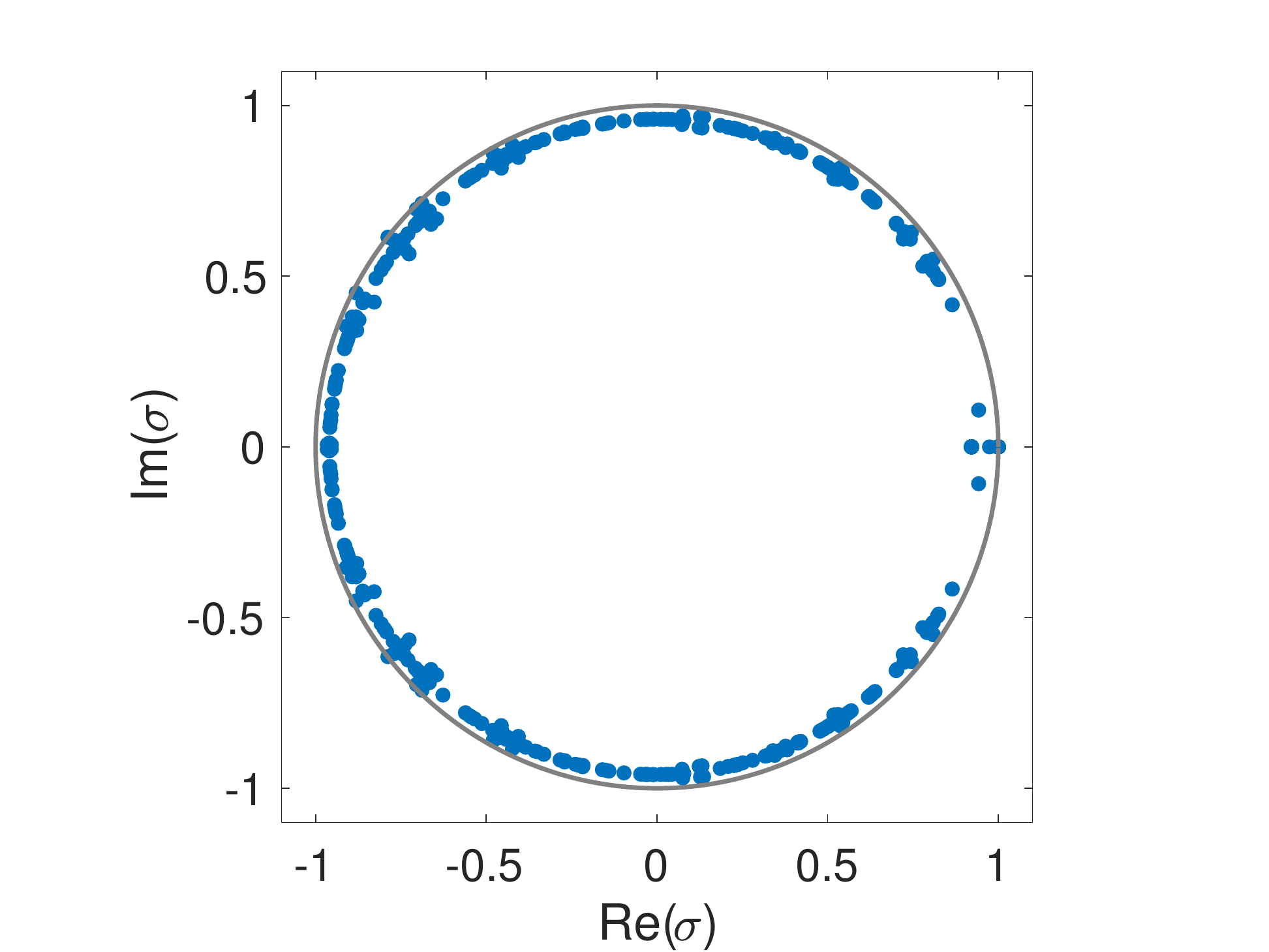} &
\subfigimg[width=\linewidth]{\bf (d)}{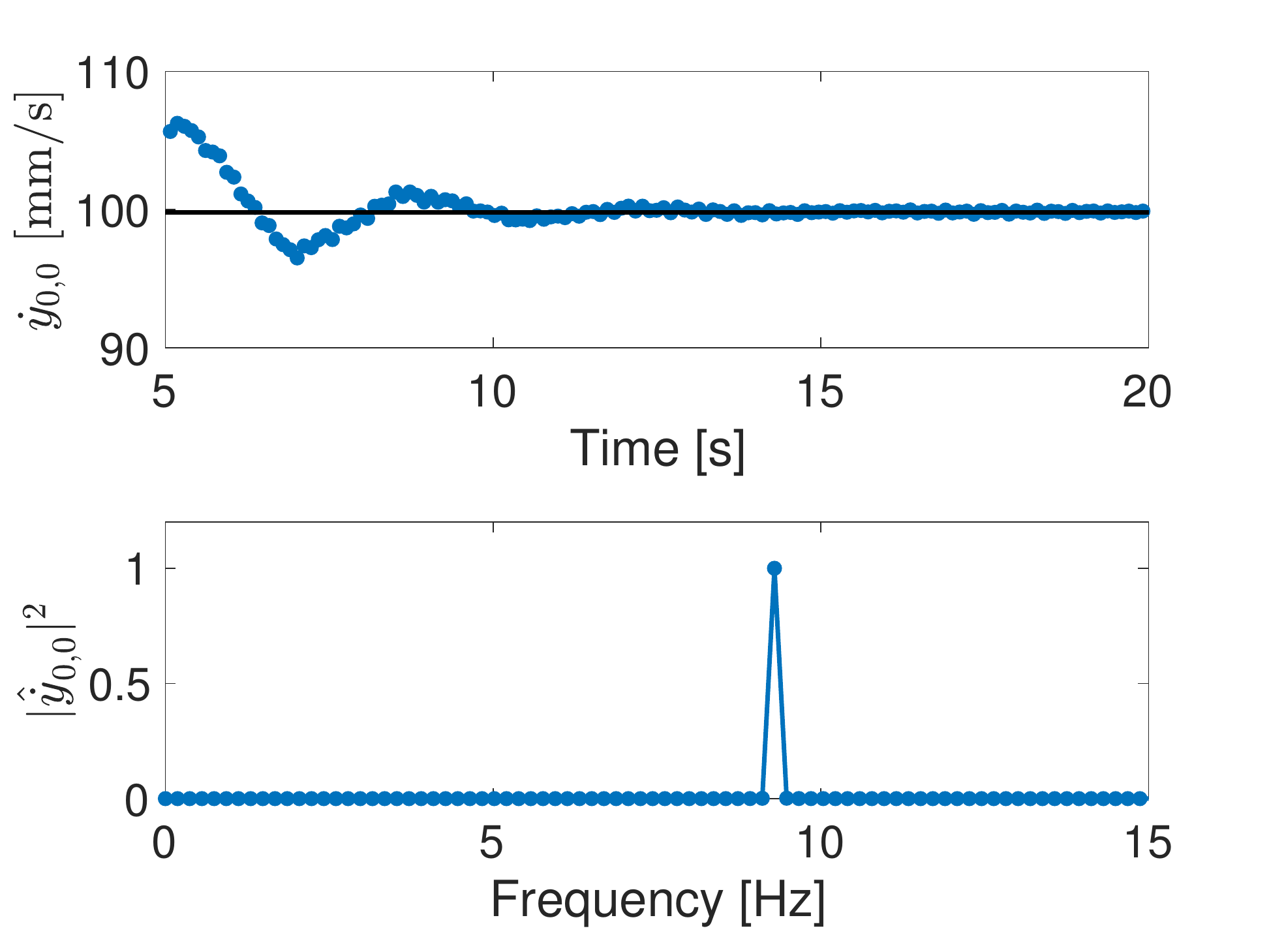} 
  \end{tabular}
 \caption{Nonlinear localized mode of Eq.~\eqref{eq:model} that we obtain using a Newton--Krylov method for $a=4$ mV, 
 %(for which $F_{\mathrm{wire}} \approx 4.9656 \cdot10^{-7 }$\,N),
  $\phi=\pi/2$, and $f = 9.3$ Hz.
 %{\bf map: is 9.3 exact or approximate? CC: exact}
  \textbf{(a)} A surface plot of the NLM. We show the RMS velocity of each magnet in the lattice.
  \textbf{(b)} Intensity plot of the NLM, where the color intensity corresponds to the RMS.
 % {\bf map: Is "color plot" the correct name for the above? I don't think so... CC: see change}
  \textbf{(c)} Floquet multipliers $\sigma$ (blue marks) that are associated with the NLM in the complex plane. We show the unit circle in gray.
  All multipliers lie within (or on) the unit circle, indicating that this solution is stable.  
%  {\bf map: see my query in the main text regarding being on the unit circle. }
  \textbf{(d)} In the top panel, we show local maxima of the time series of $\dot{y}_{0,0}$ when evolving zero initial data (i.e., the initial values of all variables are equal to $0$) for $a=4$ mV, $\phi=\pi/2$, and $f = 9.3$ Hz.
%  {\bf map: what does "zero initial data" mean?}
     %{\bf map: is 9.3 exact or approximate?}
  % {\bf map: what is meant by "zero initial data"? please make this more precise}
  We approach the value $\dot{y}_{0,0} \approx 99.8$ mm/s of the stable NLM. (See the black line.)
In the bottom panel, we show the Fourier transform of the final $5$ seconds of the time series of $\dot{y}_{0,0}$. This reveals
a single large peak at frequency $f \approx 9.3$ Hz.
   %{\bf map: is 9.3 exact or approximate? [I would guess approximate, but we write it as if it's exact] in this case it is approximate}
  }
 \label{fig:breather}
\end{figure}

%%%%

\subsection{Numerical NLMs}

Using the Newton--Krylov method that we described in Sec.~\ref{sec:nummethods}, we obtain a time-periodic solution with $f=9.3$ Hz, $a = 4$ mV 
%(for which $F_{\mathrm{wire}} \approx 4.9656 \cdot10^{-7 }$\,N)
, and $\phi = \pi/2$.
%{\bf map: what exactly do you mean by "high precision"; this should be made more precise. CC: wording removed}
Additionally, because $f = 9.3 > 8.77 \approx f_c$, this solution is localized in space [see Fig.~\ref{fig:breather}(a)]. 
 %{\bf map: is 9.3 exact or approximate?}
 The dominant peak is at the center of the lattice and the next-largest
amplitudes are for the magnets that are 
%located 
adjacent to the center magnet at angles $\pi/3$, $2\pi/3$, $4\pi/3$, and $5\pi/3$ [see Fig.~\ref{fig:breather}(b)]. This is not surprising, because we are exciting the lattice along the $\phi = \pi/2$ direction. 
The Floquet multipliers that are associated with this solution each have a magnitude that is no more than $1$, indicating
that the solution is stable [see Fig.~\ref{fig:breather}(c)]. 
%{\bf map: see past query about equality with 1}
Indeed, upon simulating Eq.~\eqref{eq:model} with initial values of all variables equal to $0$
%data that consists of all zeros 
(i.e., ``zero initial data'') and $f=9.3$ Hz, $a = 4$ mV 
%(for which $F_{\mathrm{wire}} \approx 4.9656 \cdot10^{-7 }$\,N)
, and $\phi = \pi/2$, the dynamics approaches this stable NLM. 
%{\bf map: need to state precisely what is meant by "zero initial data"; this is not clear}
[See the top panel of Fig.~\ref{fig:breather}(d).]
As expected, the Fourier transform of the corresponding time series is localized around a frequency $f \approx 9.3$ Hz.

The spatial decay of the tails of the NLM depends
on which direction of observation one considers.
%PGK: Chris, as I explained in the intro in 2d we don't expect
%e^{-r}. We expect sth like e.g. e^{a r}/\sqrt{r} from the
%corresponding Green's function. Which one is it??
For example, if one measures the RMS velocity
%speed
%amplitude (specifically, the RMS)
of the magnets that lie along the $\theta=\pi/3$ direction, 
the decay appears to be exponential or faster.
%be at least exponential. 
%{\bf map: am I correct that "at least" means what I wrote above? (I find the phrasing 'at least' in this context to be confusing.)}
See the solid blue squares 
%(which are connected by lines) 
in Fig.~\ref{fig:compare}(a), which shows
the RMS velocity
%speed 
versus distance from the origin (following the $\theta=\pi/3$ direction) in a semilog plot for the NLM from Fig.~\ref{fig:breather} [i.e., for the NLM with $f=9.3$ Hz, $a = 4$ mV 
%(for which $F_{\mathrm{wire}} \approx 4.9656 \cdot10^{-7 }$\,N)
, and $\phi = \pi/2$].
This is consistent with the spatial decay properties of breathers in continuous-space settings,
such as the ones in the quintic Ginzburg--Landau equation that were studied in \cite{Boris2Dtail}. The tails of the breathers in \cite{Boris2Dtail} decay at rate
$e^{-b r}/\sqrt{r}$, where $b > 0$ is a constant.
%the decay constant.
%{\bf map: 
%%"continuous settings": continuous in space, I guess? 
%What type of system? I think we need to be more precise about which system it is; I think we have just enough details here to raise more questions than it helps answer}
%{\bf map: I changed "$a$" (which we use immediately above to mean something else in our setting) to "$b$", although we should make sure that we don't use "$b$" to signify anything else}
%{\bf map: the decay rate from Boris's paper was very fishy as written before; it has "$\sqrt{-r}$", which doesn't make sense for a radial variable $r$; from the paper, it looks (I \emph{think}) to be what I wrote above, but please double-check this. The paper isn't using 'breather' language.}
The solid yellow circles exhibit a similar decay for the magnets along the $\theta = 0$ direction for our solution above,
%same 
%solution,
although 
%in this case 
we observe some 
%form of 
modulation in the decay profile, in contrast to the  $\theta = \pi/3$ case.
%{\bf map: the modulation is in contrast with Boris's paper? (or in contrast with our wave in the other direction?) 
%[I need to look at Boris's paper when I get a chance to connect via VPN...]
%}
%Although 
%such 
Modulations in spatial decay have been studied in other settings, such as in the biharmonic $\phi^4$ model \cite{modulated2Dtail}.
%it is interesting that the decay properties of the NLMs that we consider depend on the direction of descent. 
We observe similar decay properties
for an NLM with $f=9.3$ Hz, $a = 5.5$ mV 
%(for which $F_{\mathrm{wire}} \approx 6.8277 \cdot10^{-7 }$\,N),
%(for which $F_{\mathrm{wire}} \approx ???$\,N), 
and $\phi = \pi/2$ [see Fig.~\ref{fig:compare}(b)].
%PGK: It is a little surprising that the jaggedness occurs along a
%lattice direction like that, no? This seems like some sort of modulation of the
%monotonic tails. One would need to look at the spatial eigenvectors
%to examine that, but I understand if this is not considered now.

%{\bf map: the one about was the only one for which we didn't give $F_{\mathrm{wire}}$, so it seems that we should include it for consistency}

%%%%%

\subsection{Experimental NLMs}

\begin{figure}[tb]
   \begin{tabular}{@{\quad}p{0.3\linewidth}@{\quad}p{0.3\linewidth}@{\quad}p{0.3\linewidth}@{}   }
\subfigimg[width=\linewidth]{\bf (a)}{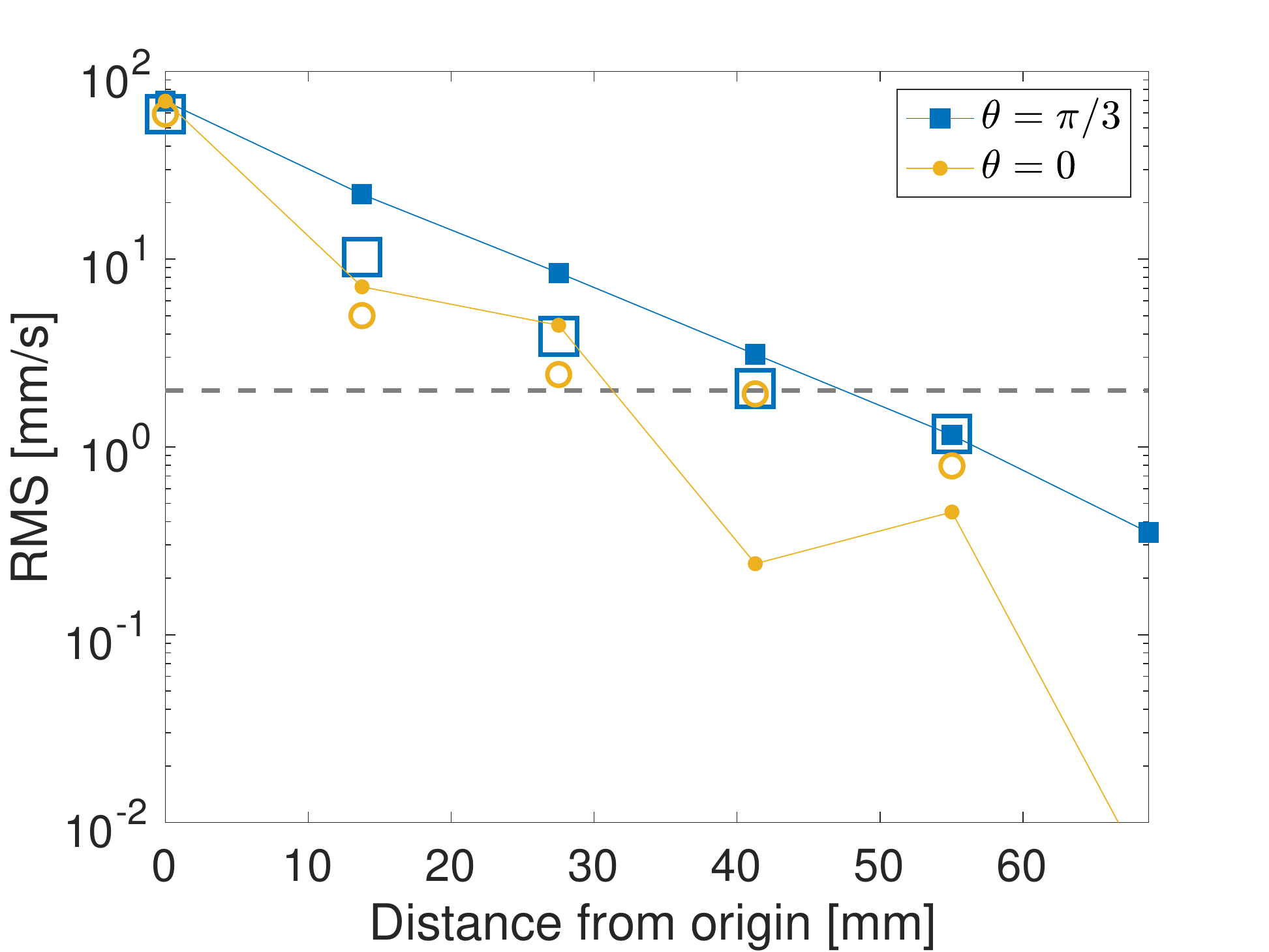} &
\subfigimg[width=\linewidth]{\bf (b)}{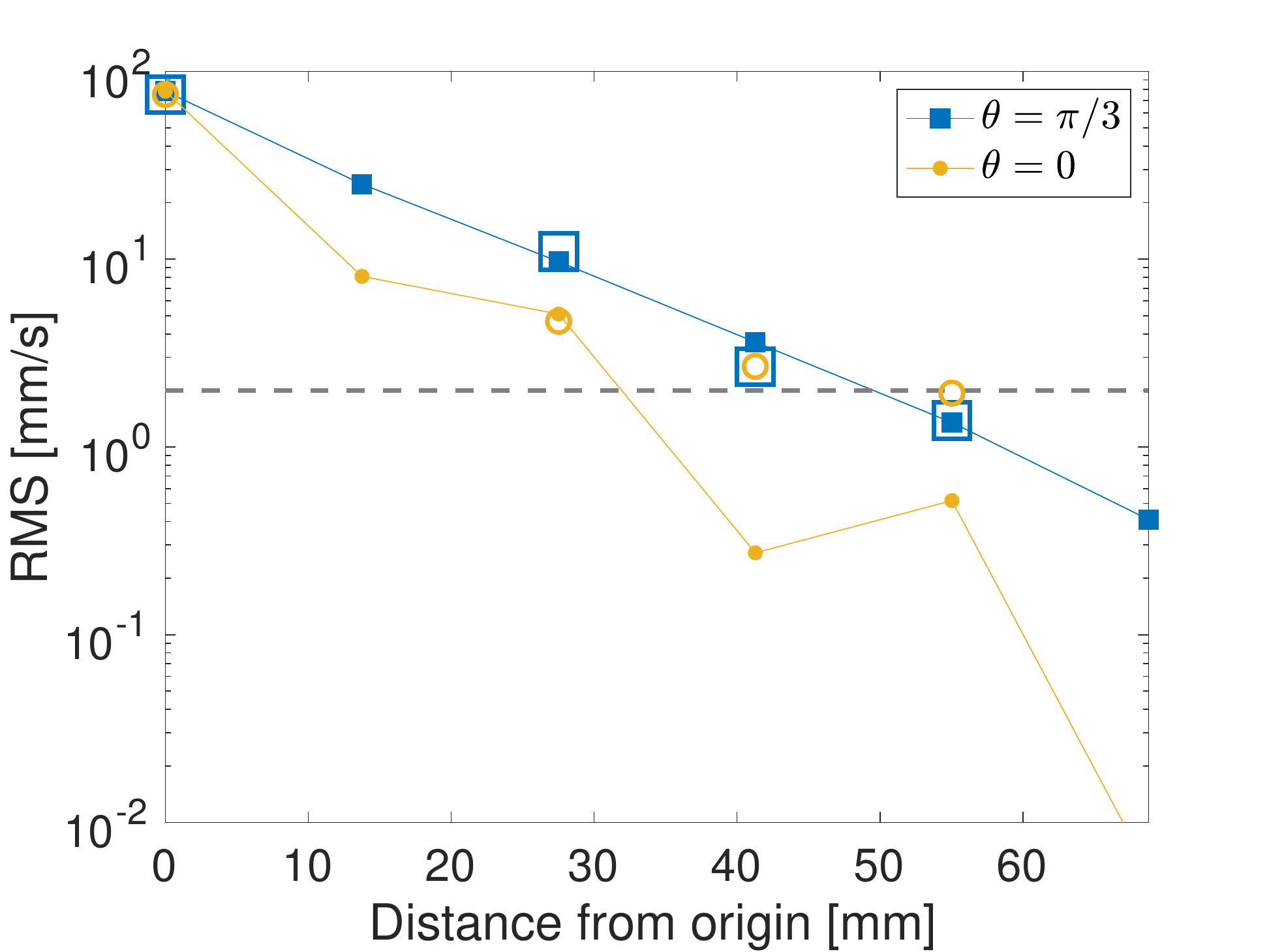} & 
\subfigimg[width=\linewidth]{\bf (c)}{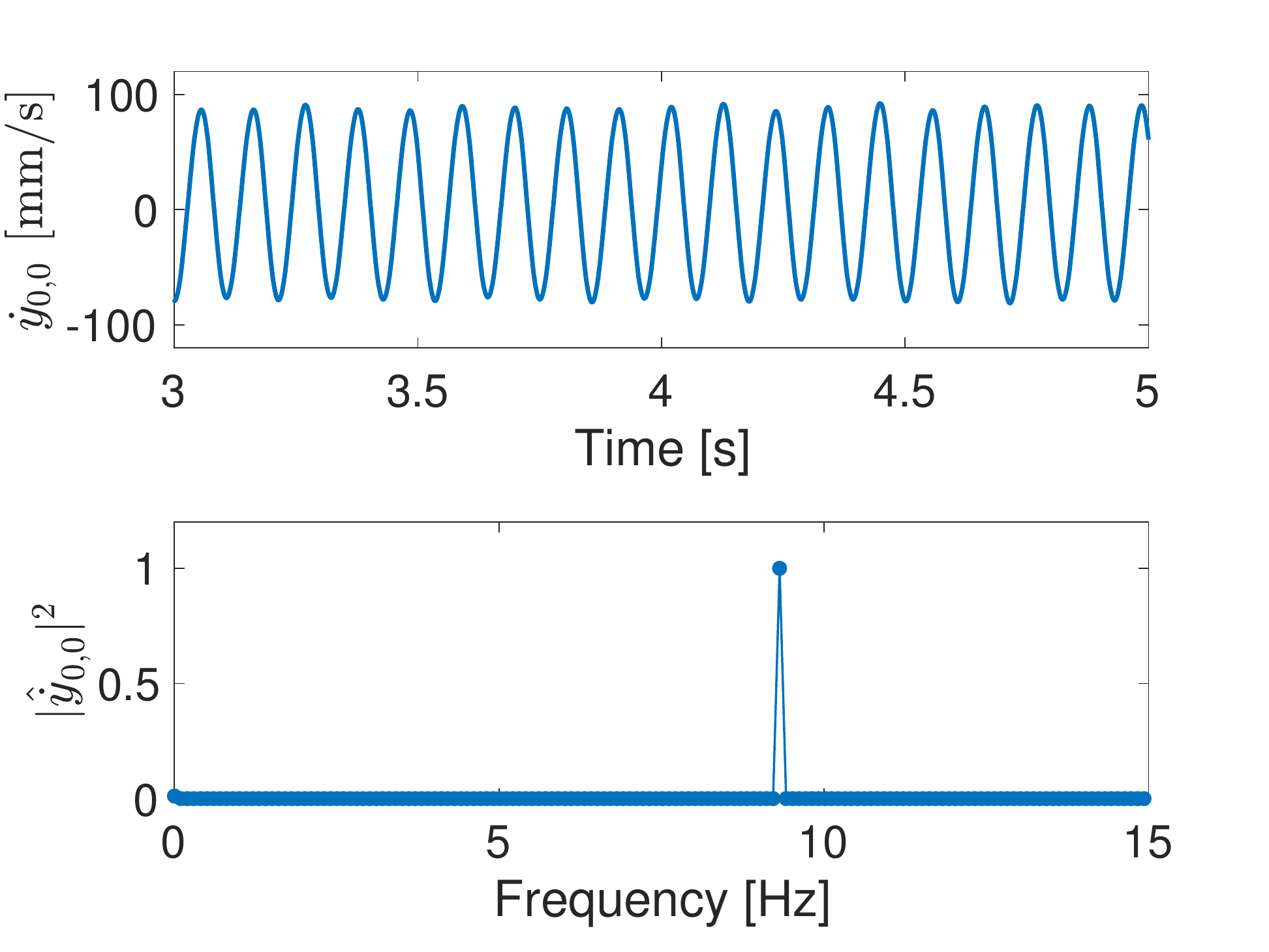} 
  \end{tabular}
 \caption{
 \textbf{(a)} Decay of the NLM in the $\theta = \pi/3$ (blue squares) and $\theta = 0$ (yellow circles) directions
 for a drive amplitude of $a=4$ mV,
 % (for which $F_{\mathrm{wire}} \approx 4.9656 \cdot10^{-7 }$\,N)
drive angle of $\phi=\pi/2$, and drive frequency of $f = 9.3$ Hz. We show the RMS velocity 
 %speed 
 versus the distance to the origin of the lattice. We show experimental results as
 open markers and the numerical results as solid markers that are connected by lines. \textbf{(b)} The same as panel
 (a), but for a drive amplitude of $a=5.5$ mV.
 % (for which $F_{\mathrm{wire}} \approx 6.8277 \cdot10^{-7 }$\,N). 
 \textbf{(c)} Time series (top panel) of the center particle of the experimental NLM
 and corresponding Fourier transform (bottom panel)
 with $\phi = \pi/2$, $f = 9.3$ Hz, and $a = 4$ mV.
 % (for which $F_{\mathrm{wire}} \approx 4.9656 \cdot10^{-7 }$\,N).
 Despite the presence of some noise, the solution is predominantly periodic
 %harmonic 
 in time.
 %{\bf map: above was written "harmonic", but isn't it instead "periodic" that was meant?. CC: OK with me}
  Indeed, the Fourier transform 
 of the time signal is highly localized around the frequency $f=9.3$ Hz.
 %{\bf map: the labels (a) and (b) are too close to the picture}
 }
 \label{fig:compare}
\end{figure}

In our experiments, it is difficult to initialize the system with predetermined positions and velocities.
To obtain the NLM, we excite the system with a small amplitude ($a=1$ mV) 
%for which $F_{\mathrm{wire}} \approx 1.2414 \cdot10^{-7 }$\,N),
 which we increase gradually to the value $a=4$ mV 
 %(for which $F_{\mathrm{wire}} \approx 4.9656 \cdot10^{-7 }$N)
  over about $3$ minutes. Because we predict the NLM to be stable at the resulting parameter values,
we record data after 90 periods of motion have elapsed once we attain the value $a=4$ mV. 
%{\bf map: how long is "sufficient time"? can we give a number here? CC: see change}
We track the velocities at the center particle with a laser vibrometer (see Sec.~\ref{sec:exp_setup}), and we record the time series of the magnet velocities using an oscilloscope. [See the top panel of Fig.~\ref{fig:compare}(c).] As expected, we obtain dynamics
that are periodic in time, as we can see not only with the time series but also via its Fourier transform [which we show in the bottom panel of Fig.~\ref{fig:compare}(c)]. We obtain similar experimental results for an amplitude of $a=5.5$ mV. 
%(for which $F_{\mathrm{wire}} \approx 6.8277 \cdot10^{-7 }$\,N). 

To examine the spatial decay of the experimental NLM, we record the positions
of the magnets in half of the lattice using a digital camera (see Sec.~\ref{sec:exp_setup}). By numerically
differentiating the positions, we obtain an estimate for the velocities of these magnets. We were unable to do a complete
full-field realization, because the DIC loses track of some magnets (if, e.g., the
magnets begin to spin). 
%{\bf map: the acronym "DIC" has not been defined; please define this acronym explicitly (and seemingly in the section on the experimental setup) CC: added}
However, we captured enough data to compute the decay along the two primary 
%diagonals 
directions ($\theta = \pi/3$ and $\theta = 0$) that we examined in our numerical NLMs. %(The missing points are due to the DIC losing track of magnets.) 
The open blue squares of Fig.~\ref{fig:compare}(a) show the decay along the $\theta=\pi/3$ direction
and the open yellow circles show the decay along the $\theta=0$ direction for data that we obtained with
$f=9.3$ Hz, $a = 4$ mV,
% (for which $F_{\mathrm{wire}} \approx 4.9656 \cdot10^{-7 }$\,N), 
and $\phi = \pi/2$. The horizontal dashed line
is our estimated mean value of the noise (see Sec.~\ref{sec:exp_setup}). We show the experimental
%ly observed 
decay rates along with our numerical results.
%the theoretical prediction. 
Although the numerical values overestimate
%theoretical value over estimates 
the RMS velocity,
%speed, 
the agreement is still reasonable, especially for the center magnet. 
We find similar decay properties in our experiment with $f=9.3$ Hz, $a = 5.5$ mV 
%(for which $F_{\mathrm{wire}} \approx 6.8277 \cdot10^{-7 }$\,N)
, and $\phi = \pi/2$. [See the open markers of Fig.~\ref{fig:compare}(b).] 

Recall that we do not tune the numerical results to fit the experimentally obtained NLM solution. Instead, 
we determine each of the parameter values beforehand, as described in Sec.~\ref{sec:exp_setup}.

 %%%%%
 
\subsection{Frequency Continuation}

    \begin{figure}
    \begin{tabular}{@{\quad}p{0.33\linewidth}@{}p{0.33\linewidth}@{}p{0.33\linewidth}@{}  }
  %\hspace{.2cm}
\subfigimg[width=\linewidth]{\bf (a)}{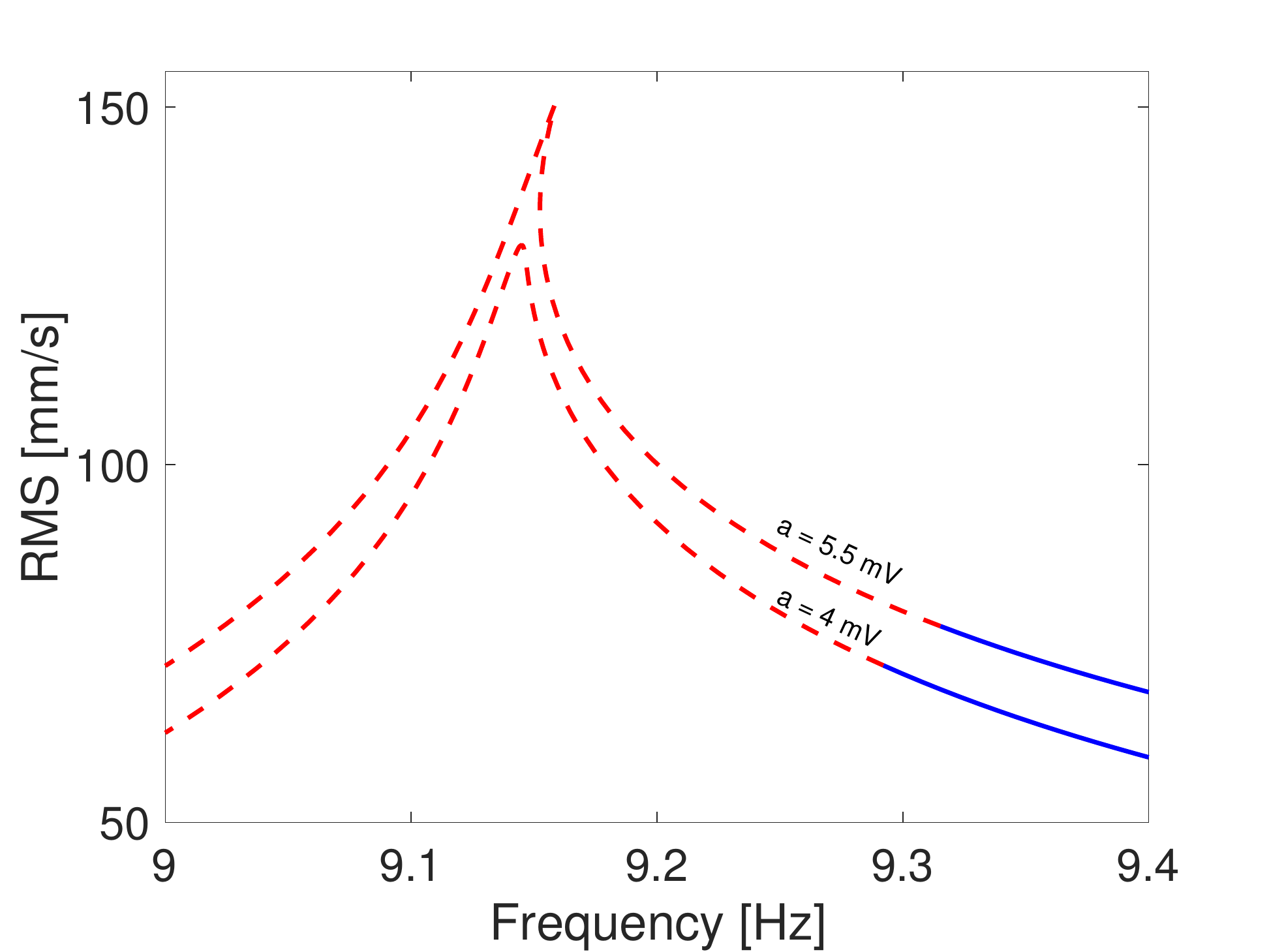} &
\subfigimg[width=\linewidth]{\bf (b)}{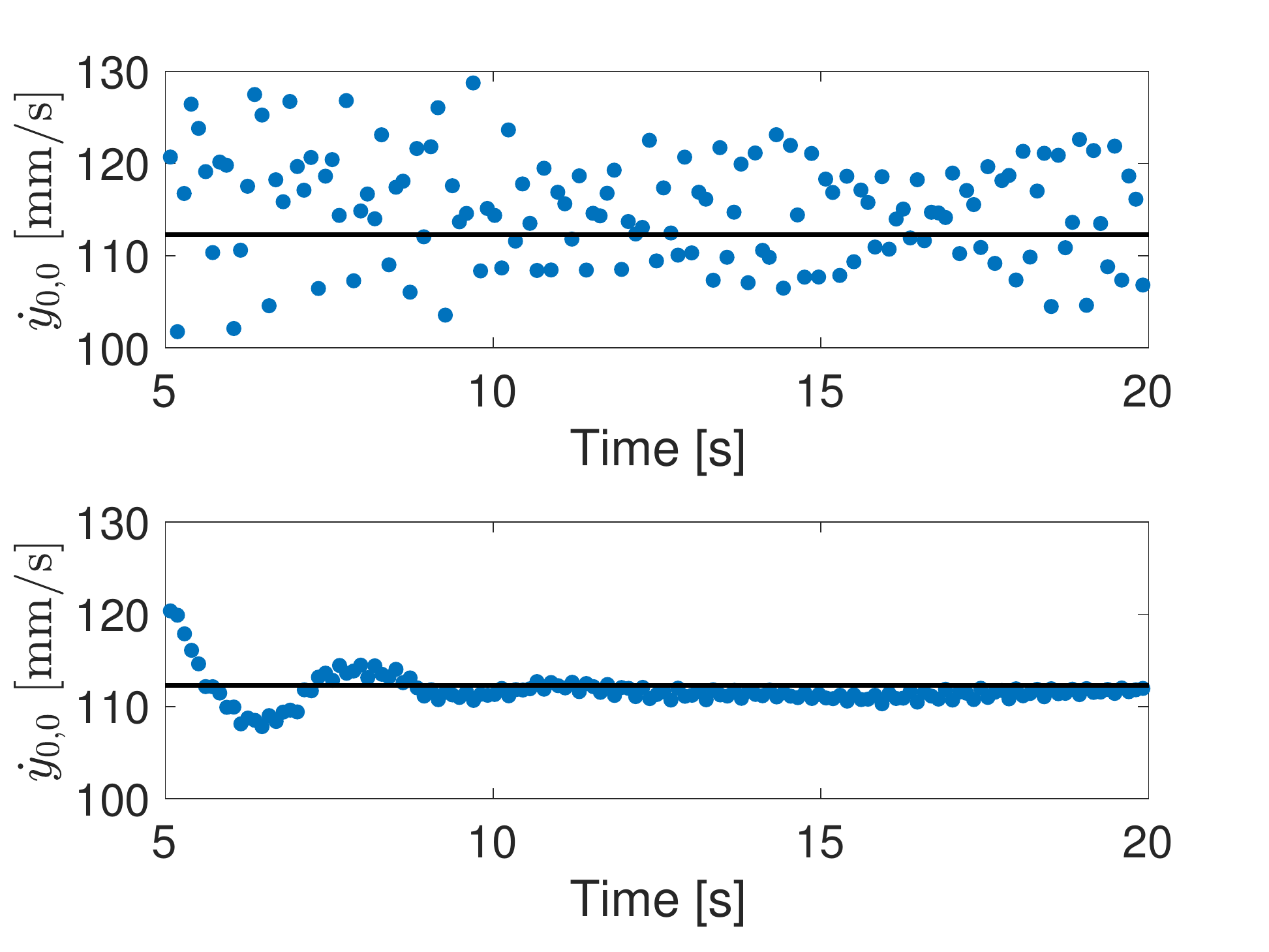} &
\subfigimg[width=\linewidth]{\bf (c)}{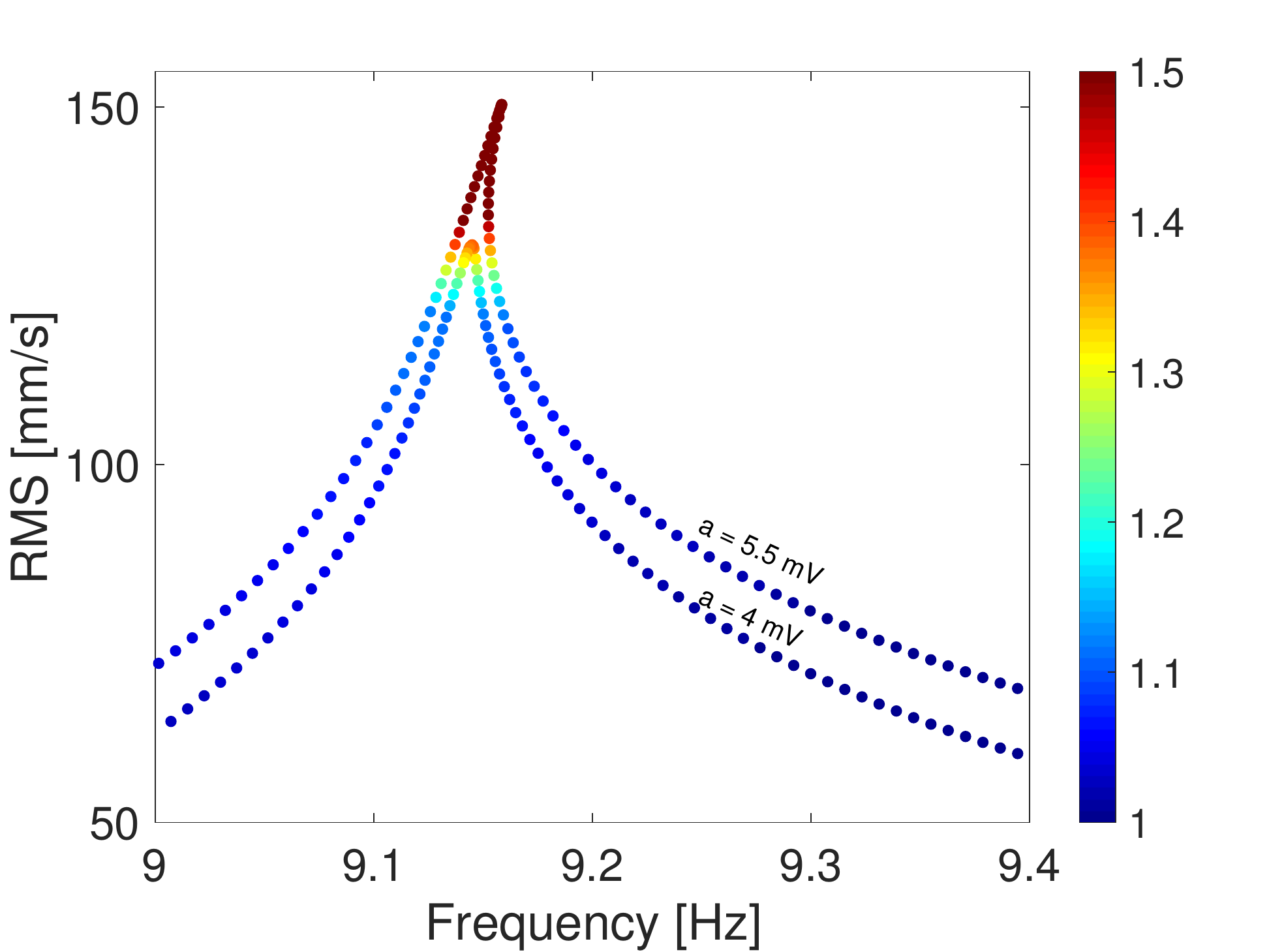} 
  \end{tabular}
 \caption{Frequency continuation of NLMs. \textbf{(a)} Continuation with respect to frequency for fixed excitation angle $\phi=\pi/2$ for
% fixed for
the two amplitudes, $a=4$ mV and $a=5.5$ mV, that we considered in Fig.~\ref{fig:compare}.
% The associated forces that are exerted by the wire are $F_{\mathrm{wire}} \approx 4.9656 \cdot10^{-7 }$\,N and $F_{\mathrm{wire}} \approx 6.8277 \cdot10^{-7 }$\,N, respectively.
 \textbf{(b)} In the top panel, we plot the local maxima of the time series of $\dot{y}_{0,0}$ when we perturb the NLM
 with $f=9.3$ Hz and $a=5.5$ mV 
 %(for which $F_{\mathrm{wire}} \approx 6.8277 \cdot10^{-7 }$\,N) 
 along the eigenvector that is associated with the largest-magnitude Floquet multiplier.
 The size of this perturbation is equal to 5\% of the amplitude of the solution.
 %{\bf map: can we be more precise; how much less than 5\%?}
  We show the corresponding value of the local maximum of the NLM that we obtain via a Newton--Krylov method as the solid black line.
  %{\bf map: for the sentence above, you mean the local maximum, I suppose?}
In the bottom panel, we plot local maxima of the time series when evolving zero initial data with a fixed frequency $f=9.3$ Hz and amplitude drive $a=5.5$.
% (for which $F_{\mathrm{wire}} \approx 6.8277 \cdot10^{-7 }$\,N).
% {\bf map: what does "zero initial data" mean? this needs to be made precise. CC: defined in text now}
 \textbf{(c)} The same as panel (a), but with color intensity corresponding to the magnitude of the largest
 Floquet multiplier. Except near the peaks of the resonant curves, the instabilities are fairly weak.
 %{\bf map: the label for (b) is too close to the picture}
  }
 \label{fig:angle90}
\end{figure}

In Figs.~\ref{fig:finite_spec}(d) and \ref{fig:compare}(a,b), we demonstrate that our 
%simplified 
model~\eqref{eq:model} agrees reasonably well with our experimental data.
% {\bf map: simplified from what? please also point explicitly to the more complicated model here}
 We now conduct a series of numerical computations, with parameter continuation (see Sec.~\ref{sec:nummethods} for a description of our procedure), to
 get a better sense of the role of nonlinearity in
 Eq.~\eqref{eq:model} and its interplay with the disorder (at the
 central magnet) and the
 discreteness of the model.
 %{\bf map: which system? the simplified model? a more complicated one? we should cite equations of motion here}
 We return to our experiments in Sec.~\ref{sec:amp} to see what nonlinear effects we are able to capture in the laboratory.
 
We first perform continuation with respect to the excitation frequency $f$ for a fixed excitation angle
 $\phi = \pi/2$ for various values of the excitation amplitude $a$. 
 We thereby generate nonlinear analogs of the linear resonant peak that we showed in Fig.~\ref{fig:finite_spec}(c). 
In Fig.~\ref{fig:angle90}(a), we show frequency continuations for our two
%with the amplitude fixed to the 
drive amplitudes, $a=4$ mV and $a=5.5$, of the NLMs from Fig.~\ref{fig:breather} and Figs.~\ref{fig:compare}(a,b). 
%For these amplitudes, recall that the forces that are exerted by the wire are $F_{\mathrm{wire}} \approx 4.9656 \cdot10^{-7 }$\,N and $F_{\mathrm{wire}} \approx 6.8277 \cdot10^{-7 }$\,N, respectively. 
From comparing these frequency continuations to the linear case in Fig.~\ref{fig:finite_spec}(c), we see that the nonlinearity deforms the peak,
%; the peak 
which becomes narrower and starts to bend towards higher frequencies. The nonlinearity also destabilizes the solutions at some critical frequency; this occurs at $f \approx 9.29$ Hz for $a=4$ mV and at $f \approx 9.31$ Hz for $a=5.5$ mV.
% This implies 
 We therefore see that the NLM in Fig.~\ref{fig:compare}(c) is unstable. Although our numerical computations predicted this NLM to be unstable, it is notable that we are able to access it in our experiments. Although this seems to imply that our theory is inconsistent with our experiments for the parameter values $f \approx 9.31$ Hz and $a=5.5$ mV, the instability of the NLM for these parameter values is rather weak (with $\mathrm{max}_i(|\sigma_i|) \approx 1.007$). We observe instability after many periods 
 %of motion 
 when we perturb the NLM along the eigenvector that is associated to the unstable Floquet multiplier $\sigma \approx 1.007$.
 See the top panel of Figure~\ref{fig:angle90}(b). However, if we initialize the dynamics with zero
 initial data, we approach and stay close to the NLM solution that we obtained via a Newton--Krylov method 
 %and stays close to it, 
 even after 200 periods of motion. 
 %{\bf map: what is meant by "zero initial data"?}
 See the bottom panel of Figure~\ref{fig:angle90}(b). This suggests that solutions with weak instabilities can still attract nearby points in phase space, at least initially, and that our numerical prediction for  
 $f \approx 9.31$ Hz and $a=5.5$ mV is consistent with our experimental observations.
% dynamics.
  Figure~\ref{fig:angle90}(c) is the same as Figure~\ref{fig:angle90}(a), but now the color scale corresponds to the magnitude of the maximum Floquet multiplier. This illustrates that the magnitude of the instability is 
  %actually quite 
 %quite  {\bf map: "quite" has opposite meanings in UK vs US English, so we should avoid this descriptor}
 weak throughout most of the solution branch. The instability
 is at its largest growth rate (with $\mathrm{max}_i(|\sigma_i|) \approx 2.15$) at the peak of the resonant curve.

 %\subfigimg[width=\linewidth]{\bf (a)}{freq_angle60} 

\begin{figure}
  %  \begin{tabular}{@{\quad}p{0.25\linewidth}@{}p{0.25\linewidth}@{}p{0.25\linewidth}@{} p{0.25\linewidth}@{}  }
   \begin{tabular}{@{\quad}p{0.33\linewidth}@{}p{0.33\linewidth}@{}p{0.33\linewidth}@{} }
  %\hspace{.2cm}
\subfigimg[width=\linewidth]{\bf (a)}{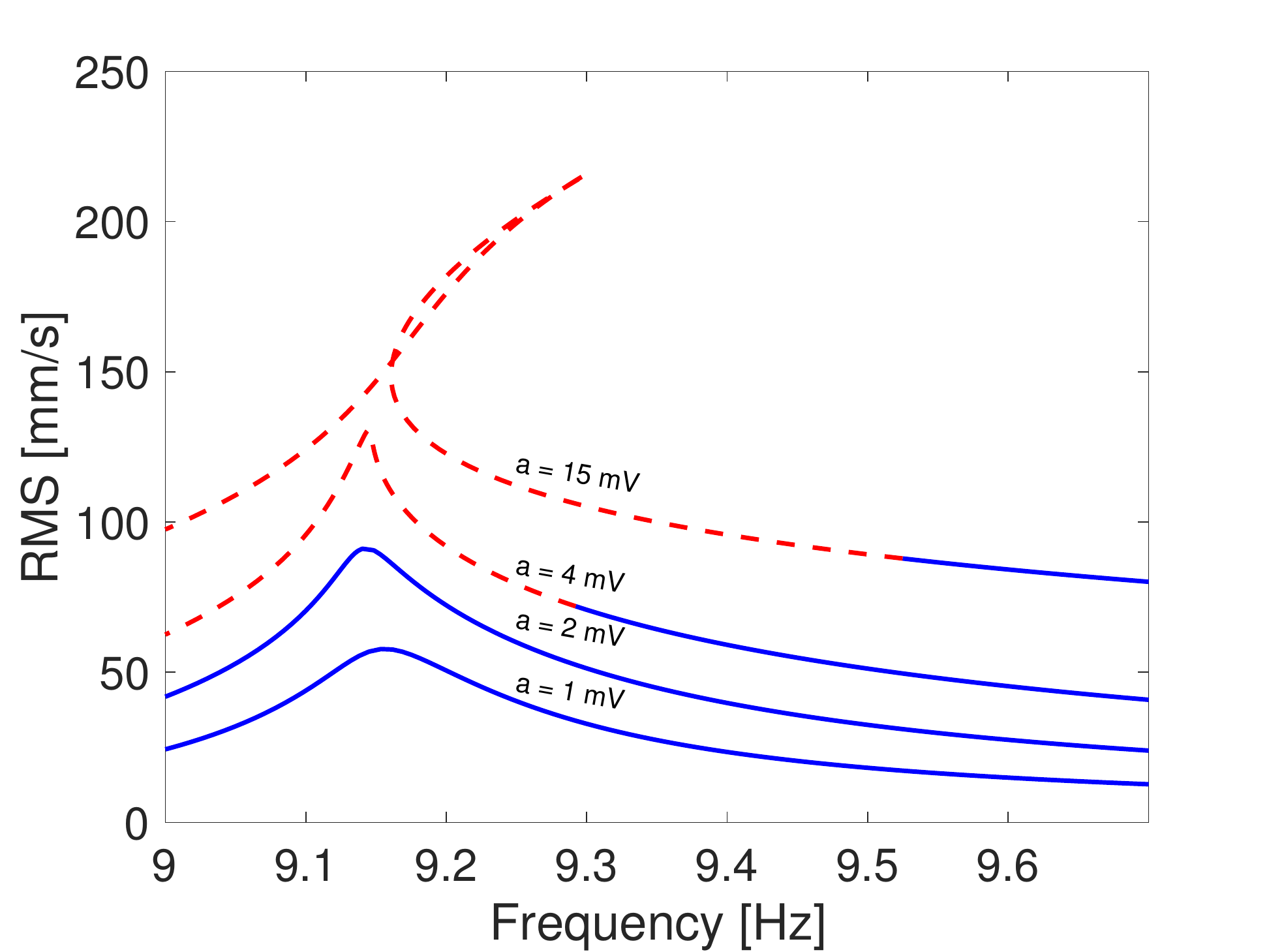} &
\subfigimg[width=\linewidth]{\bf (b)}{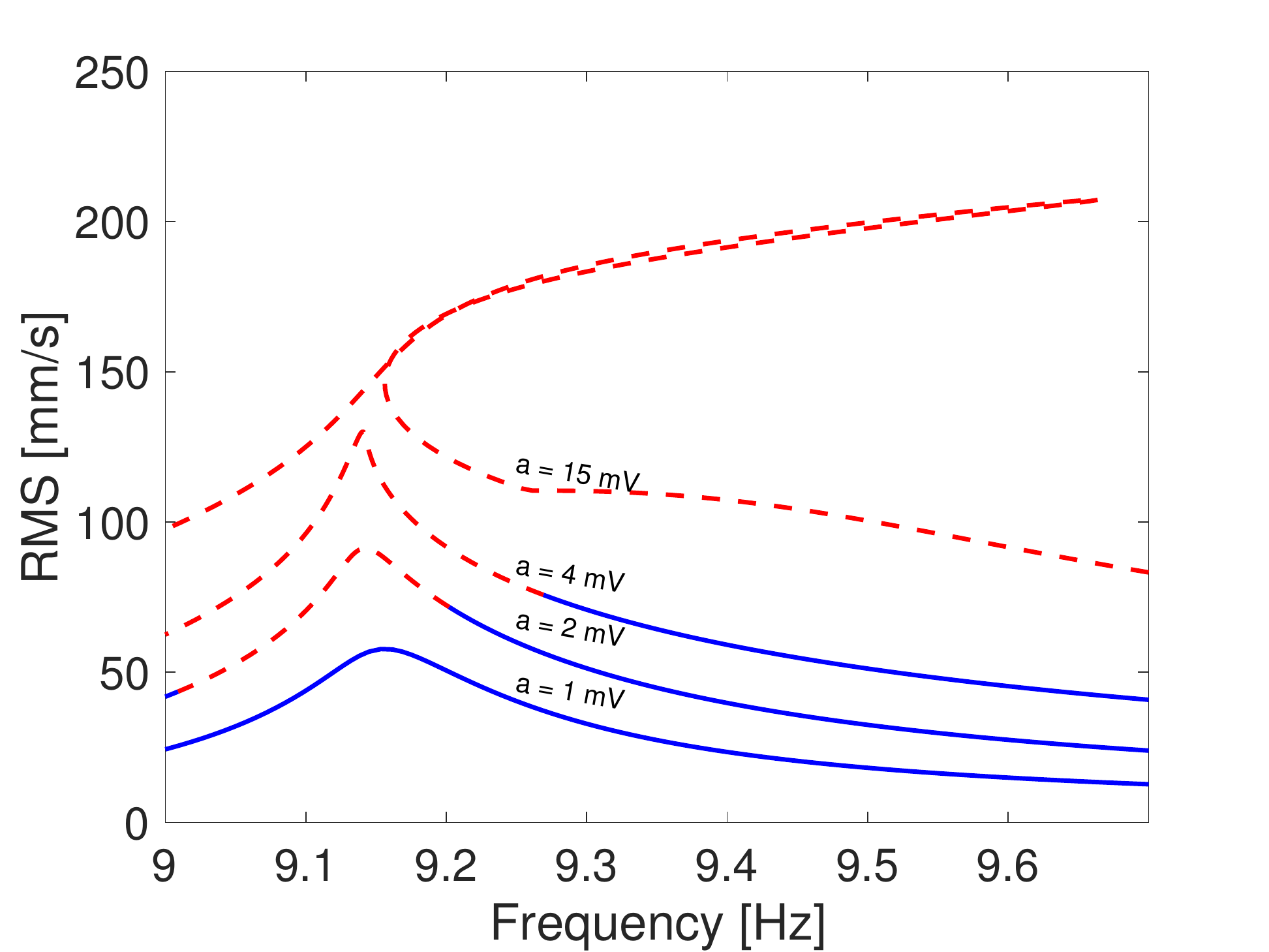}  &
\subfigimg[width=\linewidth]{\bf (c)}{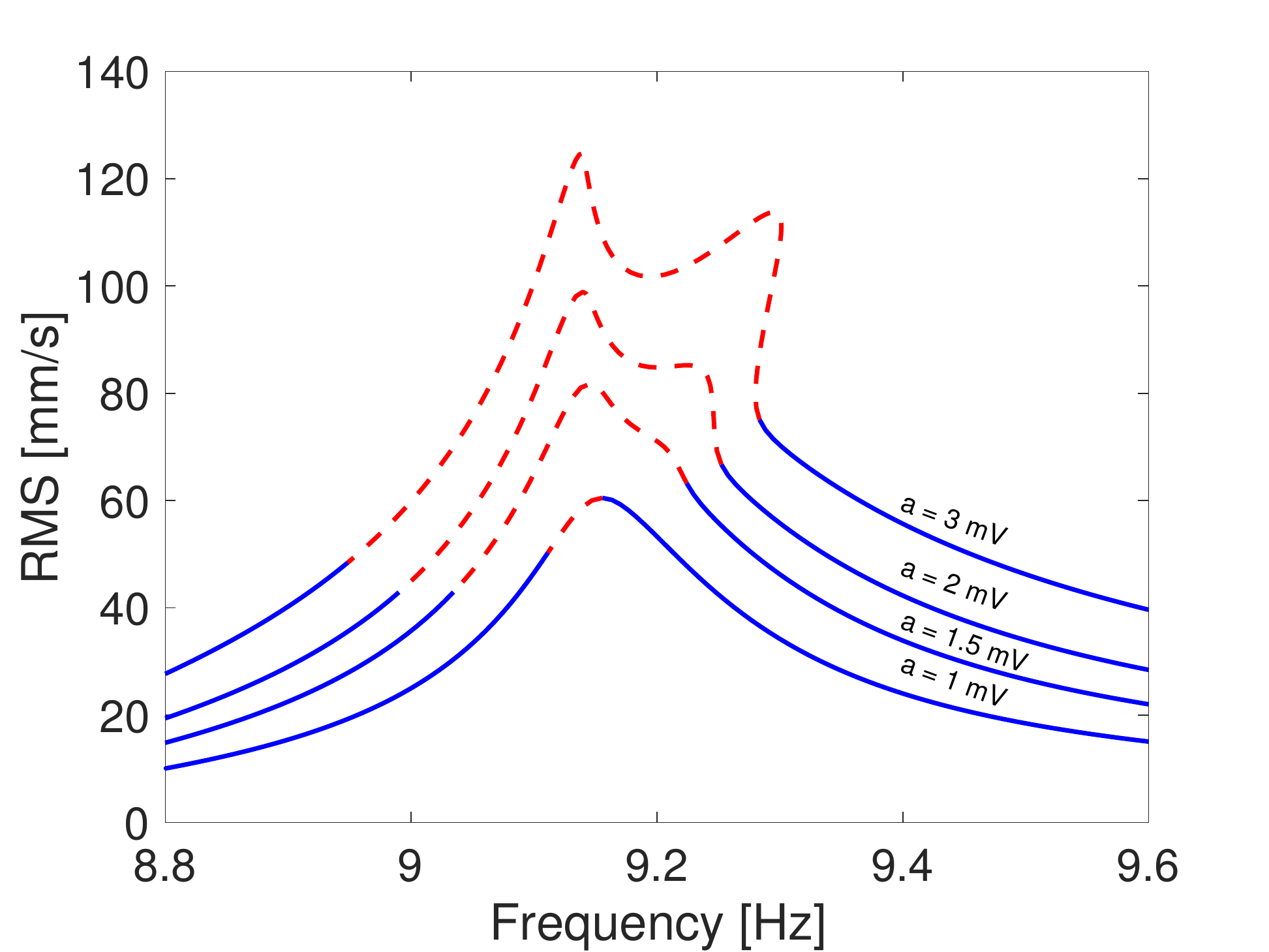} 
  \end{tabular}
 \caption{\textbf{(a)} Frequency continuation with excitation angle $\phi = \pi/2$
  for excitation amplitudes $a = 1$ mV, $a = 2$ mV, $a = 4$ mV, and $a = 15$ mV 
  %(for which the forces that are exerted by the wire are $F_{\mathrm{wire}} \approx 1.2414$\,N, $F_{\mathrm{wire}} \approx 2.4828$\,N, $F_{\mathrm{wire}} \approx 4.9656$\,N, and $F_{\mathrm{wire}} \approx 18.621 \cdot10^{-7 }$\,N, respectively). 
  \textbf{(b)}  Frequency continuation with $\phi = 0$
  for $a = 1$ mV, $a = 2$ mV, $a = 4$ mV, and $a = 15$ mV 
  %(for which $F_{\mathrm{wire}} \approx 1.2414$\,N, $F_{\mathrm{wire}} \approx 2.4828$\,N, $F_{\mathrm{wire}} \approx 4.9656$\,N, and $F_{\mathrm{wire}} \approx 18.621 \cdot10^{-7 }$\,N, respectively). 
  \textbf{(c)} Frequency continuation with  $\phi = \pi/3$
  for $a = 1$ mV, $a = 1.5$ mV, $a = 2$ mV, and $a = 3$ mV.
  % (for which $F_{\mathrm{wire}} \approx 1.2414$\,N, $F_{\mathrm{wire}} \approx 1.8621$\,N, $F_{\mathrm{wire}} \approx 2.4828$\,N, and $F_{\mathrm{wire}} \approx 3.7242 \cdot10^{-7 }$\,N, respectively).   
 % {\bf map: (1) the $a = 2$ text in panels (a) and (b) are a bit too close to the curve; (2) it would also be good to slightly raise the labels (a), (b), and (c) so that they aren't so close to the axis number}
 %{\bf map: panel (b): the two unstable branches are almost on top of each other at some point; do we want to do anything about distinguishing these two curves to make them visually more distinguishable? (I guess the issue is that this would entail changes in other panels as well...)}
  }
 \label{fig:otherangle}
\end{figure}

In Fig.~\ref{fig:otherangle}(a), we show  the gradual bending of the resonant peak
 for progressively larger excitation amplitudes with $\phi = \pi/2$. In particular, for $a=15$ mV, the peak bends so far
 that additional solutions at $f \approx 9.3$ Hz emerge. However,
 these large amplitude solutions are very unstable,
 and we were not able to access them either in our direct numerical simulations or in our experiments. Indeed, as we will discuss in Sec.~\ref{sec:amp},
  we observe different types of dynamics at such 
 %large
 excitation amplitudes. We can also tune the excitation angle $\phi$ and thereby deform
 %, which deforms 
 the resonant peak in a different way. 
For example, when we fix the excitation angle to $\phi = 0$ (i.e., an excitation along $n=0$),
the resonant curves are qualitatively similar to those for $\phi=\pi/2$ for excitation amplitudes $a = 1$, $a = 2$, and $a = 4$ [see Fig.~\ref{fig:otherangle}(b)], although the stability properties are slightly different. For the large excitation of
 $a=15$ mV, the resonant curve bends even further toward higher frequencies. Even greater qualitative
differences occur for $\phi = \pi/3$ (i.e., an excitation along $m=0$); see Fig.~\ref{fig:otherangle}(c). In this case, 
for small-amplitude excitations, the resonant peak has a unimodal shape, as expected. 
%{\bf map: this is the first time that 'bell-shaped form' is mentioned or that it is called 'usual'; this terminology should also be mentioned in the first example in which it occurs (no actual change in the present spot)}
%{\bf map: why is this form expected? Sorry, but I don't have the intuition that it should be specifically 'bell-shaped' [I do expect it to have one hump, so it's the bell-shaped part specifically that is giving me pause]}
However, as we consider gradually larger excitation amplitudes, an additional peak begins to emerge
from the main solution branch, leading to a two-humped profile in the
dependence of the RMS velocity
%speed 
%dependence 
on the frequency.
This deformation is noticeable for relatively small excitation
 amplitudes. Specifically, we observe the existence of multiple solutions even for excitation amplitudes that are as small as $a=3$ mV, where there is bifurcation at frequency $f\approx 9.28$ Hz.

\begin{figure}
  %  \begin{tabular}{@{\quad}p{0.25\linewidth}@{}p{0.25\linewidth}@{}p{0.25\linewidth}@{} p{0.25\linewidth}@{}  }
   \begin{tabular}{@{\quad}p{0.35\linewidth}@{}p{0.35\linewidth}@{}p{0.35\linewidth}@{} }
  %\hspace{.2cm}
 \subfigimg[width=\linewidth]{\bf (a)}{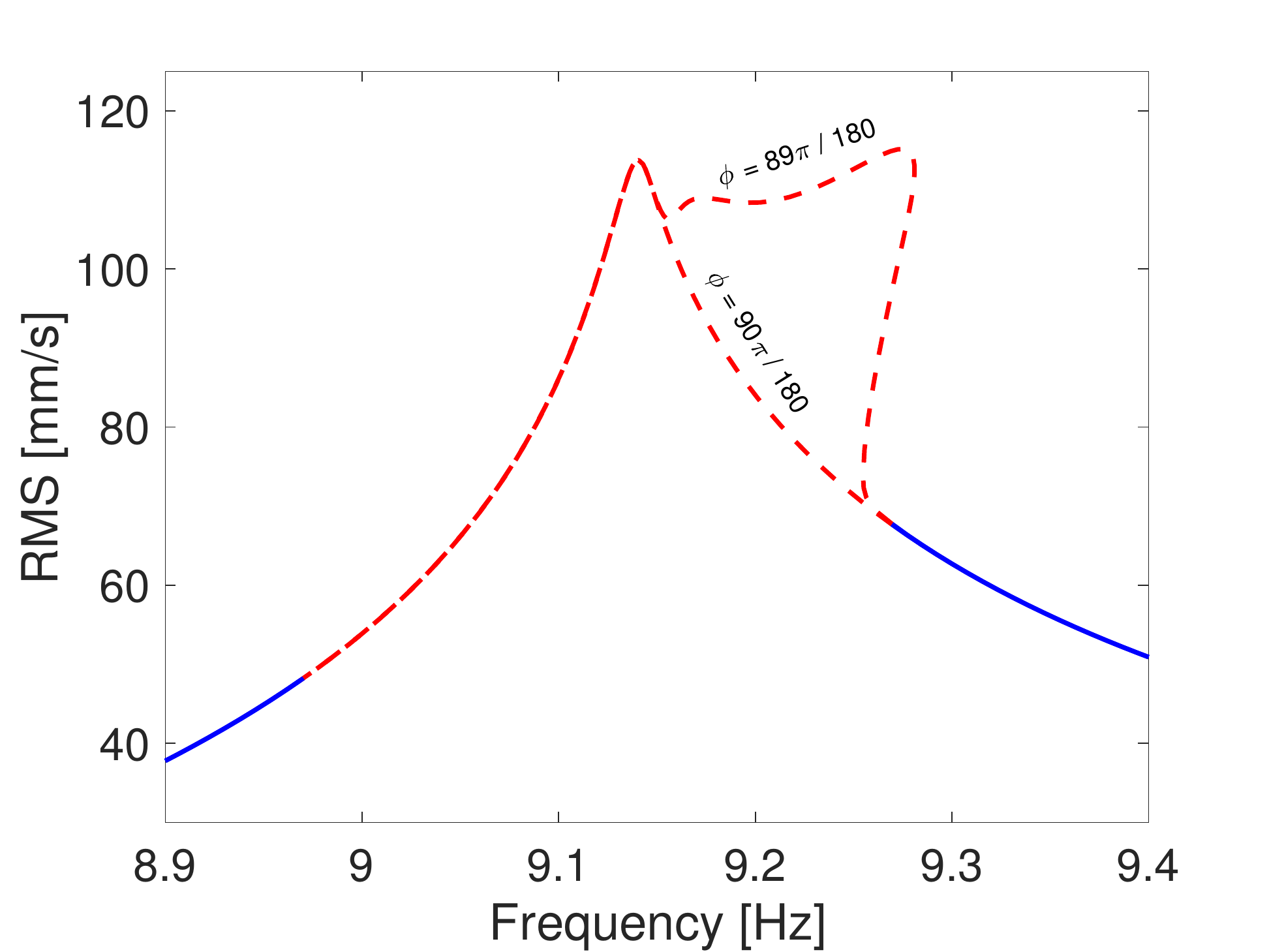} &
\subfigimg[width=\linewidth]{\bf (b)}{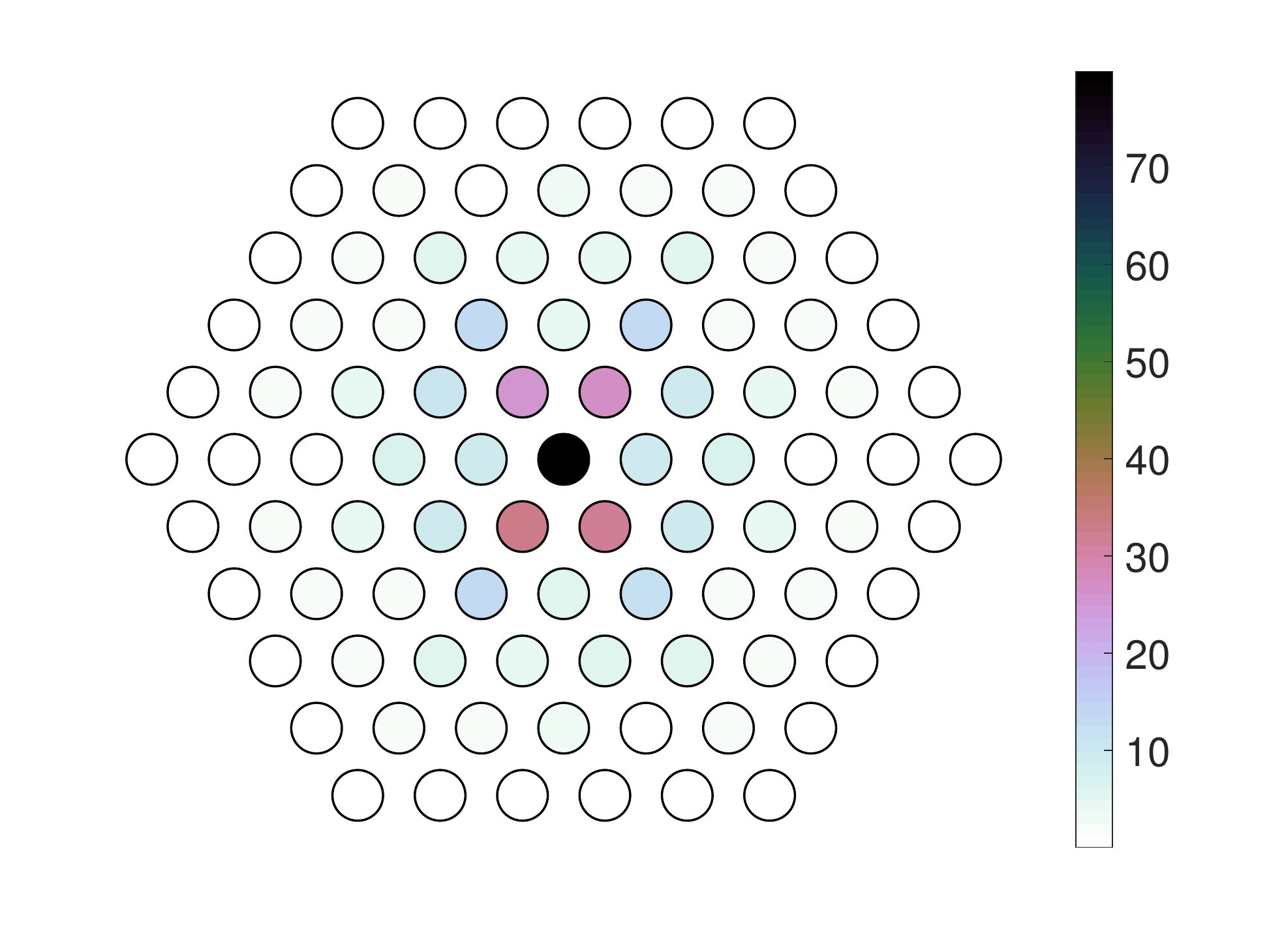} &
\subfigimg[width=\linewidth]{\bf (c)}{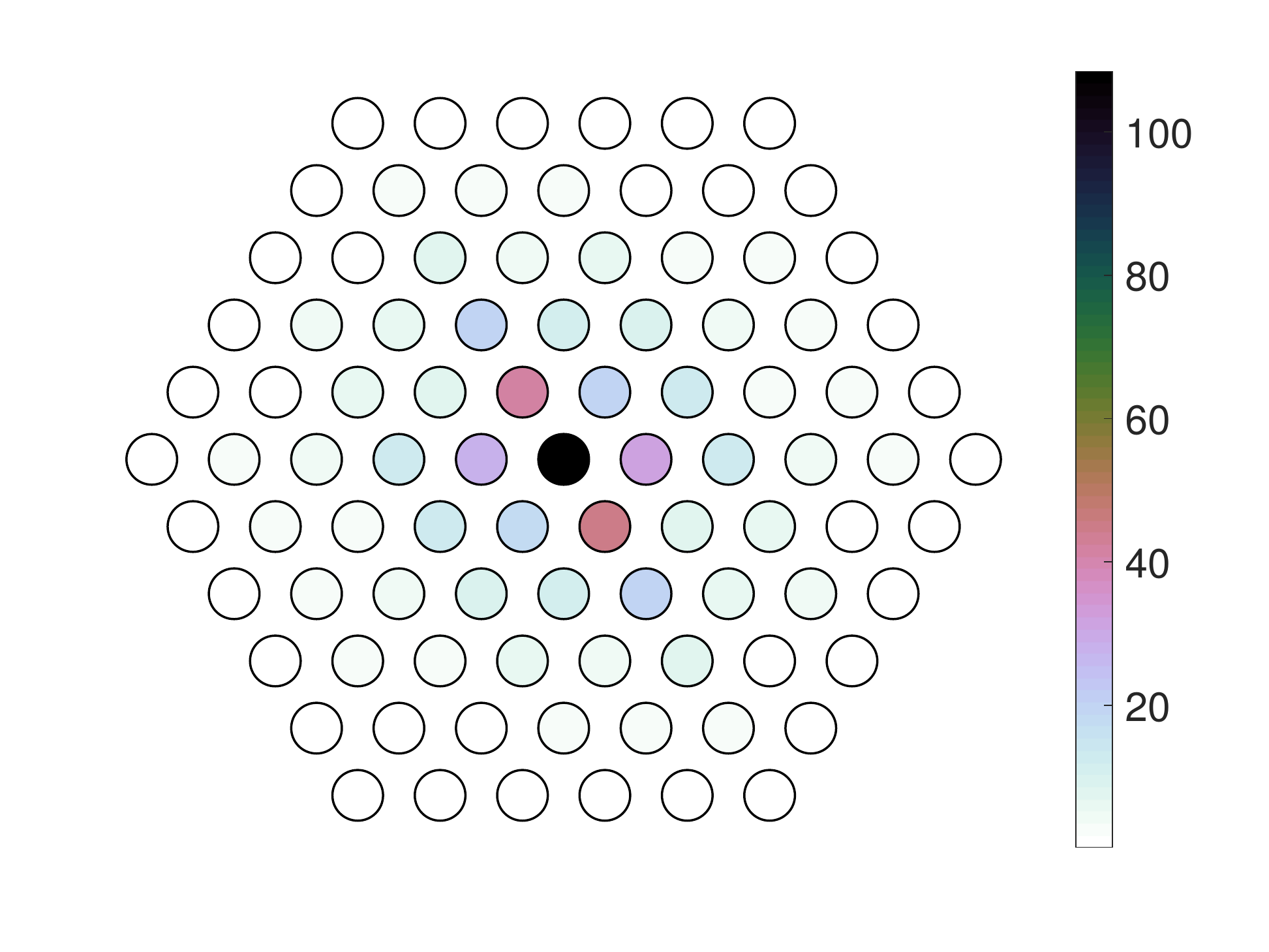} 
  \end{tabular}
 \caption{ \textbf{(a)} Frequency continuation with an excitation amplitude of $a=3$ mV 
 %(for which $F_{\mathrm{wire}} \approx 3.7242 \cdot10^{-7 }$\,N)
 for two excitation angles, $\phi = 89\pi/180$ and $\phi = \pi/2$.
 %Note that o
 Our continuations for $\phi = 89\pi/180$ and $\phi = \pi/2$ are indistinguishable when we are
 %one is 
 outside the parameter region
 in which the $\phi = 89\pi/180$ continuation has an additional branch of NLM solutions.
 \textbf{(b)} Profile of the NLM for $f=9.2$ Hz that belongs to the lower branch (i.e., the smaller-amplitude NLM branch) of the $\phi = 89\pi/180$ continuation.
  \textbf{(b)} Profile of the NLM for $f=9.2$ Hz that belongs to the upper branch of the $\phi = 89\pi/180$ continuation.
}
 \label{fig:89}
\end{figure}

The bifurcation diagram that we obtain when we use the excitation angle $\phi = \pi/3$ is more representative of
% the other 
 ``typical'' angles than the special cases $\phi = \pi/2$ and $\phi = 0$. For example, even by decreasing
 the angle slightly from $\phi=\pi/2$ to $\phi=89 \pi / 180$, we observe the additional branch 
 in the frequency continuation [see Fig.~\ref{fig:89}(a)].
We show a plot of an NLM at frequency $f=9.2$ Hz 
that belongs to the main branch (i.e., the branch with smaller-amplitude NLMs) of the $\phi = 89 \pi / 180$ continuation in Fig.~\ref{fig:89}(b). It has a similar profile
to the NLM in Fig.~\ref{fig:breather}. We show a plot of an NLM from the additional (i.e., larger-amplitude) branch at the same frequency $f=9.2$ Hz in Fig.~\ref{fig:89}(c). The solutions along this branch
have secondary amplitudes in the $-\pi/3$ direction.

%{\bf map: "dominant secondary amplitudes": what does this mean? some clarification would be good. CC: see change. I simply meant "next largest peak"}

%%%%%

\subsection{Drive-Amplitude Sweeps}\label{sec:amp}

 \begin{figure}
    \begin{tabular}{@{\quad}p{0.33\linewidth}@{}p{0.33\linewidth}@{} p{0.33\linewidth}@{}  }
  %\hspace{.2cm}
\subfigimg[width=\linewidth]{\bf (a)}{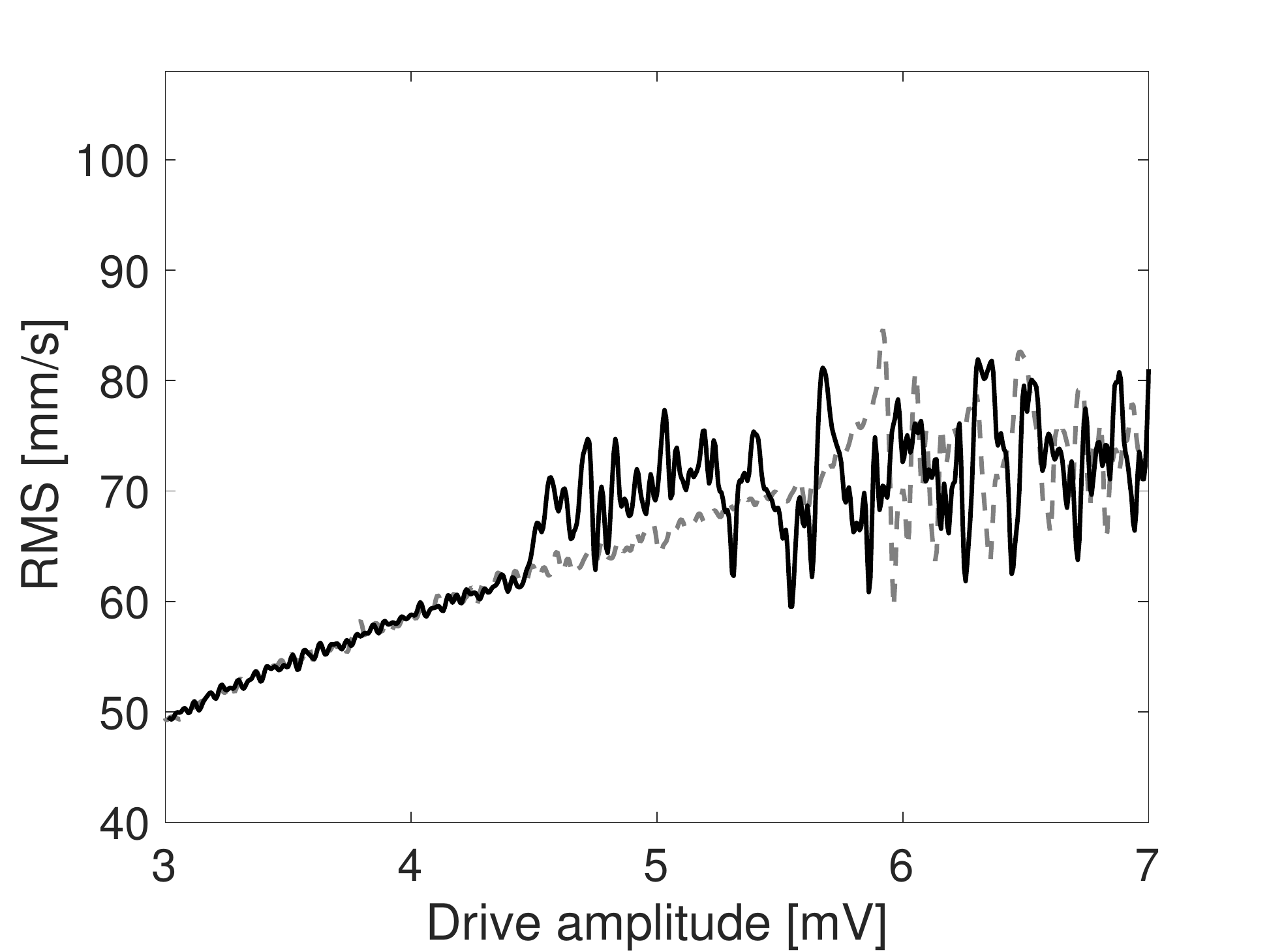} &
\subfigimg[width=\linewidth]{\bf (b)}{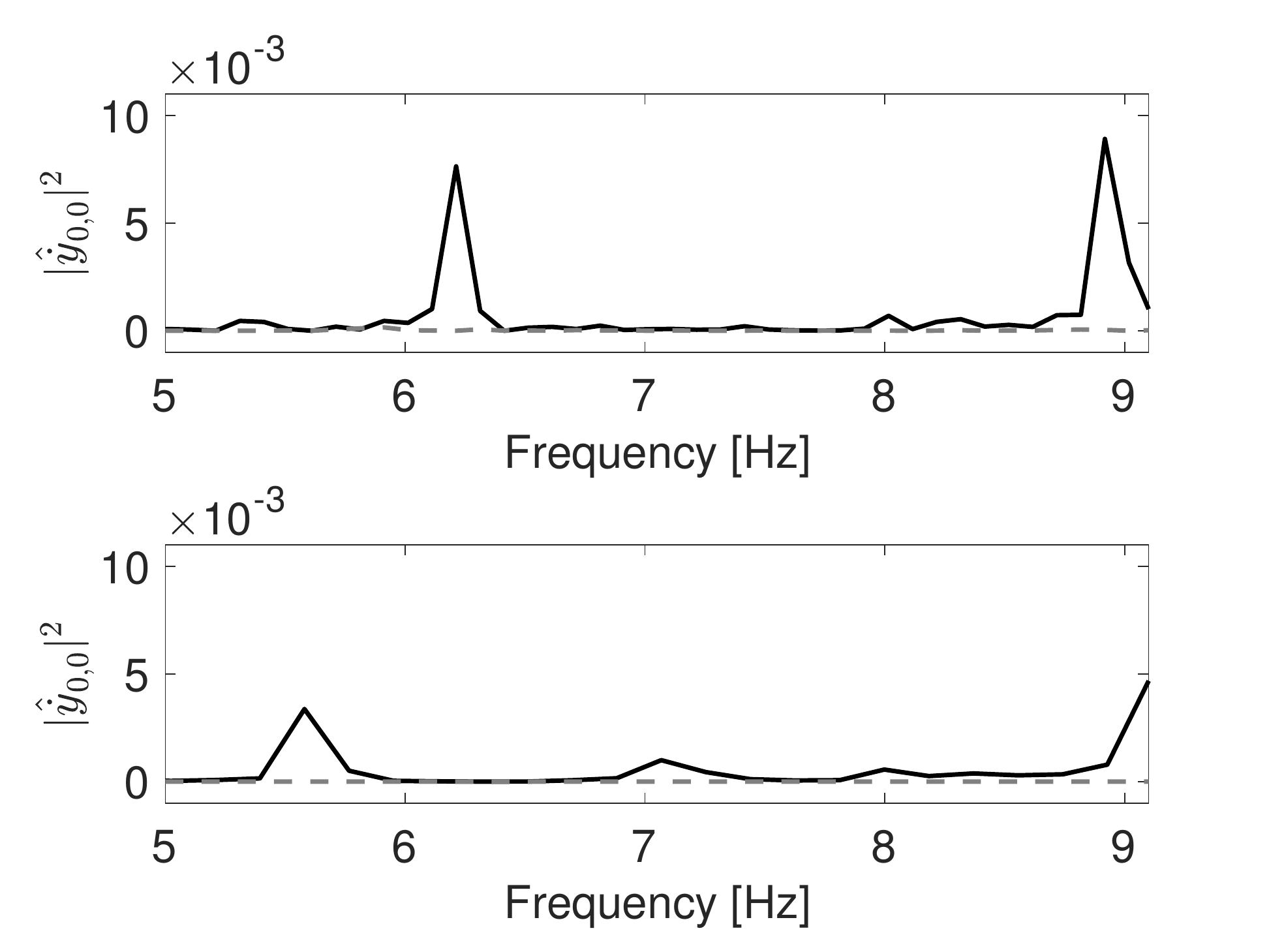} &
\subfigimg[width=\linewidth]{\bf (c)}{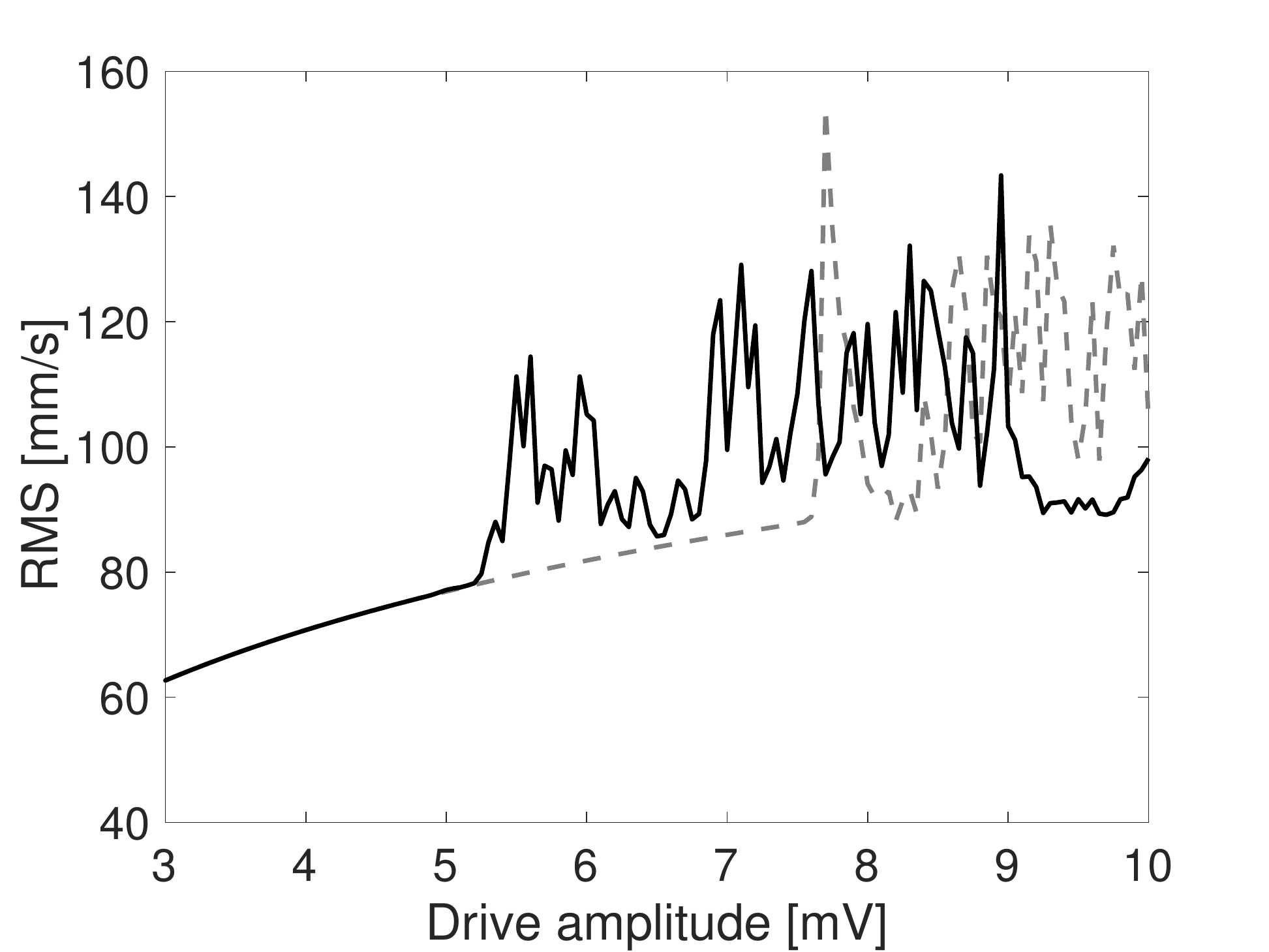} 
  \end{tabular}
 \caption{\textbf{(a)} Our upsweep (dashed gray curve) and downsweep (solid black curve) of the drive amplitude $a$ in our experiment.
 \textbf{(b)} The magnitude of the Fourier transform of the time
 series for the velocity of the center particle normalized by the
 height of the peak at $f \approx 9.3$ Hz
 for (top panel) our experiment and (bottom panel) numerical computation for an excitation amplitude $a=5.5$ mV. 
 %(for which $F_{\mathrm{wire}} \approx 6.8277 \cdot10^{-7 }$\,N). 
 The dashed gray curve is the small-amplitude state (in the form of an NLM) that we obtain from the upsweep, and the solid black curve is the large-amplitude state that we obtain from the downsweep.
 \textbf{(c)} The upsweep (dashed gray curve) and downsweep (solid black curve) of the 
 %NLM 
 amplitude in our numerical simulations.
  }
 \label{fig:dynamic_sweep}
\end{figure}

We now return to the effect of large-amplitude excitations for the parameter set --- namely, $\phi = \pi/2$ and $f \approx 9.3$ Hz --- in our laboratory experiments.
%al setup.  
Our bifurcation analysis 
%above 
revealed that the NLM solution at this parameter set destabilizes for larger amplitudes, although sometimes the instability is so weak that the NLMs are effectively stable on short enough time scales [see Fig.~\ref{fig:angle90}(b)]. We also observed
that a large-amplitude branch of NLM solutions emerges at the parameter values $\phi = \pi/2$ and $f \approx 9.3$ Hz for sufficiently large excitation amplitudes [see Fig.~\ref{fig:otherangle}(a)]. 

%{\bf map: the sentence above was a bit weird, as it literally said that large-amplitude solutions emerge for sufficiently large amplitudes; from the figure, it looks like we're referring to a solution branch, so I changed the phrasing to use the word "branch" --- although I still think it comes across as tautological at first, before one stops to think about what it means; so I wonder if there is further rephrasing that we ought to do?CC: rephrasing worked}

To study the dynamics at larger amplitudes in experiments, we initialize the system with a small-amplitude excitation (of $a=1$ mV)
%, for which $F_{\mathrm{wire}}\approx 1.2414 \cdot10^{-7 }$\,N) 
and gradually increase the amplitude in increments of $0.05$ mV.
% (which corresponds to a force increment of $F_{\mathrm{wire}} \approx 6.2070 \cdot10^{-9 }$\,N).
% {\bf map: what is the precise value of $\Delta a$ for these steps? this needs to be stated explicitly}
For each step, we run the system for 90 periods, which allows sufficient
time to settle to a steady state if there is one. 
%{\bf map: I noticed various comments about 'steady state' being a misnomer at some point, so I thought it would be relevant to put a flag here regarding the use of the term above. Panos was referring to something in the introduction, and not this}
We record the RMS
of the velocity of the center magnet
%particle 
for the final 5 seconds.
% PGK: this is again a little confusing. It seems to suggest that the
% 90 periods throughout this sweep correspond to 10s. But this doesn't
% seem to me to be very likely to be true.
We call this procedure an amplitude ``upsweep''. We use an analogous procedure when we start with a large excitation amplitude, 
%(of $a=15$ mV, for which $F_{\mathrm{wire}} \approx 1.8621 \cdot10^{-6 }$\,N),
 which we gradually decrease using the same increments. 
%{\bf map: is the increment value the same as for the upsweeps? presumably yes, but we should state this explicitly here}
We call this procedure an amplitude ``downsweep''. 
 We show our experimental results for the upsweep and downsweep in Fig.~\ref{fig:dynamic_sweep}(a). For sufficiently
 small amplitudes (specifically, for $a \lessapprox 4.5$ mV)
 %, for which $F_{\mathrm{wire}} \lessapprox 5.5863 \cdot10^{-7 }$\,N),
  both the upsweep and downsweep approach the same NLM, suggesting that there is a single stable branch of NLMs for $a \lessapprox 4.5$ mV. However, for $a \gtrapprox 4.5$ mV, there appear to be two different states; we obtain the small-amplitude states when we perform an upsweep 
 %approach 
 and the large-amplitude states when we perform a downsweep.
   The small-amplitude states have the form of an NLM. The experimental result in Fig.~\ref{fig:compare}(b) is an example
 of the small-amplitude state for $a=5.5$ mV. 
 %(for which $F_{\mathrm{wire}} \lessapprox 6.8277 \cdot10^{-7 }$\,N). 
 The large-amplitude states are also localized,
 but they are no longer periodic in time. An inspection of the Fourier transform
 of the large-amplitude state for $a=5.5$ mV reveals other peaks in the spectrum (in addition
 to a peak at the excitation frequency $f \approx 9.3$ Hz). In the top panel of Fig.~\ref{fig:dynamic_sweep}(b),
 we show the Fourier transform of both the large-amplitude state 
 %(solid black curve) 
 and the small-amplitude state. 
 %(dashed gray curve). 
 The small-amplitude state (i.e., the NLM) has no peaks for lower frequencies,
 whereas the large-amplitude state has peaks at approximately $f =
 8.9$ Hz and $f=6.2$ Hz; this is
 %being 
 suggestive of quasiperiodic behavior.
 %PGK: although BTW the 6.2 is 2/3 of the 9.3.
 %The higher-amplitude state thus resemble more a quasiperiodic state.

 We obtain qualitatively similar results when we perform analogous upsweeps and downsweeps in
 numerical computations. We also observe the emergence of two states in these computations [see
 Fig.~\ref{fig:dynamic_sweep}(c)]. In our computations,
 %these simulations, 
 the large-amplitude state departs from the branch of NLMs for excitation amplitudes that are slightly larger (specifically, for $a\gtrapprox5.1$ mV) than in our experiments. The amplitude $a \approx 5.1$ mV is roughly where the numerical NLM branch destabilizes. As in our experiments, these large-amplitude states
 %the solutions in the large-amplitude branch 
 are not periodic in time. One can also observe the presence of secondary peaks in their
  %states' 
  Fourier transforms 
  %can also be observed 
  in our numerical solutions, although the locations
 of these peaks are slightly different than in our experiments. See the bottom panel of Fig.~\ref{fig:dynamic_sweep}(b).
Although these numerical large-amplitude states have features that are similar to those of time-quasiperiodic states
(given the multiple incommensurate peaks in the Fourier transform), it is also possible 
that these large-amplitude states are weakly chaotic. One issue is that
we were unable to detect asymptotically stable time-quasiperiodic orbits for parameter values that correspond to the experiments. If such solutions were dynamic attractors, it would be straightforward to classify the large-amplitude states as time-quasiperiodic ones by plotting Poincar\'e sections of the orbits.

%{\bf map: what exactly do you mean by 'relevant' above?}

%%Solutions in the large-amplitude branch of Fig.~\ref{fig:dynamic_sweep}(a,c) feature transient
%%dynamics that result from the destabilization of the main branch of time-periodic
%%NLMs. 
%%%{\bf map: check rephrasing in the sentence above; I am not sure if I accidentally changed the meaning. CC: OK}
%%Inspection of the Fourier transforms of these solutions
%%suggests that they are influenced by nearby solutions that are temporally quasiperiodic or chaotic.
%%However, we are unable to tell with our present numerics whether the
%%large-amplitude states themselves share these properties.

%PGK: Chris I think one needs to say a little more clearly what it is
% that is oberved. Right now, it is not sufficiently clear. The
% expression "influenced by" is something that is not easy to
% understand -- it is not for me anyway. And I am not sure, what is
% the end observation. Is it quasiperiodicity, is it chaoticity? Is
% it that we cannot decide between the two because we see many
% frequencies? Could you please write sth clearer?

%With present results it is not possible
%to conclude what exactly the larger-amplitude state is.
%Additional numerics and experiments are needed.

\begin{figure}
   \begin{tabular}{@{}p{0.33\linewidth}@{}p{0.33\linewidth}@{} p{0.33\linewidth}@{}  }
%  %\hspace{.2cm}
  \subfigimg[width=\linewidth]{\bf (a)}{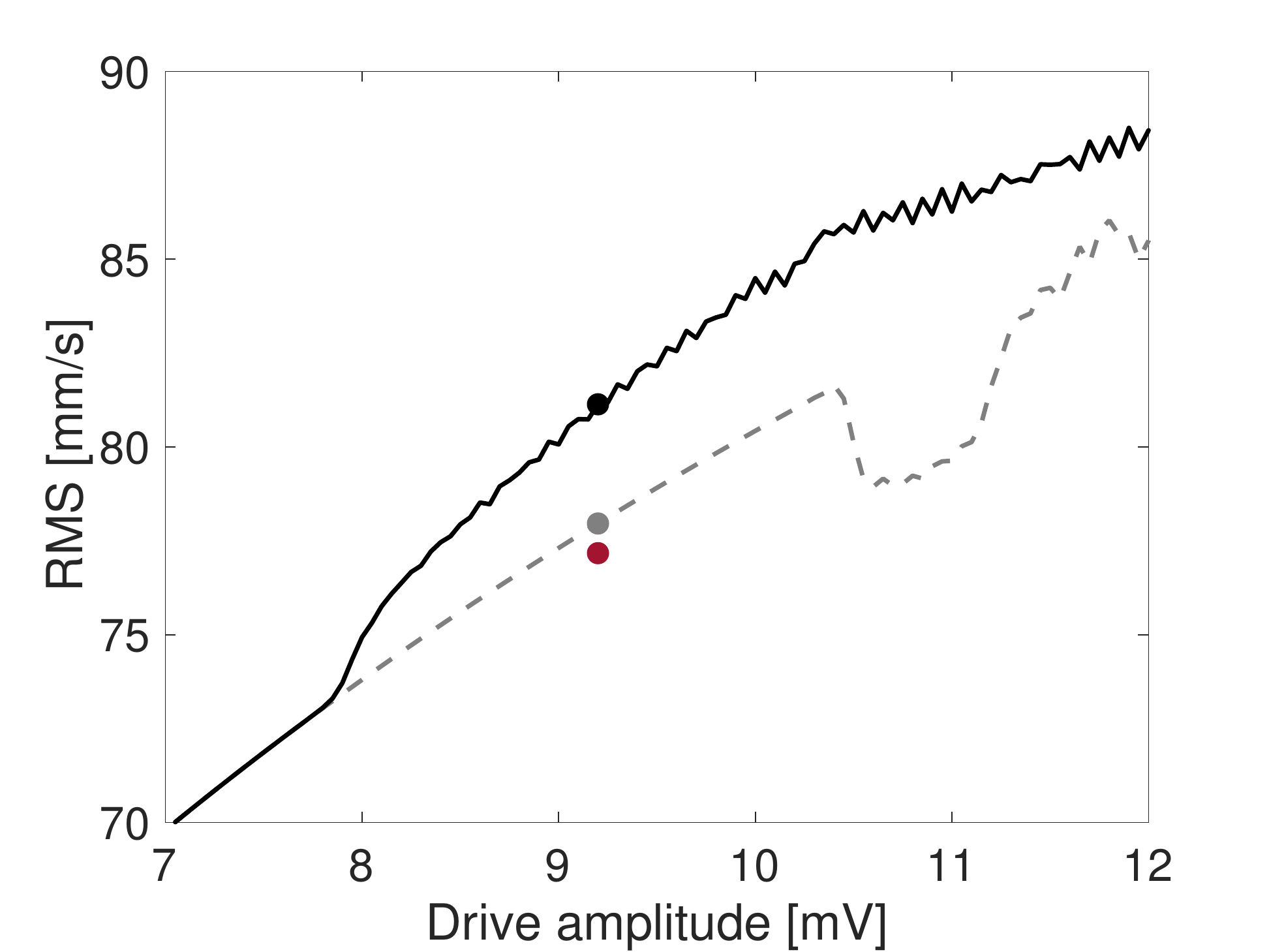} &
  \subfigimg[width=\linewidth]{\bf (b)}{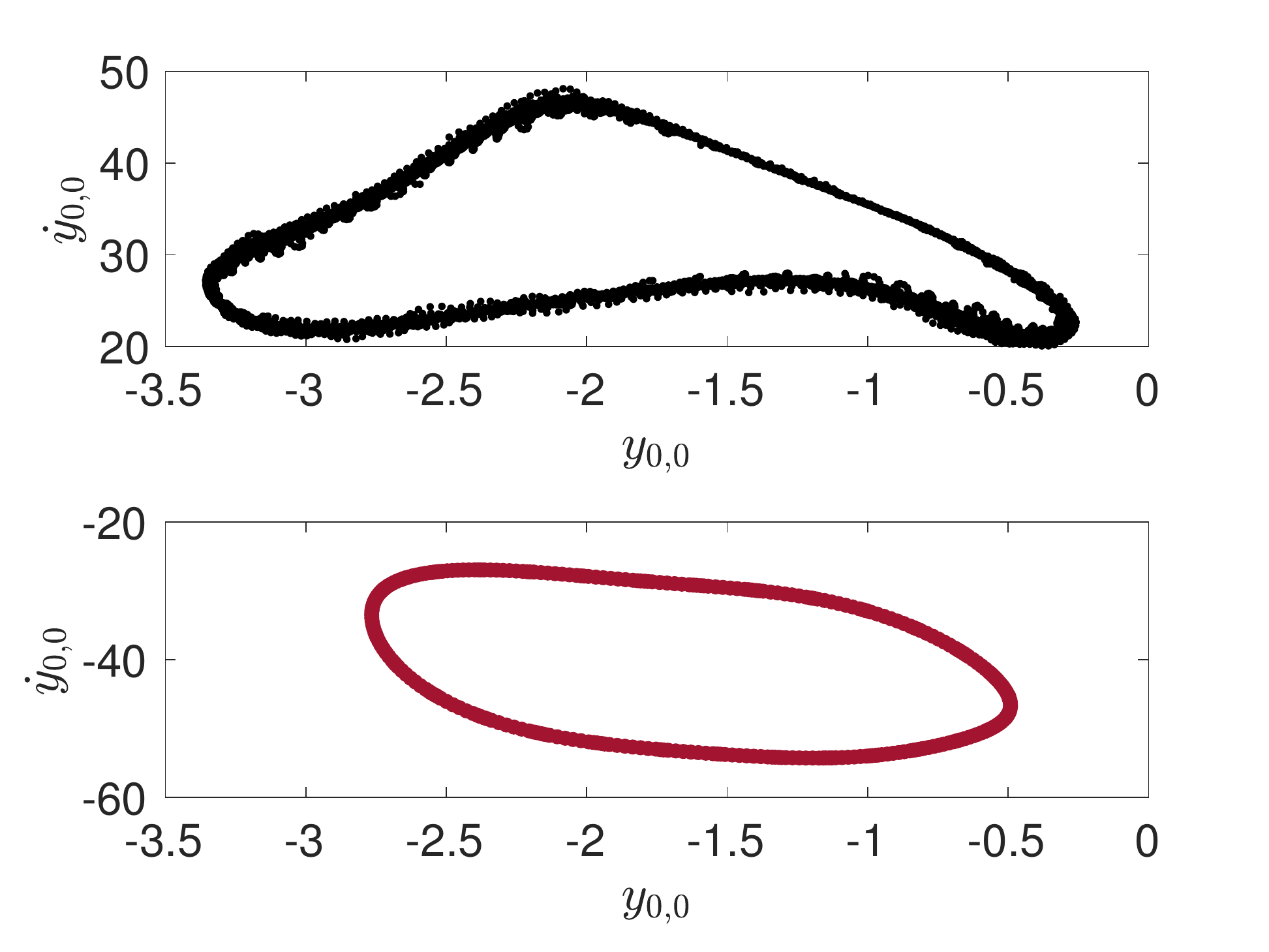} &
  \subfigimg[width=\linewidth]{\bf (c)}{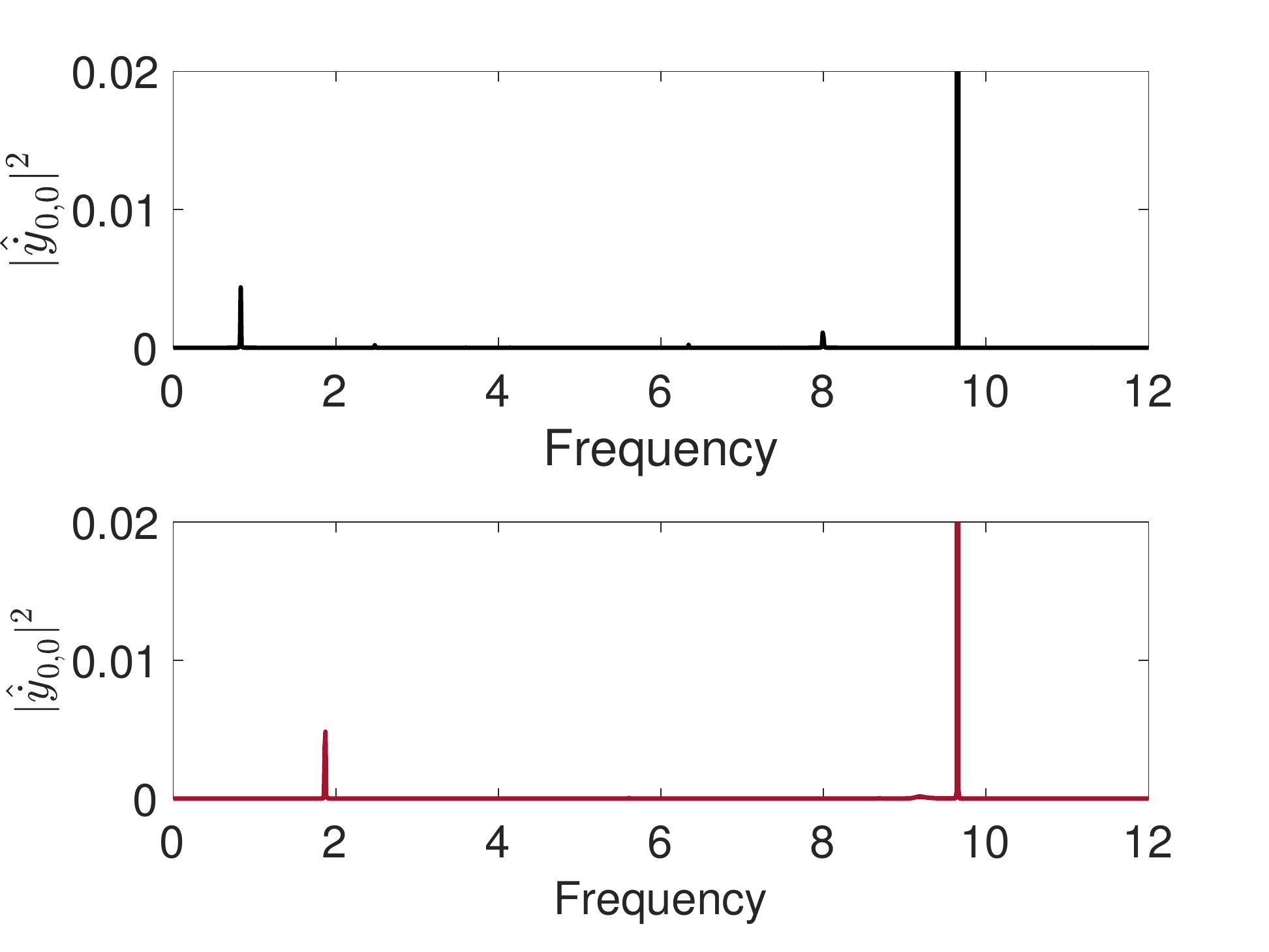} 
   \end{tabular}
  \begin{tabular}{@{\quad}p{0.33\linewidth}@{}p{0.33\linewidth}@{} p{0.33\linewidth}@{}  }
 \subfigimg[width=\linewidth]{\bf (d)}{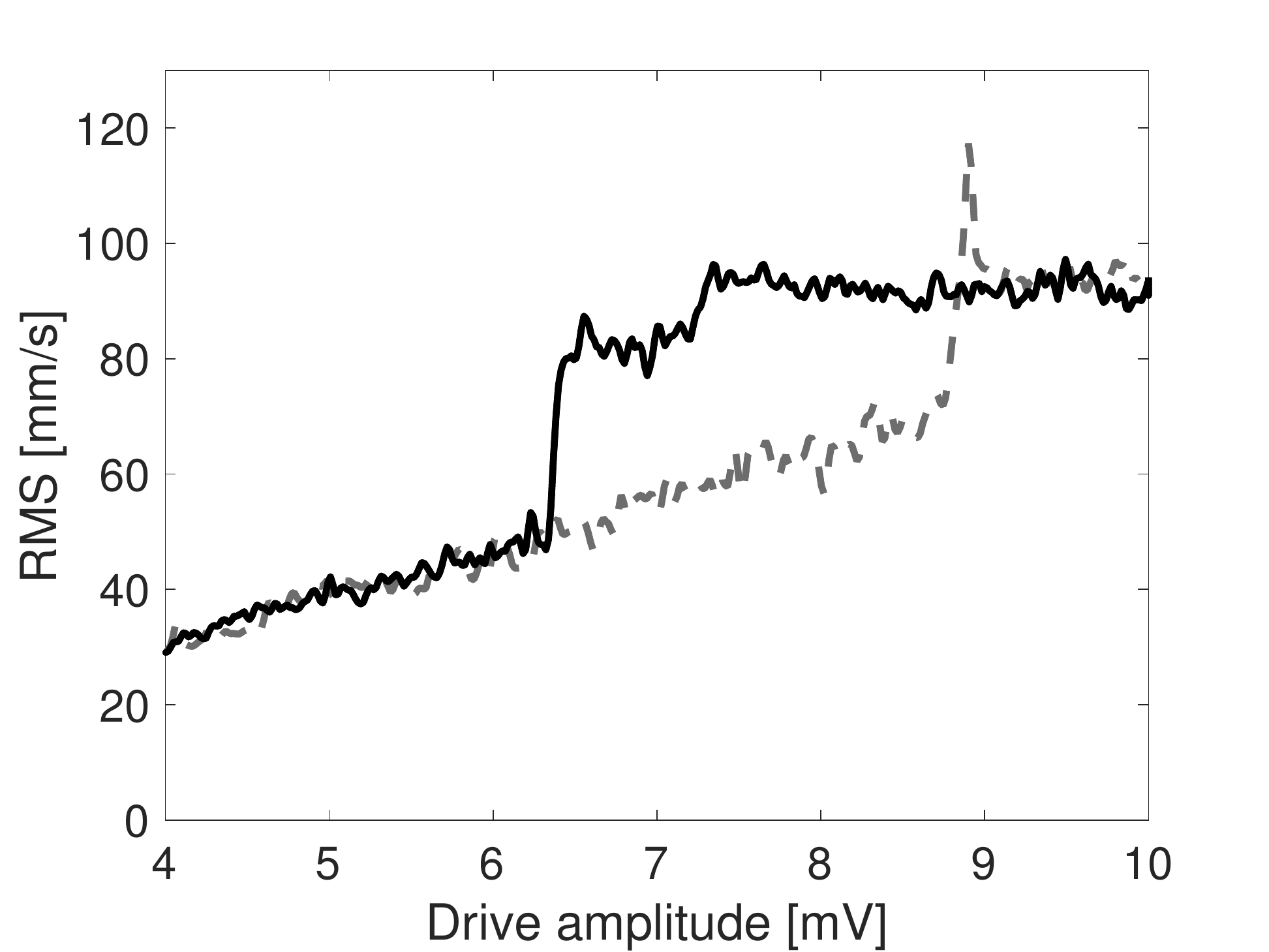} &
  \subfigimg[width=\linewidth]{\bf (e)}{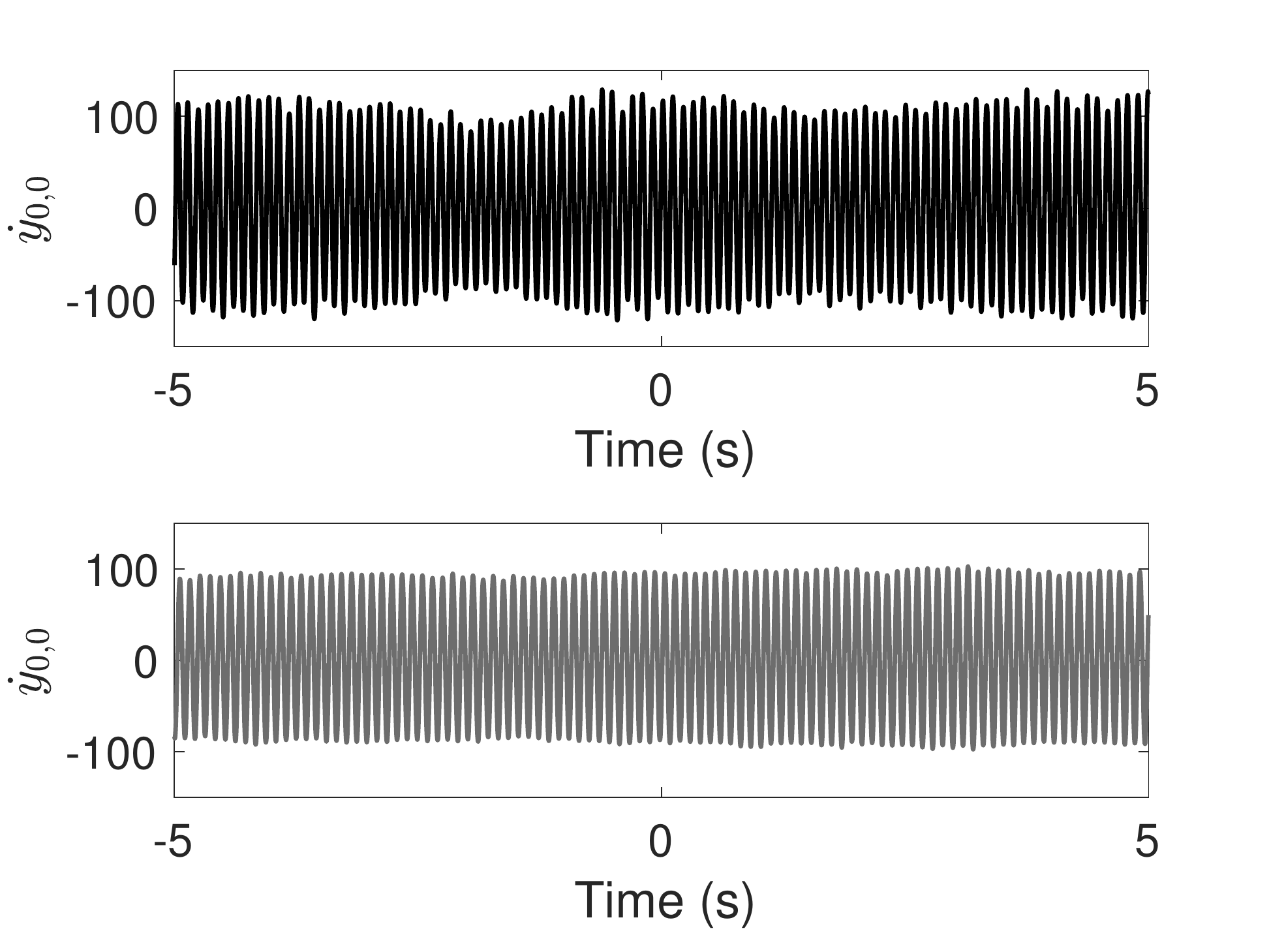} &
\subfigimg[width=\linewidth]{\bf (f)}{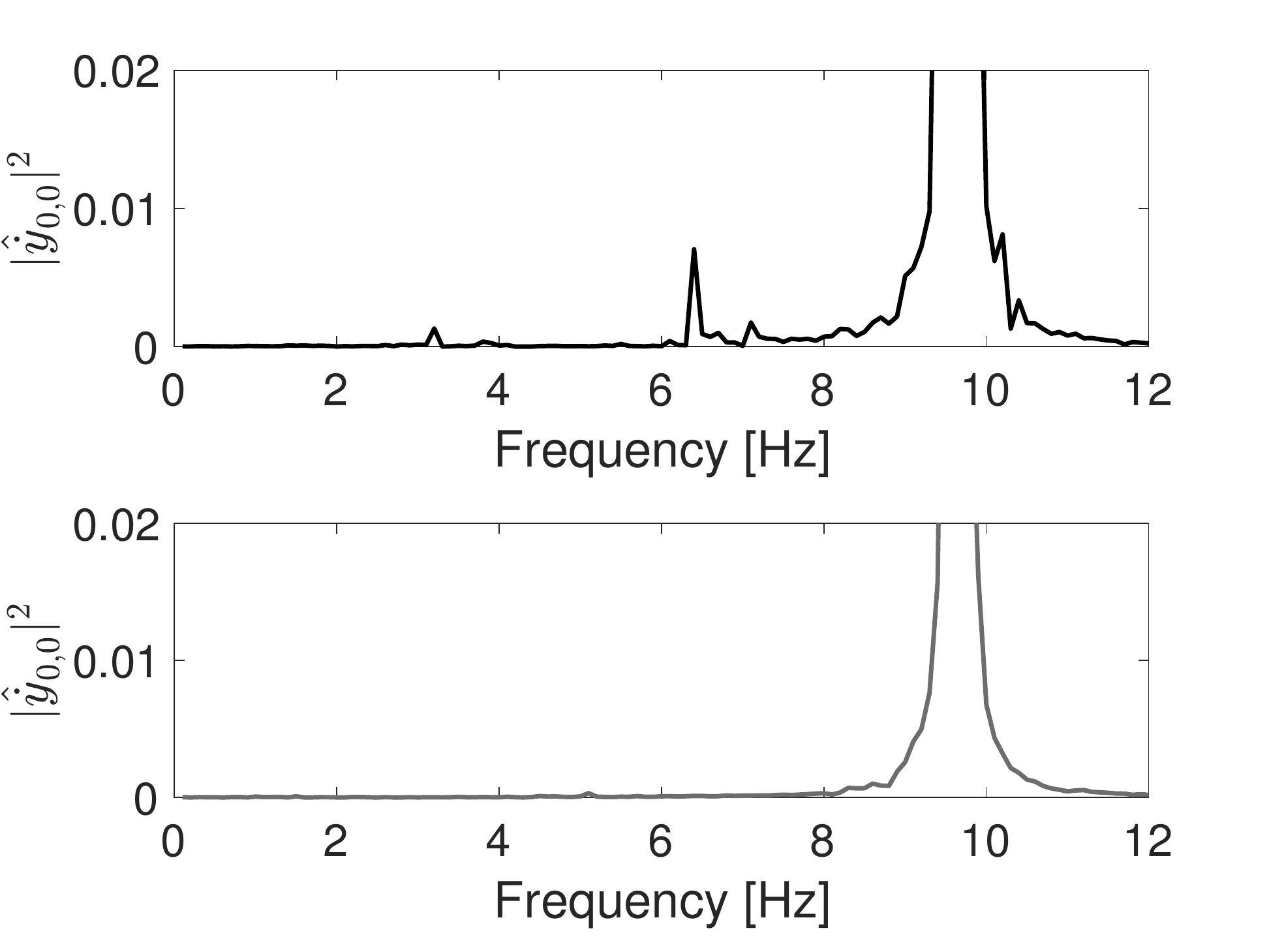} 
  \end{tabular}
\caption{\textbf{(a)} The upsweep (dashed gray curve) and downsweep (solid black curve) for our numerical simulations
 with $M_b=125$ g and $f=9.65$ Hz. Three states exist when we use an excitation amplitude of $a=9.2$ mV. Two of the states (the black and red dots) are not periodic in time, and the other (the gray dot) is time-periodic. We obtain the state that is indicated by the red dot by simulating zero initial data for 90 periods of motion. 
 %{\bf map: (1) what is "zero initial data" ? ; (2) how many periods of motion is "many"?}
 \textbf{(b)} In the top panel, we show a projection of the Poincar\'e section of the solution that is represented by the black dot in panel (a).
 %{\bf map: but sure this is a projection of a Poincar\'e section? need to be more precise in this description}
  In the bottom panel, we show a projection of the Poincar\'e section of the solution that is represented by the red dot in panel (a).
 \textbf{(c)} The magnitude of the Fourier transforms of the solutions in panel (b).
  \textbf{(d)} The upsweep (dashed gray curve) and downsweep (solid black curve) for our experiments.
  \textbf{(e)}  The gray curve is the experimental time series of the small-amplitude state that we obtain from the upsweep for an excitation amplitude of $a=8$ mV. The solid black curve is the time series of the large-amplitude state that we obtain from the downsweep for an excitation amplitude of $a=8$ mV.
 \textbf{(f)} The magnitude of the Fourier transform of the time series of panel (e) normalized by the height of the peak at $f=9.5$. The gray curve corresponds to the small-amplitude state, and the solid black curve corresponds to the large-amplitude state.
% {\bf map: several of the labels ("(a)" etc.) are too close to the associated pictures and should be moved a bit}
  }
 \label{fig:dynamic_sweep2}
\end{figure}

To clarify the nature of the large-amplitude states,
we modify the parameter values slightly to obtain
stable time-quasiperiodic solutions. We perform amplitude upsweeps and downsweeps for the parameter set $\phi = \pi/2$, $f=9.65$ Hz, and $M_b = 125$ g. 
We use a different drive frequency from our prior calculations, because the smaller mass $M_b = 125$ g leads to a cutoff frequency of $f_c \approx 9.01$ Hz and a defect frequency of $f_d \approx 9.36$ Hz. The amplitude sweeps with these
%modified 
parameter values lead to a well-defined large-amplitude branch of solutions that bifurcates from the main branch of periodic ones (the NLMs) [see Fig.~\ref{fig:dynamic_sweep2}(a)]. 
%{\bf map: I don't know what "more well-defined" means; please rephrase. CC: done}
In Fig.~\ref{fig:dynamic_sweep2}(a), we show three
solid markers to point out the locations of three solutions: a large-amplitude state that appears to be either time-quasiperiodic or time-chaotic in black,
the NLM (in gray), and a stable time-quasiperiodic orbit (in red).
%The latter does not appear to be approached in the dynamics of either
%the upsweep or the downsweep.
%PGK: is that right? CC: note quite...
We show a plot of a projection of the Poincar\'e section in the $(y,\dot{y})$
plane 
%(i.e. plotting $(y(nT),\dot{y}(nT))$ where $T$ is the period of the drive and $n = 1,2,3,\ldots$)
in Fig.~\ref{fig:dynamic_sweep2}(b) for the two states that are not time-periodic.
 %{\bf map: is $y$ the same notation as earlier in the paper?}
The orbit in the bottom panel reveals a well-defined invariant curve, illustrating the quasiperiodic nature of the solution.
 %The orbit in the top panel appears to be a weakly-unstable quasiperiodic, since the curve is less-defined. {\bf map: this sentence claims things (e.g., 'weakly unstable') that we are not scientifically justified to claim}
We show the Fourier transforms of these two non-periodic states in
Fig.~\ref{fig:dynamic_sweep2}(c). Both have a secondary peak
in the spectrum, demonstrating that the solutions are indeed non-periodic in time (because of the incommensurate peaks in the frequency spectra). 
Laboratory experiments for this modified parameter set
yield similar results. In particular, there is a well-defined large-amplitude branch of solutions 
that bifurcate from a branch of time-periodic solutions [see
Fig.~\ref{fig:dynamic_sweep2}(d,e)]. The Fourier transform of one of the large-amplitude
states also has a secondary peak
in the spectrum, an indication that the state is 
%(nearly)
nearly quasiperiodic. For simplicity, we henceforth call them simply ``quasiperiodic''.

Because the numerical computations and experiments with the modified mass $M_b = 125$ g
reveal the existence of time-quasiperiodic orbits that bifurcate from the main branch of time-periodic
NLMs, it is reasonable to conclude that these
quasiperiodic solutions persist when we continue the parameters
to the original parameter set $M_b = 138.2$ g and $f_d=9.3$.
This suggests that the large-amplitude branch in Fig.~\ref{fig:dynamic_sweep}(a,c)
consists of time-quasiperiodic solutions.

%{\bf map: I am confused by the above sentence: (1) I'm not sure how to parse it; and (2) I am not really following what claim is being made; this sentence needs to be changed; I am lost here. CC: see re-wording}

 %%%%%%

\section{Conclusions} \label{sec:theend}

We have demonstrated, both experimentally and numerically, the existence of nonlinear localized modes in a 2D hexagonal lattice
of repelling magnets. By exploring the effects of nonlinearity numerically using frequency continuation and experimentally
using amplitude sweeps, we revealed the emergence of both time-periodic NLMs and time-quasiperiodic localized states.
% not only the time-periodic NLMs but also of quasiperiodic ones.
%{\bf map: part of my rephrasing above is because the above definition of NLMs directly contradicted the one from the introduction; that said, I would prefer to change the definition in the intro to allow time-quasiperiodicity and then change all appropriate language throughout the paper. CC: I would leave the definition of breather as is}
We have also established that our experimental setup is a viable approach for fundamental studies in nonlinear lattice
systems that go beyond what can occur in 1D chains. We found that the smaller-amplitude NLMs that we considered are stable, whereas progressively larger excitation amplitudes led to instabilities and more complicated dynamics,
including time-quasiperiodic and potentially time-chaotic behavior. We also explored the anisotropy of the
hexagonal lattice by considering different excitation angles and
examining the nature and decay of the states
%modes 
along these angles.

Our work paves the way for many future studies. For example, although our parameter continuation in frequency revealed several families of solutions,
%bifurcations, 
there are undoubtedly --- given the complexity of the studied system ---
several other ones (including possibly exotic ones) to discover.
%In our paper, we provided heuristic evidence for the existence of time-quasiperiodic localized states. 
%One can demonstrate more definitively that time-quasiperiodic orbits exist by implementing a numerical method
%for the computation of quasiperiodic states (including unstable ones).
%{\bf map: the above is so generic that I think we weaken the conclusion section substantially if we keep it}
Other avenues for future work include the study of refined models --- such as ones
that account for nonlinear damping (or, more generally, a more elaborate
form of damping~\cite{rosas2007,vergara,khatriprl}), rotational effects (which
are often considered to be important~\cite{merkel,yuli}), and/or long-range
interactions~\cite{magBreathers} --- of our lattice system.
Each of these aspects will add elements of complexity, but they also may lead to
other types of interesting dynamics, such as the possibility of
breather solutions with algebraically decaying tails of waves in space.
%long-range systems. 
It is also certainly possible that the inclusion of
rotational effects and/or more
sophisticated damping models 
%concerning the damping  
may help improve matches with laboratory
experiment. Such models have an associated cost of being more complicated and hence more cumbersome to analyze and simulate.
% come at the cost of leading to a more
%complicated (and progressively more cumbersome to analyze/simulate)
%model. 
Our attempt in the present paper has been to explore the principal features of the
interplay of discreteness, local disorder, and nonlinearity in a hexagonal lattice of magnets.
%the framework of this
%setting
%and to illustrate its availability and tractability as a useful tool for this type of studies.
%{\bf map: I don't understand the last line above (e.g., what tool do you have in mind? I'm lost here), and the simplest modification seemed to be to commented it out, as I don't see a point that needs to be made in this case}
Breathers in heterogenous hexagonal magnetic lattices (e.g., ones with a repeating pattern of two masses)
may lead to the existence of intrinsic localized modes and are also
worthy of future study. In that context, the study of band gaps, instabilities, and
% as well as
nonlinear modes and their propagation 
%therein 
is another topic of
substantial ongoing interest.
%~\cite{granularBook}. 
%, as are dark breathers in homogenous lattices.
%{\bf map: I removed the last one above, as we don't say what dark breathers are, so I feel that we're not providing context and going into laundry-list mode}

%%%%%

\section*{Acknowledgements}

The present paper is based on work that was supported by the US National Science
Foundation under Grant Nos. DMS-1615037 (CC), DMS-1809074
(PGK), and EFRI-1741565 (CD). AJM acknowledges support
from the Agencia Nacional de Investigaci\'on y Desarrollo de Chile
(ANID) under Grant No. 3190906.
EGC thanks Bowdoin College for the kind hospitality 
where the initial stages of this work were carried out.
%This material is based upon work supported by the US National Science
%Foundation under Grant DMS-1809074
%(PGK). 
PGK also acknowledges support from the Leverhulme Trust via a
Visiting Fellowship and thanks the Mathematical Institute of the University
of Oxford for its hospitality during part of this work. 
%Y.W. and C.D. acknowledge the support from the National Science Foundation under 
We give special thanks to Bowdoin undergraduates Ariel
Gonzales, Patrycja Pekala, Anam Shah, and Steven Xu
for help with simulations and
data management. 

%{\bf map: (1) other funding sources to thank? (2) any people to thank?}

%{\bf map: MAP has no funding sources to thank}

%%%%

\newpage

\section*{Appendix}

\begin{figure}[h]
  
      \centering
   \begin{tabular}{@{}p{0.25\linewidth}@{}p{0.35\linewidth}@{}  }
  \rlap{\hspace*{5pt}\raisebox{\dimexpr\ht1-.1\baselineskip}{\bf (a)}}
\includegraphics[width=.9\linewidth]{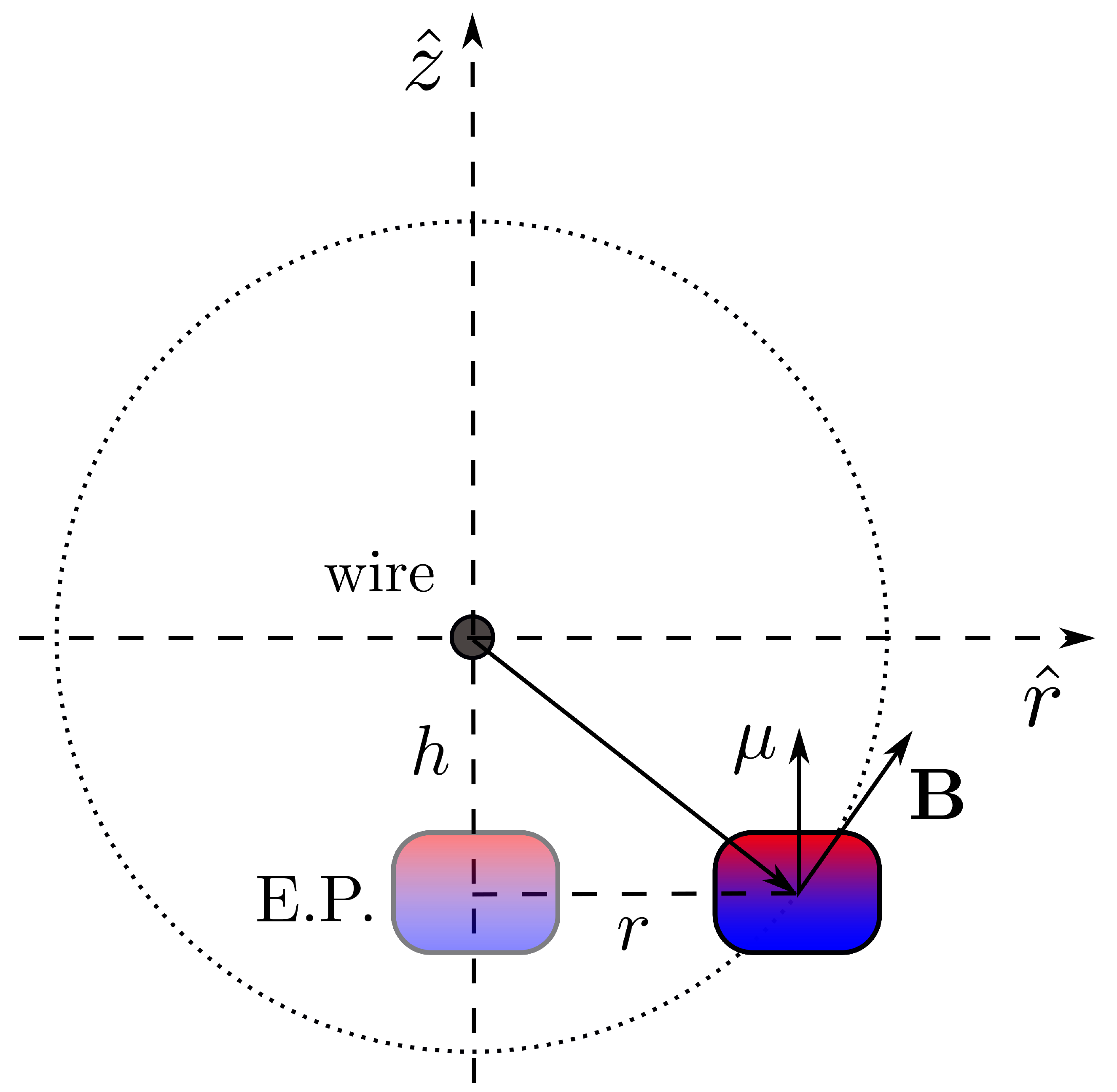} &
  \rlap{\hspace*{5pt}\raisebox{\dimexpr\ht1-.1\baselineskip}{\bf (b)}}
\includegraphics[width= .9\linewidth]{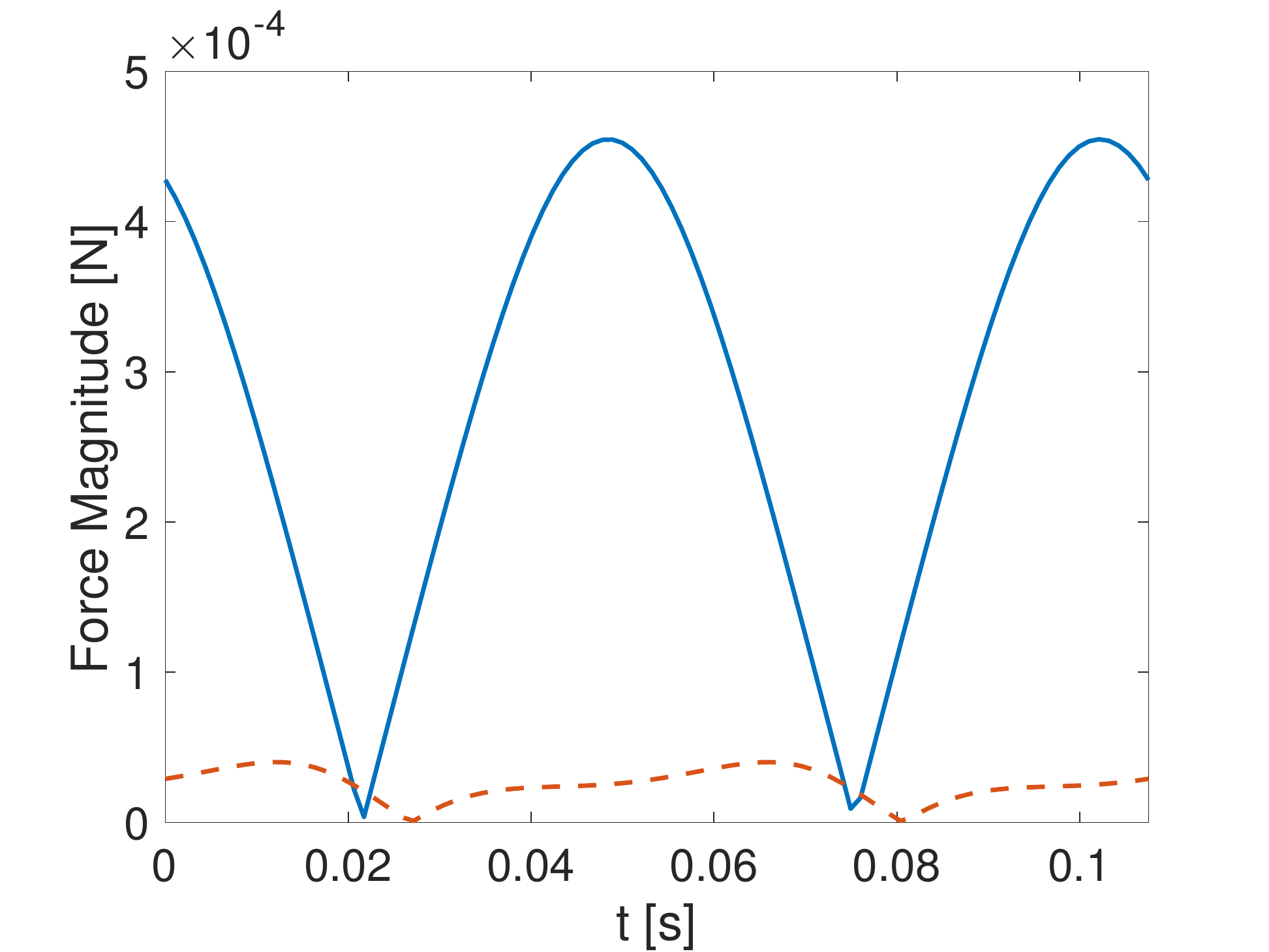} 
  \end{tabular}
\caption{\textbf{(a)} Schematic of the interaction between a single magnet and a wire in the $(\hat{r},\hat{z})$ plane, which is orthogonal to the direction of the wire. We use ``E.P.'' to denote the equilibrium position of the magnet.
\textbf{(b)} Squared magnitudes of the forces that result from lattice interactions (solid blue curve) and the external drive from the wire  
(dashed red curve) for the NLM from Fig.~\ref{fig:breather}(a) during one period of motion. Specifically, we plot
 $|\mathbf{F}^{\mathrm{lattice}}_{0,0}(t)|^2$ using the solid blue curve and $|\mathbf{F}^{\mathrm{ext}}_{0,0}(t)|^2$ using the dashed red curve.
}
 \label{fig:wirepics}
\end{figure}

%\subsection{Derivation of external force from wire}
To derive the external force that the wire exerts on a magnet, we
first define our coordinate system. We choose a set of orthogonal unit
vectors $\hat{r}$, $\hat{z}$, and $\hat{s}$ that are centered on the
wire and oriented such that the wire is aligned with the $\hat{s}$ axis [see Fig.~\ref{fig:wirepics}(a)]. 
%Moreover, 
%A standard way to 
We model the magnetic moment $\mu$ of the magnet using
%to use 
the Gilbert model of a magnetic dipole~\cite{jackson_classical_1999}, 
\begin{equation}
    {\bf F}_{\mathrm{wire}} = \left(\mu\cdot \nabla \right){\bf B}: 
\label{bforce}
\end{equation}
which describes the force that acts on the dipole
%over the dipole $\mu$ 
due to the magnetic field ${\bf B}$. In our setup, the wire carries an electric current $I$ that generates the magnetic field ${\bf B}(r,z) =
B_r(r,z)\hat{r}+B_z(r,z)\hat{z}$. Evaluating ${\bf B}$ at the position $-h\hat{z}+r\hat{r}$ of the magnet yields
\begin{equation}
	{\bf B}(r,-h) = \frac{I\mu_0}{2\pi\sqrt{h^2+r^2}}\left(\sin\theta \,\hat{r} + \cos\theta \,\hat{z} \right)\,,
\label{bfield}
\end{equation}
where $\mu_0$ is the magnetic permeability and $\theta$ is the angle between ${\bf B}$ and $\mu$.
%PGK: I do worry about the accuracy of this a little too. Should we
% really be thinking of this as a loop or a solid cylinder?? Obviously
% this will not be changed now, but I am wondering why we didn't
% question this assumption earlier. If you have a response as to why
% the loop is a good approximation it would be useful to know.                                                            
We are interested only in the $\hat{r}$ component of the force, because we are assuming that we can neglect the dynamics along the $\hat{z}$ axis. 
Therefore, inserting Eq.~\eqref{bfield} into Eq.~\eqref{bforce} and taking
%considering 
$\mu =\mathcal{M}\hat{z}$, we obtain
\begin{align}
	{\bf F}_{\mathrm{wire}} = \mathcal{M}\,\partial_rB_z(r,z) \hat{r}
= \frac{I \mu_0 \mathcal{M} }{2 \pi}  \frac{h^2 - r^2}{(h^2 + r^2)^2}\hat{r}\,, 
\label{eq:wireappendix}
\end{align} 
which corresponds to Eq.~\eqref{eq:wire}.
In our coordinates, we place the wire in an
%a generic 
orientation in the plane that is spanned by the lattice basis vectors $e_1 = (1,0)$ and $e_2=(1/2,\sqrt{3}/2)$. However, it is straightforward to write Eq.~\eqref{eq:wireappendix} in coordinates in the $\{e_1, e_2\}$ basis by including a parameter $\phi$ that accounts for the excitation angle. For instance, for the central magnet, 
%particle, 
we may write
\begin{equation} \label{eq:wireappendix2}
	\mathbf{F}^{\mathrm{ext}}_{0,0} =  \ \  \frac{I \mu_0 \mathcal{M}}{2\pi} \begin{pmatrix} \cos(\phi)   \frac{h^2 - x_{0,0}^2}{(h^2 + x_{0,0}^2)^2}  \\ \sin(\phi)  \frac{h^2 - y_{0,0}^2}{(h^2 + y_{0,0}^2)^2}  \end{pmatrix}\,,
\end{equation}
which corresponds to Eq.~\eqref{eq:excite}.

Although the form of the force that describes the external drive is specific to our experimental setup, the dynamics
are dominated by the lattice forces. For example, in Fig.~\ref{fig:wirepics}(b), we compare the forces that result from the
wire (see Eq.~\eqref{eq:wireappendix2}) and the force from the lattice for the NLM of Fig.~\ref{fig:breather}(a). The lattice forces are
\begin{equation} \label{ext}
	\begin{array}{ll} \displaystyle
\mathbf{F}^{\mathrm{lattice}}_{m,n}= 
&- \mathbf{F}_0(\mathbf{q}_{m+1,n} - \mathbf{q}_{m,n}) 
 - \mathbf{F}_1(\mathbf{q}_{m,n+1} - \mathbf{q}_{m,n}) 
+\mathbf{F}_{-1}(\mathbf{q}_{m,n} - \mathbf{q}_{m-1,n+1}) 
 \\
& +\mathbf{F}_0(\mathbf{q}_{m,n} -\mathbf{q}_{m-1,n})
+\mathbf{F}_1( \mathbf{q}_{m,n} - \mathbf{q}_{m,n-1})
- \mathbf{F}_{-1}(\mathbf{q}_{m+1,n-1} -\mathbf{q}_{m,n})
	\end{array}
\end{equation}
%That being said, it should also be acknowledged 
However, it is also important to acknowledge that the effective
``defect'' that is produced by the force \eqref{ext}
%this force 
is responsible for the presence of
the corresponding linear defect frequency and hence for the associated
NLMs in the presence of nonlinearity.

\bibliography{ChongBibFeb2020-v2} 

\begin{thebibliography}{10}

\bibitem{moti}
F.~Lederer, G.~I. Stegeman, D.~N. Christodoulides, G.~Assanto, M.~Segev, and
  Y.~Silberberg, ``Discrete solitons in optics,'' {\em Phys. Rep.}, vol.~463,
  p.~1, 2008.

\bibitem{alex}
P.~Binder, D.~Abraimov, A.~V. Ustinov, S.~Flach, and Y.~Zolotaryuk,
  ``Observation of breathers in {Josephson} ladders,'' {\em Phys. Rev. Lett.},
  vol.~84, p.~745, 2000.

\bibitem{alex2}
E.~Tr\'{\i}as, J.~J. Mazo, and T.~P. Orlando, ``Discrete breathers in nonlinear
  lattices: Experimental detection in a {J}osephson array,'' {\em Phys. Rev.
  Lett.}, vol.~84, p.~741, 2000.

\bibitem{lars3}
L.~Q. English, M.~Sato, and A.~J. Sievers, ``Modulational instability of
  nonlinear spin waves in easy-axis antiferromagnetic chains. {II}. influence
  of sample shape on intrinsic localized modes and dynamic spin defects,'' {\em
  Phys. Rev. B}, vol.~67, p.~024403, 2003.

\bibitem{lars4}
U.~T. Schwarz, L.~Q. English, and A.~J. Sievers, ``Experimental generation and
  observation of intrinsic localized spin wave modes in an antiferromagnet,''
  {\em Phys. Rev. Lett.}, vol.~83, p.~223, 1999.

\bibitem{swanson}
B.~I. Swanson, J.~A. Brozik, S.~P. Love, G.~F. Strouse, A.~P. Shreve, A.~R.
  Bishop, W.-Z. Wang, and M.~I. Salkola, ``Observation of intrinsically
  localized modes in a discrete low-dimensional material,'' {\em Phys. Rev.
  Lett.}, vol.~82, p.~3288, 1999.

\bibitem{Peybi}
M.~Peyrard, ``Nonlinear dynamics and statistical physics of {DNA},'' {\em
  Nonlinearity}, vol.~17, p.~R1, 2004.

\bibitem{2Dreview2015}
J.~{Bajars}, J.~C. {Eilbeck}, and B.~{Leimkuhler}, {\em {Numerical Simulations
  of Nonlinear Modes in Mica: Past, Present and Future}}, vol.~221, p.~35.
\newblock Cham, Switzerland: Springer International Publishing, 2015.

\bibitem{Morsch}
O.~Morsch and M.~Oberthaler, ``Dynamics of {B}ose--{E}instein condensates in
  optical lattices,'' {\em Rev. Mod. Phys.}, vol.~78, p.~179, 2006.

\bibitem{Flach2007}
S.~Flach and A.~Gorbach, ``Discrete breathers: {A}dvances in theory and
  applications,'' {\em Phys. Rep.}, vol.~467, p.~1, 2008.

\bibitem{pgk:2011}
P.~G. Kevrekidis, ``Non-linear waves in lattices: {P}ast, present, future,''
  {\em IMA J. Appl. Math.}, vol.~76, pp.~389--423, 2011.

\bibitem{Dmitriev_2016}
S.~V. Dmitriev, E.~A. Korznikova, Y.~A. Baimova, and M.~G. Velarde, ``Discrete
  breathers in crystals,'' {\em Physics-Uspekhi}, vol.~59, pp.~446--461, may
  2016.

\bibitem{FPU55}
E.~Fermi, J.~Pasta, and S.~Ulam, ``{Studies of Nonlinear Problems. I.},'' {\em
  (Los Alamos National Laboratory, Los Alamos, NM, USA)}, vol.~Tech. Rep.,
  pp.~LA--1940, 1955.

\bibitem{FPUreview}
G.~Gallavotti, {\em The Fermi--Pasta--Ulam Problem: A Status Report}.
\newblock Heidelberg, Germany: Springer-Verlag, 2008.

\bibitem{Nester2001}
V.~F. Nesterenko, {\em Dynamics of Heterogeneous Materials}.
\newblock Heidelberg, Germany: Springer-Verlag, 2001.

\bibitem{granularBook}
C.~Chong and P.~G. Kevrekidis, {\em Coherent Structures in Granular Crystals:
  From Experiment and Modelling to Computation and Mathematical Analysis}.
\newblock Heidelberg, Germany: Springer-Verlag, 2018.

\bibitem{yuli_book}
Y.~Starosvetsky, K.~R. Jayaprakash, M.~A. Hasan, and A.~F. Vakakis, {\em Topics
  On The Nonlinear Dynamics And Acoustics Of Ordered Granular Media}.
\newblock World Scientific, Singapore, 2017.

\bibitem{gc_review}
C.~Chong, M.~A. Porter, P.~G. Kevrekidis, and C.~Daraio, ``Nonlinear coherent
  structures in granular crystals,'' {\em J. Phys. Cond. Matt.}, vol.~29,
  p.~413003, 2017.

\bibitem{moleron}
M.~Moler\'{o}n, A.~Leonard, and C.~Daraio, ``Solitary waves in a chain of
  repelling magnets,'' {\em J. App. Phys.}, vol.~115, no.~18, p.~184901, 2014.

\bibitem{Mehrem2017}
A.~Mehrem, N.~Jim\'enez, L.~J. Salmer\'on-Contreras, X.~Garc\'{\i}a-Andr\'es,
  L.~M. Garc\'{\i}a-Raffi, R.~Pic\'o, and V.~J. S\'anchez-Morcillo, ``Nonlinear
  dispersive waves in repulsive lattices,'' {\em Phys. Rev. E}, vol.~96,
  p.~012208, 2017.

\bibitem{Marc2017}
M.~Serra-Garcia, M.~Moler\'on, and C.~Daraio, ``Tunable, synchronized frequency
  down-conversion in magnetic lattices with defects,'' {\em Phil. Trans. Roy.
  Soc. A: Math. Phys. Eng. Sci.}, vol.~376, no.~2127, p.~20170137, 2018.

\bibitem{Marin2000}
J.~L. {Mar{\'\i}n}, J.~C. {Eilbeck}, and F.~M. {Russell}, {\em {2-D Breathers
  and Applications}}, vol.~542, p.~293.
\newblock Heidelberg, Germany: Springer-Verlag, 2000.

\bibitem{English2013}
L.~Q. English, F.~Palmero, J.~F. Stormes, J.~Cuevas, R.~Carretero-Gonz\'alez,
  and P.~G. Kevrekidis, ``Nonlinear localized modes in two-dimensional
  electrical lattices,'' {\em Phys. Rev. E}, vol.~88, p.~022912, 2013.

\bibitem{DustyPlasma}
S.~Vladimirov and K.~Ostrikov, ``Dynamic self-organization phenomena in complex
  ionized gas systems: {N}ew paradigms and technological aspects,'' {\em Phys.
  Rep.}, vol.~393, no.~3, pp.~175--380, 2004.

\bibitem{Koukouloyannis2010}
V.~Koukouloyannis, P.~G. Kevrekidis, K.~J.~H. Law, I.~Kourakis, and D.~J.
  Frantzeskakis, ``Existence and stability of multisite breathers in honeycomb
  and hexagonal lattices,'' {\em J. Phys. A: Math. Theor.}, vol.~43, no.~23,
  p.~235101, 2010.

\bibitem{Wattis_2Dreview}
J.~A.~D. {Wattis}, {\em {Asymptotic Approximation of Discrete Breather Modes in
  Two-Dimensional Lattices}}, vol.~221, p.~179.
\newblock Heidelberg, Germany: Springer-Verlag, 2015.

\bibitem{Flach2D_1997}
S.~Flach, K.~Kladko, and S.~Takeno, ``Acoustic breathers in two-dimensional
  lattices,'' {\em Phys. Rev. Lett.}, vol.~79, pp.~4838--4841, 1997.

\bibitem{Marin98}
J.~L. Mar\'{i}n, J.~C. Eilbeck, and F.~M. Russell, ``Localized moving breathers
  in a {2D} hexagonal lattice,'' {\em Phys. Lett. A}, vol.~248, no.~2,
  pp.~225--229, 1998.

\bibitem{Qiang2009}
B.-B. L{\"u} and T.~Qiang, ``Discrete gap breathers in a two-dimensional
  diatomic face-centered square lattice,'' {\em Chinese Physics B}, vol.~18,
  no.~10, p.~4393, 2009.

\bibitem{Koukouloyannis2009}
V.~Koukouloyannis and I.~Kourakis, ``Discrete breathers in hexagonal dusty
  plasma lattices,'' {\em Phys. Rev. E}, vol.~80, p.~026402, 2009.

\bibitem{2Dreview2007}
B.-F. Feng and T.~Kawahara, ``Discrete breathers in two-dimensional nonlinear
  lattices,'' {\em Wave Motion}, vol.~45, no.~1, pp.~68 -- 82, 2007.

\bibitem{Mara}
A.~A. Maradudin, E.~W. Montroll, and G.~H. Weiss, {\em Theory of Lattice
  Dynamics in the Harmonic Approximation}.
\newblock New York City, NY, USA: Academic Press, 1963.

\bibitem{Theocharis2009}
G.~Theocharis, M.~Kavousanakis, P.~G. Kevrekidis, C.~Daraio, M.~A. Porter, and
  I.~G. Kevrekidis, ``Localized breathing modes in granular crystals with
  defects,'' {\em Phys. Rev. E}, vol.~80, p.~066601, 2009.

\bibitem{Nature11}
N.~Boechler, G.~Theocharis, and C.~Daraio, ``Bifurcation based acoustic
  switching and rectification,'' {\em Nat. Mater.}, vol.~10, no.~9, p.~665,
  2011.

\bibitem{hooge12}
C.~Hoogeboom, Y.~Man, N.~Boechler, G.~Theocharis, P.~G. Kevrekidis, I.~G.
  Kevrekidis, and C.~Daraio, ``Hysteresis loops and multi-stability: {F}rom
  periodic orbits to chaotic dynamics (and back) in diatomic granular
  crystals,'' {\em Euro. Phys. Lett.}, vol.~101, p.~44003, 2013.

\bibitem{cantilevers}
C.~Chong, A.~Foehr, E.~G. Charalampidis, P.~G. Kevrekidis, and C.~Daraio,
  ``Breathers and other time-periodic solutions in an array of cantilevers
  decorated with magnets,'' {\em Math. Eng.}, vol.~1, pp.~489--507, 2019.

\bibitem{magBreathers}
M.~Moler\'on, C.~Chong, A.~J. Mart\'inez, M.~A. Porter, P.~G. Kevrekidis, and
  C.~Daraio, ``Nonlinear excitations in magnetic lattices with long-range
  interactions,'' {\em New J. Phys.}, vol.~21, p.~063032, 2019.

\bibitem{Flach_long}
S.~Flach, ``Breathers on lattices with long range interaction,'' {\em Phys.
  Rev. E}, vol.~58, pp.~R4116--R4119, 1998.

\bibitem{SteidelBook}
R.~F. Steidel, {\em An Introduction to Mechanical Vibrations}.
\newblock New York City, NY, USA: John Wiley \& Sons, Inc., 1989.

\bibitem{nsoli}
C.~T. Kelley, {\em Solving Nonlinear Equations with Newton's Method}.
\newblock Philadelphia, PA, USA: Society for Industrial and Applied
  Mathematics, 2003.

\bibitem{Doedel}
E.~Doedel and L.~S. Tuckerman, {\em Numerical Methods for Bifurcation Problems
  and Large-Scale Dynamical Systems}.
\newblock Heidelberg, Germany: Springer-Verlag, 2000.

\bibitem{HirschSmaleDevaney}
M.~W. Hirsch, S.~Smale, and R.~L. Devaney, {\em Differential Equations,
  Dynamical Systems, and an Introduction to Chaos}.
\newblock Amsterdam, The Netherlands: Elsevier, 2004.

\bibitem{Boris2Dtail}
B.~A. Malomed, ``Potential of interaction between two- and three-dimensional
  solitons,'' {\em Phys. Rev. E}, vol.~58, pp.~7928--7933, Dec 1998.

\bibitem{modulated2Dtail}
R.~J. Decker, A.~Demirkaya, N.~S. Manton, and P.~G. Kevrekidis,
  ``Kink--antikink interaction forces and bound states in a biharmonic $\phi^4$
  model,'' {\em arXiv:2005.04523}, 2020.

\bibitem{rosas2007}
A.~Rosas, A.~H. Romero, V.~F. Nesterenko, and K.~Lindenberg, ``Observation of
  two-wave structure in strongly nonlinear dissipative granular chains,'' {\em
  Phys. Rev. Lett.}, vol.~98, p.~164301, 2007.

\bibitem{vergara}
L.~Vergara, ``Model for dissipative highly nonlinear waves in dry granular
  systems,'' {\em Phys. Rev. Lett.}, vol.~104, p.~118001, 2010.

\bibitem{khatriprl}
R.~Carretero-Gonz\'alez, D.~Khatri, M.~A. Porter, P.~G. Kevrekidis, and
  C.~Daraio, ``Dissipative solitary waves in granular crystals,'' {\em Phys.
  Rev. Lett.}, vol.~102, p.~024102, 2009.

\bibitem{merkel}
A.~Merkel, V.~Tournat, and V.~Gusev, ``Experimental evidence of rotational
  elastic waves in granular phononic crystals,'' {\em Phys. Rev. Lett.},
  vol.~107, p.~225502, 2011.

\bibitem{yuli}
K.~Vorotnikov, M.~Kovaleva, and Y.~Starosvetsky, ``Emergence of non-stationary
  regimes in one- and two-dimensional models with internal rotators,'' {\em
  Phil. Trans. A: Math. Phys. Eng. Sci.}, vol.~376, p.~20170134, 2018.

\bibitem{jackson_classical_1999}
J.~D. Jackson, {\em Classical Electrodynamics}.
\newblock New York City, NY, USA: John Wiley \& Sons, Inc., 3rd ed.~ed., 1999.

\end{thebibliography}
\bibliographystyle{ieeetr}

%%%%%%

\end{document}